\pdfoutput=1

\documentclass[12pt,a4paper]{article}
\usepackage{ifthen} 
\newboolean{pdflatex}
\setboolean{pdflatex}{true} 

\newboolean{articletitles}
\setboolean{articletitles}{true} 

\newboolean{uprightparticles}
\setboolean{uprightparticles}{false} 

\newboolean{inbibliography}
\setboolean{inbibliography}{false} 

\def\paperauthors{LHCb collaboration} 
\def\paperasciititle{Studies of the resonance structure in D to K pi pi pi decays  } 
\def\papertitle{Studies of the resonance structure in $D^{0} \to K^\mp \pi^\pm \pi^\pm \pi^\mp$ decays}
\def\paperkeywords{{High Energy Physics}, {LHCb}} 
\def\papercopyright{CERN on behalf of the LHCb collaboration}
\def\paperlicence{CC-BY-4.0}
\def\paperlicenceurl{https://creativecommons.org/licenses/by/4.0/}

\usepackage[top=1in, bottom=1.25in, left=1in, right=1in]{geometry}

\usepackage{microtype}
\usepackage{lineno}  
\usepackage{xspace} 
\usepackage{caption} 

\usepackage{graphicx}  
\usepackage{color}
\usepackage{colortbl}
\graphicspath{{./figs/}} 

\usepackage{amsmath} 
\usepackage{amssymb}
\usepackage{amsfonts}
\usepackage{upgreek} 

\newcommand*\patchAmsMathEnvironmentForLineno[1]{%
\expandafter\let\csname old#1\expandafter\endcsname\csname #1\endcsname
\expandafter\let\csname oldend#1\expandafter\endcsname\csname
end#1\endcsname
 \renewenvironment{#1}%
   {\linenomath\csname old#1\endcsname}%
   {\csname oldend#1\endcsname\endlinenomath}%
}
\newcommand*\patchBothAmsMathEnvironmentsForLineno[1]{%
  \patchAmsMathEnvironmentForLineno{#1}%
  \patchAmsMathEnvironmentForLineno{#1*}%
}
\AtBeginDocument{%
\patchBothAmsMathEnvironmentsForLineno{equation}%
\patchBothAmsMathEnvironmentsForLineno{align}%
\patchBothAmsMathEnvironmentsForLineno{flalign}%
\patchBothAmsMathEnvironmentsForLineno{alignat}%
\patchBothAmsMathEnvironmentsForLineno{gather}%
\patchBothAmsMathEnvironmentsForLineno{multline}%
\patchBothAmsMathEnvironmentsForLineno{eqnarray}%
}

\usepackage{hyperxmp}

\usepackage[pdftex,
            pdfauthor={\paperauthors},
            pdftitle={\paperasciititle},
            pdfkeywords={\paperkeywords},
            pdfcopyright={Copyright (C) \papercopyright},
            pdflicenseurl={\paperlicenceurl}]{hyperref}

\usepackage[all]{hypcap} 

\usepackage{xspace} 
\usepackage{upgreek}

\def\lhcb {\mbox{LHCb}\xspace}

\def\cleo   {\mbox{CLEO}\xspace}

\def\MagUp {\mbox{\em Mag\kern -0.05em Up}\xspace}

\ifthenelse{\boolean{uprightparticles}}%
{

 \def\Pnu         {\ensuremath{\upnu}\xspace}                 
                  
 \def\Ppi         {\ensuremath{\uppi}\xspace}

 \def\PDelta      {\ensuremath{\Delta}\xspace}                 
 \def\PXi      {\ensuremath{\Xi}\xspace}                 
 \def\PLambda      {\ensuremath{\Lambda}\xspace}                 
 \def\PSigma      {\ensuremath{\Sigma}\xspace}                 
 \def\POmega      {\ensuremath{\Omega}\xspace}                 
 \def\PUpsilon      {\ensuremath{\Upsilon}\xspace}                 
 
 \def\PB      {\ensuremath{\mathrm{B}}\xspace}                 
                  
 \def\PD      {\ensuremath{\mathrm{D}}\xspace}

 \def\PK      {\ensuremath{\mathrm{K}}\xspace}

 \def\Pb      {\ensuremath{\mathrm{b}}\xspace}                 
 \def\Pc      {\ensuremath{\mathrm{c}}\xspace}                 
 \def\Pd      {\ensuremath{\mathrm{d}}\xspace}

 \def\Pi      {\ensuremath{\mathrm{i}}\xspace}

 \def\Pp      {\ensuremath{\mathrm{p}}\xspace}

 \def\Pu      {\ensuremath{\mathrm{u}}\xspace}

}
{

 \def\Pnu         {\ensuremath{\nu}\xspace}                 
                  
 \def\Ppi         {\ensuremath{\pi}\xspace}

 \mathchardef\PDelta="7101
 \mathchardef\PXi="7104
 \mathchardef\PLambda="7103
 \mathchardef\PSigma="7106
 \mathchardef\POmega="710A
 \mathchardef\PUpsilon="7107
                  
 \def\PB      {\ensuremath{B}\xspace}                 
                  
 \def\PD      {\ensuremath{D}\xspace}

 \def\PK      {\ensuremath{K}\xspace}

 \def\Pb      {\ensuremath{b}\xspace}                 
 \def\Pc      {\ensuremath{c}\xspace}                 
 \def\Pd      {\ensuremath{d}\xspace}

 \def\Pi      {\ensuremath{i}\xspace}

 \def\Pp      {\ensuremath{p}\xspace}

 \def\Pu      {\ensuremath{u}\xspace}

}

\makeatletter
\ifcase \@ptsize \relax
  \newcommand{\miniscule}{\@setfontsize\miniscule{4}{5}}
\or
  \newcommand{\miniscule}{\@setfontsize\miniscule{5}{6}}
\or
  \newcommand{\miniscule}{\@setfontsize\miniscule{5}{6}}
\fi
\makeatother

\DeclareRobustCommand{\optbar}[1]{\shortstack{{\miniscule (\rule[.5ex]{1.25em}{.18mm})}
  \\ [-.7ex] $#1$}}

\def\neub       {{\ensuremath{\overline{\Pnu}}}\xspace}

\def\neutb      {{\ensuremath{\neub_\tau}}\xspace}

\def\uquark    {{\ensuremath{\Pu}}\xspace}

\def\dquark    {{\ensuremath{\Pd}}\xspace}

\def\cquark    {{\ensuremath{\Pc}}\xspace}

\def\bquark    {{\ensuremath{\Pb}}\xspace}

\def\pion   {{\ensuremath{\Ppi}}\xspace}

\def\pip    {{\ensuremath{\pion^+}}\xspace}
\def\pim    {{\ensuremath{\pion^-}}\xspace}
\def\pipm   {{\ensuremath{\pion^\pm}}\xspace}
\def\pimp   {{\ensuremath{\pion^\mp}}\xspace}

\def\kaon    {{\ensuremath{\PK}}\xspace}
  \def\Kbar    {{\kern 0.2em\overline{\kern -0.2em \PK}{}}\xspace}

\def\KorKbar    {\kern 0.18em\optbar{\kern -0.18em K}{}\xspace}

\def\Kp      {{\ensuremath{\kaon^+}}\xspace}
\def\Km      {{\ensuremath{\kaon^-}}\xspace}
\def\Kpm     {{\ensuremath{\kaon^\pm}}\xspace}
\def\Kmp     {{\ensuremath{\kaon^\mp}}\xspace}

\def\Kstar   {{\ensuremath{\kaon^*}}\xspace}
\def\Kstarb  {{\ensuremath{\Kbar{}^*}}\xspace}

\def\Dbar    {{\kern 0.2em\overline{\kern -0.2em \PD}{}}\xspace}
\def\D       {{\ensuremath{\PD}}\xspace}

\def\DorDbar    {\kern 0.18em\optbar{\kern -0.18em D}{}\xspace}
\def\Dz      {{\ensuremath{\D^0}}\xspace}
\def\Dzb     {{\ensuremath{\Dbar{}^0}}\xspace}

\def\Dit       {{\ensuremath{D}}\xspace}
\def\Dbit    {{\kern 0.2em\overline{\kern -0.2em \Dit}{}}\xspace}

\def\Dp      {{\ensuremath{\D^+}}\xspace}

\def\Dstar   {{\ensuremath{\D^*}}\xspace}

\def\Dstarp  {{\ensuremath{\D^{*+}}}\xspace}

\def\B       {{\ensuremath{\PB}}\xspace}
\def\Bbar    {{\ensuremath{\kern 0.18em\overline{\kern -0.18em \PB}{}}}\xspace}
\def\Bb      {{\ensuremath{\Bbar}}\xspace}
\def\BorBbar    {\kern 0.18em\optbar{\kern -0.18em B}{}\xspace}

\def\Bub     {{\ensuremath{\B^-}}\xspace}

\def\Bm      {{\ensuremath{\Bub}}\xspace}

  \def\Y#1S{\ensuremath{\PUpsilon{(#1S)}}\xspace}

\def\proton      {{\ensuremath{\Pp}}\xspace}

\def\Lbar        {{\ensuremath{\kern 0.1em\overline{\kern -0.1em\PLambda}}}\xspace}
\def\LorLbar    {\kern 0.18em\optbar{\kern -0.18em \PLambda}{}\xspace}

\def\to                 {\ensuremath{\rightarrow}\xspace}

\def\grpsuthree {{\ensuremath{\mathrm{SU}(3)}}\xspace}

\def\CP                {{\ensuremath{C\!P}}\xspace}

\def\Vud  {{\ensuremath{V_{\uquark\dquark}}}\xspace}
\def\Vcd  {{\ensuremath{V_{\cquark\dquark}}}\xspace}

\def\Vubs  {{\ensuremath{V_{\uquark\bquark}^\ast}}\xspace}
\def\Vcbs  {{\ensuremath{V_{\cquark\bquark}^\ast}}\xspace}

\def\AT#1     {\ensuremath{A_{\mathrm{T}}^{#1}}\xspace}           

\def\C#1      {\ensuremath{\mathcal{C}_{#1}}\xspace}                       
\def\Cp#1     {\ensuremath{\mathcal{C}_{#1}^{'}}\xspace}                    
\def\Ceff#1   {\ensuremath{\mathcal{C}_{#1}^{\mathrm{(eff)}}}\xspace}        
\def\Cpeff#1  {\ensuremath{\mathcal{C}_{#1}^{'\mathrm{(eff)}}}\xspace}       
\def\Ope#1    {\ensuremath{\mathcal{O}_{#1}}\xspace}                       
\def\Opep#1   {\ensuremath{\mathcal{O}_{#1}^{'}}\xspace}                    

\newcommand{\ket}[1]{\ensuremath{|#1\rangle}}              

\newcommand{\tev}{\ifthenelse{\boolean{inbibliography}}{\ensuremath{~T\kern -0.05em eV}\xspace}{\ensuremath{\mathrm{\,Te\kern -0.1em V}}}\xspace}
\newcommand{\gev}{\ensuremath{\mathrm{\,Ge\kern -0.1em V}}\xspace}
\newcommand{\mev}{\ensuremath{\mathrm{\,Me\kern -0.1em V}}\xspace}
\newcommand{\kev}{\ensuremath{\mathrm{\,ke\kern -0.1em V}}\xspace}
\newcommand{\ev}{\ensuremath{\mathrm{\,e\kern -0.1em V}}\xspace}
\newcommand{\gevc}{\ensuremath{{\mathrm{\,Ge\kern -0.1em V\!/}c}}\xspace}
\newcommand{\mevc}{\ensuremath{{\mathrm{\,Me\kern -0.1em V\!/}c}}\xspace}
\newcommand{\gevcc}{\ensuremath{{\mathrm{\,Ge\kern -0.1em V\!/}c^2}}\xspace}
\newcommand{\gevgevcccc}{\ensuremath{{\mathrm{\,Ge\kern -0.1em V^2\!/}c^4}}\xspace}
\newcommand{\mevcc}{\ensuremath{{\mathrm{\,Me\kern -0.1em V\!/}c^2}}\xspace}

\def\mum  {\ensuremath{{\,\upmu\mathrm{m}}}\xspace}

\def\invfb   {\ensuremath{\mbox{\,fb}^{-1}}\xspace}

\def\gsim{{~\raise.15em\hbox{$>$}\kern-.85em
          \lower.35em\hbox{$\sim$}~}\xspace}
\def\lsim{{~\raise.15em\hbox{$<$}\kern-.85em
          \lower.35em\hbox{$\sim$}~}\xspace}

\def\ptot       {\mbox{$p$}\xspace}
\def\pt         {\mbox{$p_{\mathrm{ T}}$}\xspace}

\def\degrees{\ensuremath{^{\circ}}\xspace}

\def\evtgen     {\mbox{\textsc{EvtGen}}\xspace}

\def\geant      {\mbox{\textsc{Geant4}}\xspace}

\def\pythia     {\mbox{\textsc{Pythia}}\xspace}

\def\tell1  {TELL1\xspace}
\def\ukl1   {UKL1\xspace}

\newcommand{\vs}{\mbox{\itshape vs.}\xspace}

\usepackage{cite} 
\usepackage{mciteplus}

\usepackage{longtable} 
\usepackage[toc,page]{appendix}

\def\Eta    {\ensuremath{\eta}}
\def\Etapr  {\ensuremath{\eta^{\prime}}}
\def\bdt    {BDT\xspace}
\def\KPI    {\PK\pi\pi\pi}
\def\xp     {\mathbf{x}}

\def\RS {\ensuremath{\Dz\rightarrow\Km\pip\pip\pim}\xspace}
\def\WS {\ensuremath{\Dz\rightarrow\Kp\pim\pim\pip}\xspace}
\def\CF {\Km\pip\pip\pim}
\def\DCS {\Kp\pim\pim\pip}
\def\dof {degree of freedom\xspace}

\newcommand{\KONE}[1] { \ensuremath{K_{1}(#1)} }
\def\AONE { \ensuremath{a_{1}(1260)^{+}} }
\def\Kexc {\ensuremath{K(1460)^{-}}\xspace}

\usepackage{gensymb}
\usepackage{menukeys}
\usepackage{longtable} 
\usepackage{booktabs,siunitx}
\usepackage{multirow}
\usepackage{array}
\usepackage{placeins}

\usepackage{collcell}

\begin{document}

\renewcommand{\thefootnote}{\fnsymbol{footnote}}
\setcounter{footnote}{1}


\begin{titlepage}
\pagenumbering{roman}

\vspace*{-1.5cm}
\centerline{\large EUROPEAN ORGANIZATION FOR NUCLEAR RESEARCH (CERN)}
\vspace*{1.5cm}
\noindent
\begin{tabular*}{\linewidth}{lc@{\extracolsep{\fill}}r@{\extracolsep{0pt}}}
\ifthenelse{\boolean{pdflatex}}
{\vspace*{-1.5cm}\mbox{\!\!\!\includegraphics[width=.14\textwidth]{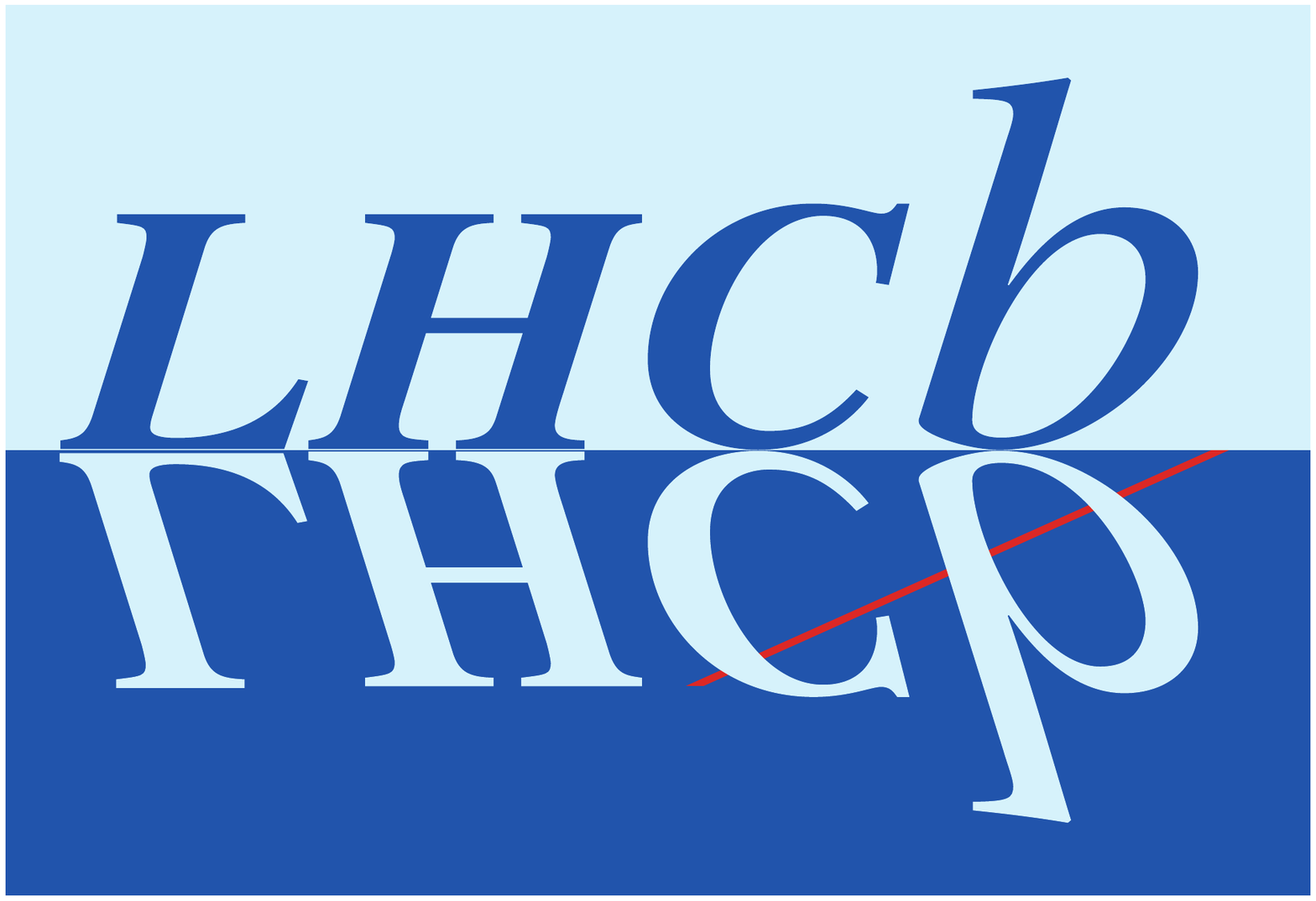}} & &}%
{\vspace*{-1.2cm}\mbox{\!\!\!\includegraphics[width=.12\textwidth]{lhcb-logo.eps}} & &}%
\\
 & & CERN-EP-2017-314 \\  
 & & LHCb-PAPER-2017-040 \\  
 & & 22 December 2017 \\ 
 & & \\
\end{tabular*}

\vspace*{2.25cm}

{\normalfont\bfseries\boldmath\huge
\begin{center}
  \papertitle 
\end{center}
}

\vspace*{1.5cm}

\begin{center}
\paperauthors\footnote{Authors are listed at the end of this paper.}
\end{center}

\vspace{\fill}

\begin{abstract}
  \noindent
  Amplitude models are constructed to describe the resonance structure of ${D^{0}\to K^{-}\pi^{+}\pi^{+}\pi^{-}}$ and ${D^{0} \to K^{+}\pi^{-}\pi^{-}\pi^{+}}$ decays using $pp$ collision data 
  collected at centre-of-mass energies of 7 and 8 Te{\kern -0.1em V} with the LHCb experiment,
  corresponding to an integrated luminosity of $3.0\mbox{\,fb}^{-1}$.
  The largest contributions to both decay amplitudes are found to come from axial resonances, with decay modes $D^{0} \to a_1(1260)^{+} K^{-}$ and $D^{0} \to K_1(1270/1400)^{+} \pi^{-}$ being prominent in ${D^{0}\to K^{-}\pi^{+}\pi^{+}\pi^{-}}$ and $D^{0}\to K^{+}\pi^{-}\pi^{-}\pi^{+}$, respectively. 
  Precise measurements of the lineshape parameters and couplings of the $a_1(1260)^{+}$, $K_1(1270)^{-}$ and $K(1460)^{-}$ resonances are made, and a quasi model-independent study of the $K(1460)^{-}$ resonance is performed.
  The coherence factor of the decays is calculated from the amplitude models to be $R_{K3\pi} = 0.459\pm 0.010\,(\mathrm{stat}) \pm 0.012\,(\mathrm{syst}) \pm 0.020\,(\mathrm{model})$, which is consistent with direct measurements.
 These models will be useful in future measurements of the unitary-triangle angle $\gamma$ and studies of charm mixing and $C\!P$ violation. 

\end{abstract}

\vspace*{1.75cm}

\begin{center}
  Published in Eur.~Phys.~J.~C \textbf{78} (2018) 443. 
\end{center}

\vspace{\fill}

{\footnotesize 
\centerline{\copyright~\papercopyright, licence \href{\paperlicenceurl}{\paperlicence}.}}
\vspace*{2mm}

\end{titlepage}


\newpage
\setcounter{page}{2}
\mbox{~}

\cleardoublepage

\renewcommand{\thefootnote}{\arabic{footnote}}
\setcounter{footnote}{0}

\pagestyle{plain} 
\setcounter{page}{1}
\pagenumbering{arabic}


\section{\label{sec:Introduction}Introduction}                     The decays\footnote{The inclusion of charge-conjugate processes is implied throughout.} $\Dz \to \Km \pip \pip \pim$ and  $\Dz \to \Kp \pim \pip \pim$ have an important role to play in improving knowledge of the Cabibbo-Kobayashi-Maskawa (CKM) unitarity-triangle angle $\gamma \equiv \arg(-\Vud\Vubs/\Vcd\Vcbs)$.  Sensitivity to this parameter can be obtained by measuring \CP-violating and associated observables in the decay $\Bm \to D \Km$, where $D$ indicates a neutral charm meson reconstructed in final states common to both $\Dz$ and $\Dzb$, of which $\Kmp \pipm \pipm \pimp$ are significant examples~\cite{Atwood:1996ci,Atwood:2000ck}. 
A straightforward approach to such an analysis is to reconstruct the four-body $D$-meson decays inclusively, 
which was performed by the \lhcb collaboration in a recent measurement~\cite{LHCb-PAPER-2016-003}.
Alternatively, additional sensitivity can be sought by studying the variation of the observables across the phase space of the $D$ decays, a strategy that requires knowledge of the variation of the decay amplitudes of the charm mesons.

Studies of charm mixing and searches for \CP violation in the \Dz-\Dzb system, which for these final states have only been performed inclusively~\cite{LHCb-PAPER-2015-057}, will also benefit from an understanding of the variation of the decay amplitudes across their phase space.
These decay modes are also a rich laboratory for examining the behaviour of the strong interaction at low energy, through studies of 
the intermediate resonances that contribute to the final states. All these considerations motivate an amplitude analysis of the two decays.

The decay $\Dz \to \Km \pip \pip\pim$ has a branching ratio of $(8.29 \pm 0.20)\%$~\cite{PhysRevD.89.072002}, which is the highest of all $\Dz$ decay modes involving only charged particles, and is predominantly mediated by Cabibbo-favoured (CF) transitions. 
The decay $\Dz \to \Kp \pim \pim\pip$ is dominated by doubly Cabibbo-suppressed (DCS) amplitudes,  with small contributions from mixing-related effects, and occurs at a rate that is suppressed by a factor of $(3.22 \pm 0.05) \times 10^{-3}$~\cite{LHCb-PAPER-2015-057} compared to that of the favoured mode. 
The favoured and suppressed modes are here termed the `right-sign' (RS) and `wrong-sign' (WS) decay, respectively, 
on account of the charge correlation between the kaon and the particle used to tag the flavour of the parent meson.

In this paper, time-integrated amplitude models of both decay modes are constructed using $pp$ collision data collected at centre-of-mass energies of 7 and 8\tev with the LHCb experiment, corresponding to an integrated luminosity of 3.0\invfb.
 The RS sample size is around 700 times larger than the data set used by the Mark III collaboration to develop the first amplitude model of this decay~\cite{PhysRevD.45.2196}. An amplitude analysis has also been performed on the RS decay by the BES III collaboration~\cite{Ablikim:2017eqz} with around 1.6\% of the sample size used in this analysis. 
This paper reports the first amplitude analysis of the WS decay.

The paper is organised as follows. In Sect.~\ref{sec:detector} the detector, data and simulation samples are described, and in Sect.~\ref{sec:selection} the signal selection is discussed. The amplitude-model formalism is presented in Sect.~\ref{sec:formalism}, and the fit method and model-building procedure in Sect.~\ref{sec:fitModel}. Section~\ref{sec:results} contains the fit results and conclusions are drawn in Sect.~\ref{sec:conclusions}.

\section{\label{sec:detector}Detector and simulation}              The \lhcb detector~\cite{LHCb-DP-2014-002} is a single-arm forward
spectrometer covering the \mbox{pseudorapidity} range $2<\eta <5$,
designed for the study of particles containing \bquark or \cquark
quarks. The detector includes a high-precision tracking system
consisting of a silicon-strip vertex detector surrounding the $pp$
interaction region, a large-area silicon-strip detector located
upstream of a dipole magnet with a bending power of about
$4{\mathrm{\,Tm}}$, and three stations of silicon-strip detectors and straw
drift tubes placed downstream of the magnet.
The tracking system provides a measurement of momentum, \ptot, of charged particles with
a relative uncertainty that varies from 0.5\% at low momentum to 1.0\% at 200\gevc.
The minimum distance of a track to a primary vertex (PV), the impact parameter, 
is measured with a resolution of $(15+29/\pt)\mum$,
where \pt is the component of the momentum transverse to the beam, in\,\gevc.
Different types of charged hadrons are distinguished using information
from two ring-imaging Cherenkov (RICH) detectors. 
Photons, electrons and hadrons are identified by a calorimeter system consisting of
scintillating-pad and preshower detectors, an electromagnetic
calorimeter and a hadronic calorimeter. Muons are identified by a
system composed of alternating layers of iron and multiwire
proportional chambers.

The trigger~\cite{LHCb-DP-2012-004} consists of a
hardware stage, based on information from the calorimeter and muon
systems, followed by a software stage, in which all charged particles
with $\pt>500\,(300)\mevc$ are reconstructed for 2011\,(2012) data.
At the hardware trigger stage, events are required to have a muon with high \pt or a
  hadron, photon or electron with high transverse energy in the calorimeters. 
  The software trigger requires a two-, three- or four-track
  secondary vertex with a significant displacement from the primary
  $pp$ interaction vertices. At least one charged particle
  must have a transverse momentum $\pt > 1.7(1.6)\gevc$ and be
  inconsistent with originating from a PV.
  A multivariate algorithm~\cite{BBDT} is used for
  the identification of secondary vertices consistent with the decay
  of a \bquark hadron.

In the simulation, $pp$ collisions are generated using
\pythia~\cite{Sjostrand:2007gs} 
 with a specific \lhcb
configuration~\cite{LHCb-PROC-2010-056}. Particle decays are described by \evtgen~\cite{Lange:2001uf}. 
The interaction of the generated particles with the detector, and its response,
are implemented using the \geant
toolkit~\cite{Allison:2006ve, *Agostinelli:2002hh} as described in
Ref.~\cite{LHCb-PROC-2011-006}.

\section{\label{sec:selection}Signal selection and backgrounds}    The decay chain $\Bbar \rightarrow \Dstar(2010)^{+} \mu^{-} X$ with $\Dstar(2010)^{+} \rightarrow \Dz \pi_{\mathrm{slow}}^{+}$ is reconstructed as a clean source of $\Dz$ mesons for analysis. 
The $\Dz$ mesons are reconstructed in the $\Kmp\pipm\pipm\pimp$ final states.
The charged pion, $\pi_{\mathrm{slow}}^{+}$, originating from the $\Dstar(2010)^{+}$ is referred to as `slow' due to the small $Q$-value of the decay.
The charge of the muon and slow pion are used to infer the flavour of the neutral $D$ meson. 
Candidates are only accepted if these charges 
lead to a consistent hypothesis for the flavour of the neutral $D$ meson. 
All other aspects of the reconstruction and selection criteria are identical between the RS and WS samples. 

The two-dimensional plane $m_{\PK\pi\pi\pi}$ vs. $\Delta m$, where $m_{\PK\pi\pi\pi}$ is the invariant mass of the $\Dz$ meson candidate, and $\Delta m = m_{\PK\pi\pi\pi\pi_{\mathrm{slow}}} - m_{\PK\pi\pi\pi}$ is mass difference between the $\D^*(2010)^{+}$ and $\Dz$ meson candidates, is used to define signal and sideband regions with which to perform the amplitude analysis and study sources of background contamination. The signal region is defined as $\pm0.75\mevcc(\pm18\mevcc)$ of the signal peak in $\Delta m (m_{\PK\pi\pi\pi})$, which corresponds to about three times the width of the peak.

It is required that the hardware trigger decision is either due to the muon candidate or is independent of the particles constituting the reconstructed decay products of the $\PB$ candidate. 
For example, a high-$p_T$ particle from the other $\PB$ meson decay in the event firing the hadron trigger. 
The software trigger decision is required to either be due to the muon candidate or a two- three- or four-track secondary secondary vertex. 

The WS sample is contaminated by a category of RS decays in which the kaon is mis-identified as a pion, and a pion as a kaon. 
To suppress this background, it is required that the kaon is well identified by the RICH detectors. 
The residual contamination from this background is removed by recalculating the mass of the \Dz candidate with the mass hypotheses of a kaon and each oppositely charged pion swapped, then vetoing candidates that fall within 12\mevcc of the nominal mass of the \Dz meson.
As the majority of particles from the PV are pions, the particle identification requirements on the kaon also reduces the background from random combinations of particles. 

Remaining background from random combinations of particles can be divided into two categories. 
Candidates where the $\Dz$ is reconstructed from a random combination of tracks are referred to as {\it combinatorial} background. 
Candidates where the $\Dz$ is correctly reconstructed but paired with an unrelated $\pi_{\mathrm{slow}}^{+}$ are referred to as {\it mistag} background. 
This latter source of background is dominated by RS decays. 
Both of these backgrounds are suppressed using a multivariate classifier based on a Boosted Decision Tree (\bdt) \cite{Breiman, Roe, AdaBoost} algorithm. 
The BDT is trained on RS data candidates from the signal region and the sidebands of the WS data, and uses 15 variables related to the quality of the reconstruction of the PV, $\PB$ and $\Dz$ decay vertices, and the consistency of tracks in the signal candidate incoming from these vertices. 
Variables pertaining to the $\Dz$ kinematics and its decay products are avoided to minimise any bias of the phase-space acceptance. 

\begin{figure}
  \includegraphics[width=0.5\textwidth]{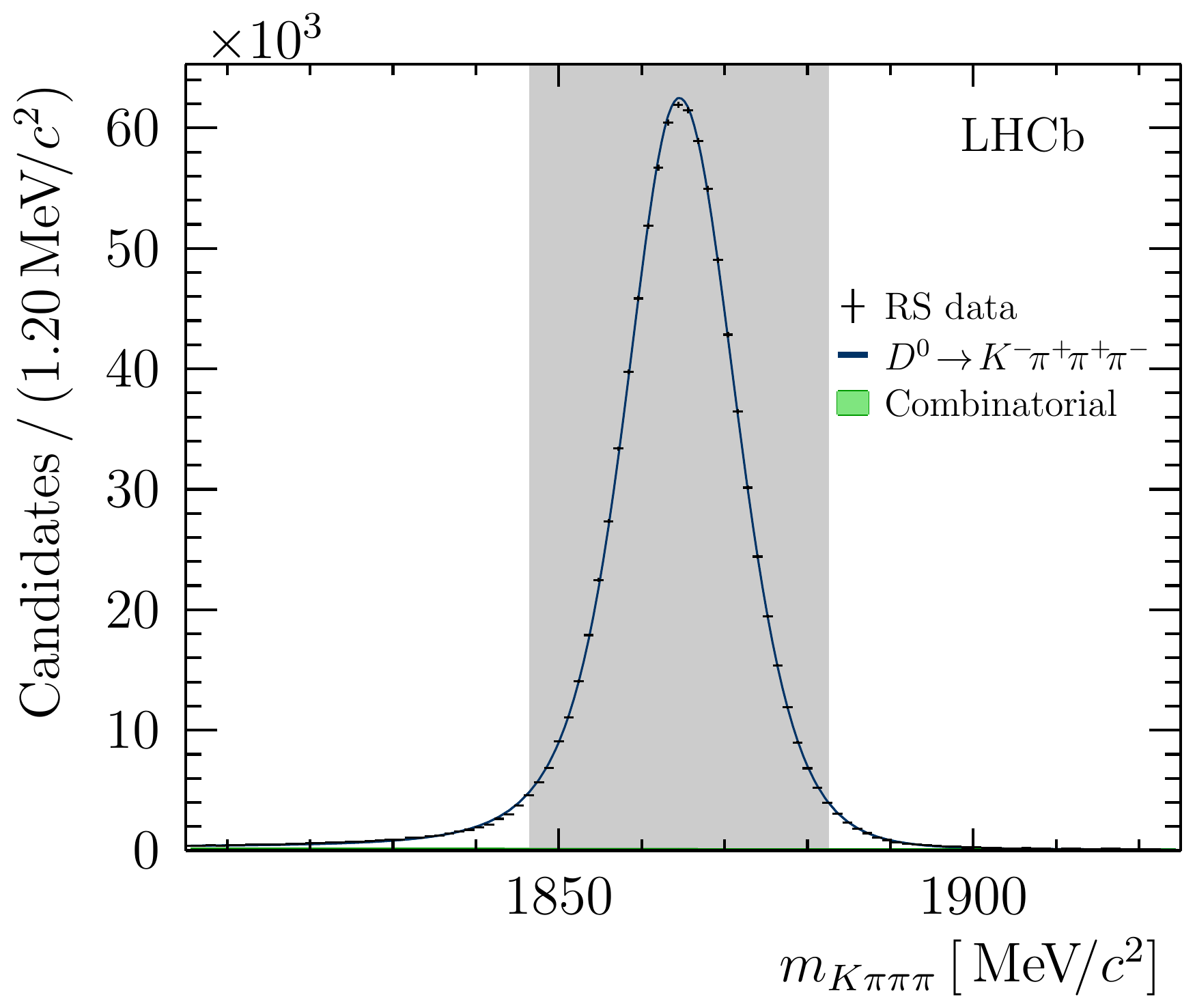}
  \includegraphics[width=0.5\textwidth]{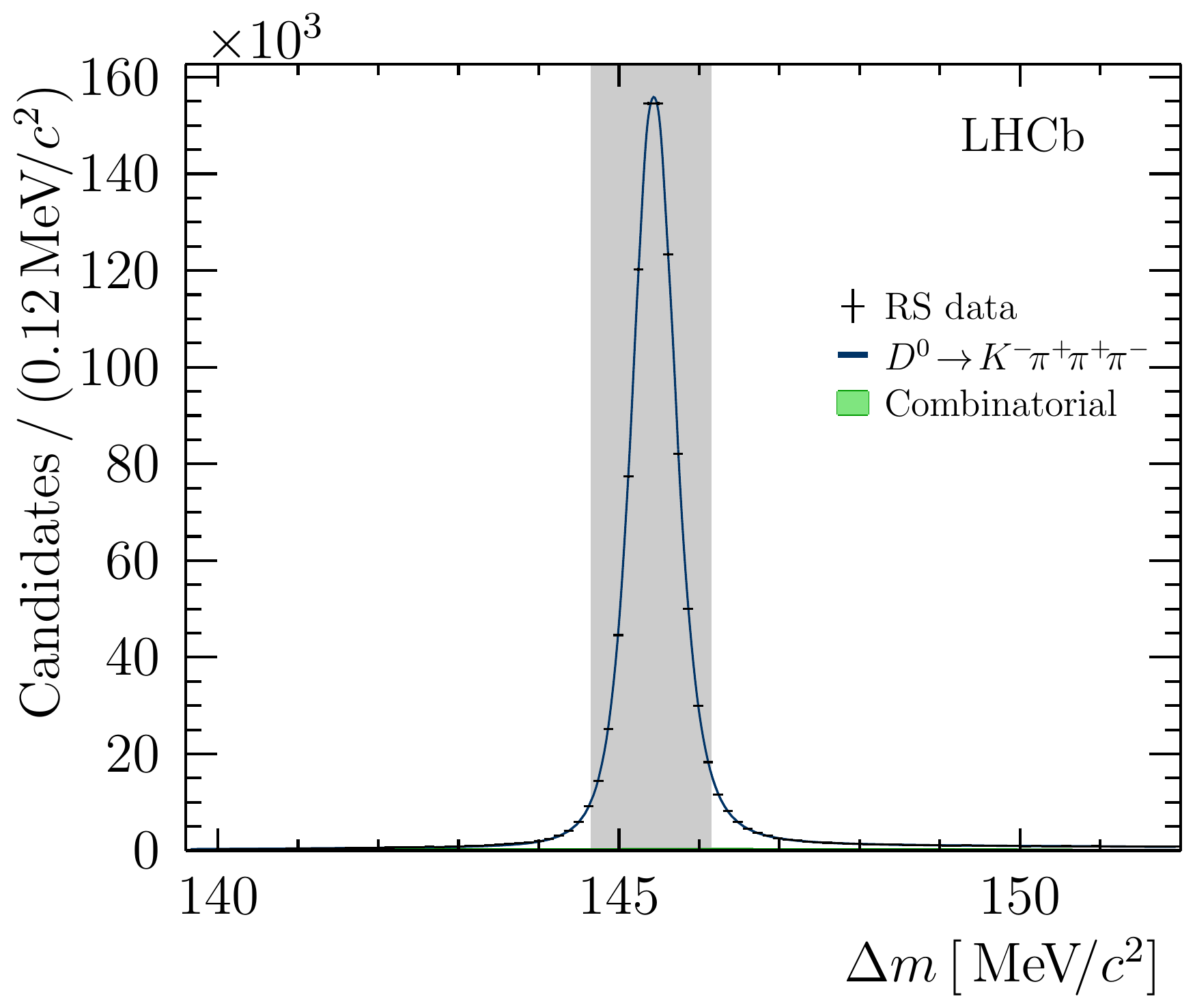}

  \includegraphics[width=0.5\textwidth]{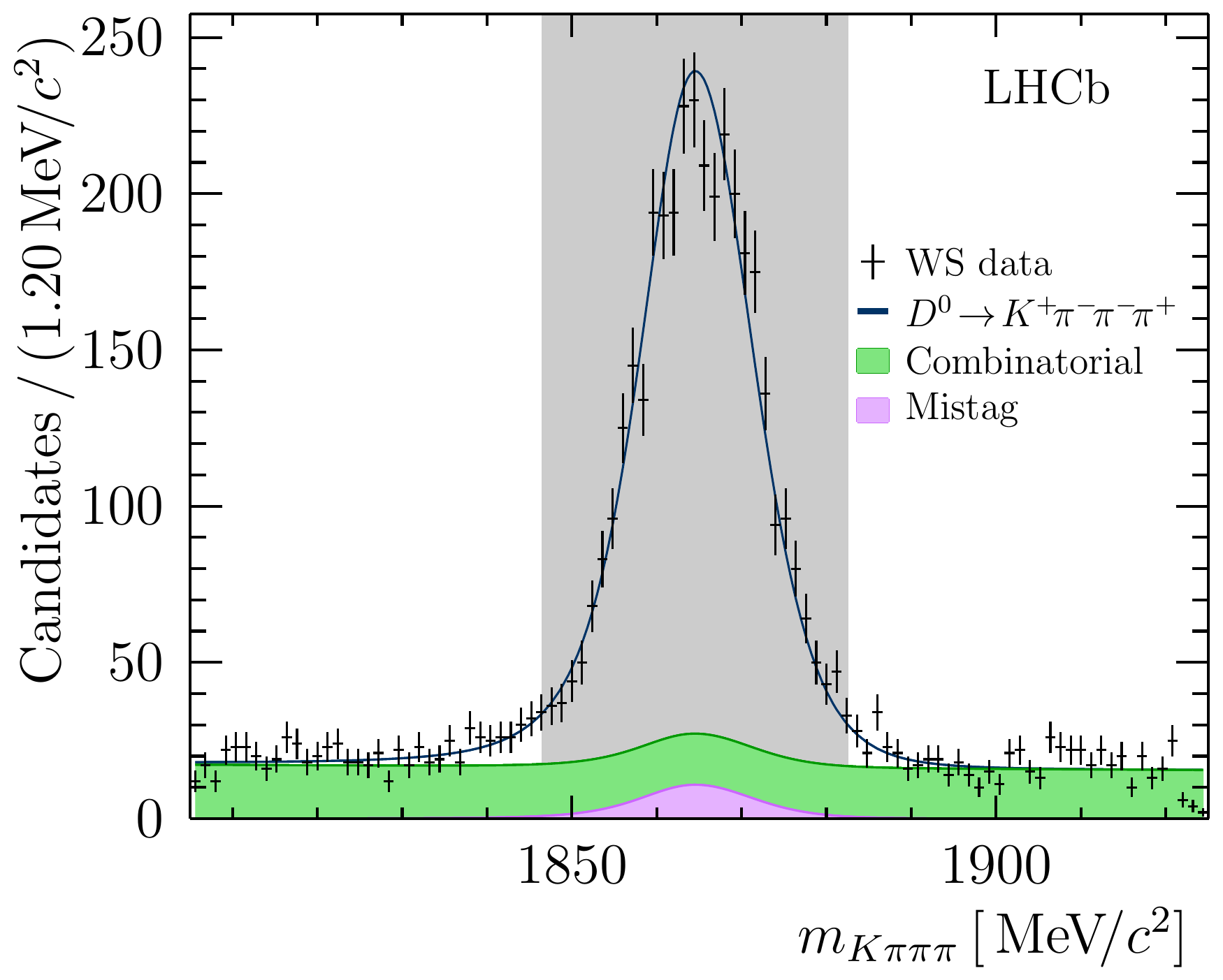}
  \includegraphics[width=0.5\textwidth]{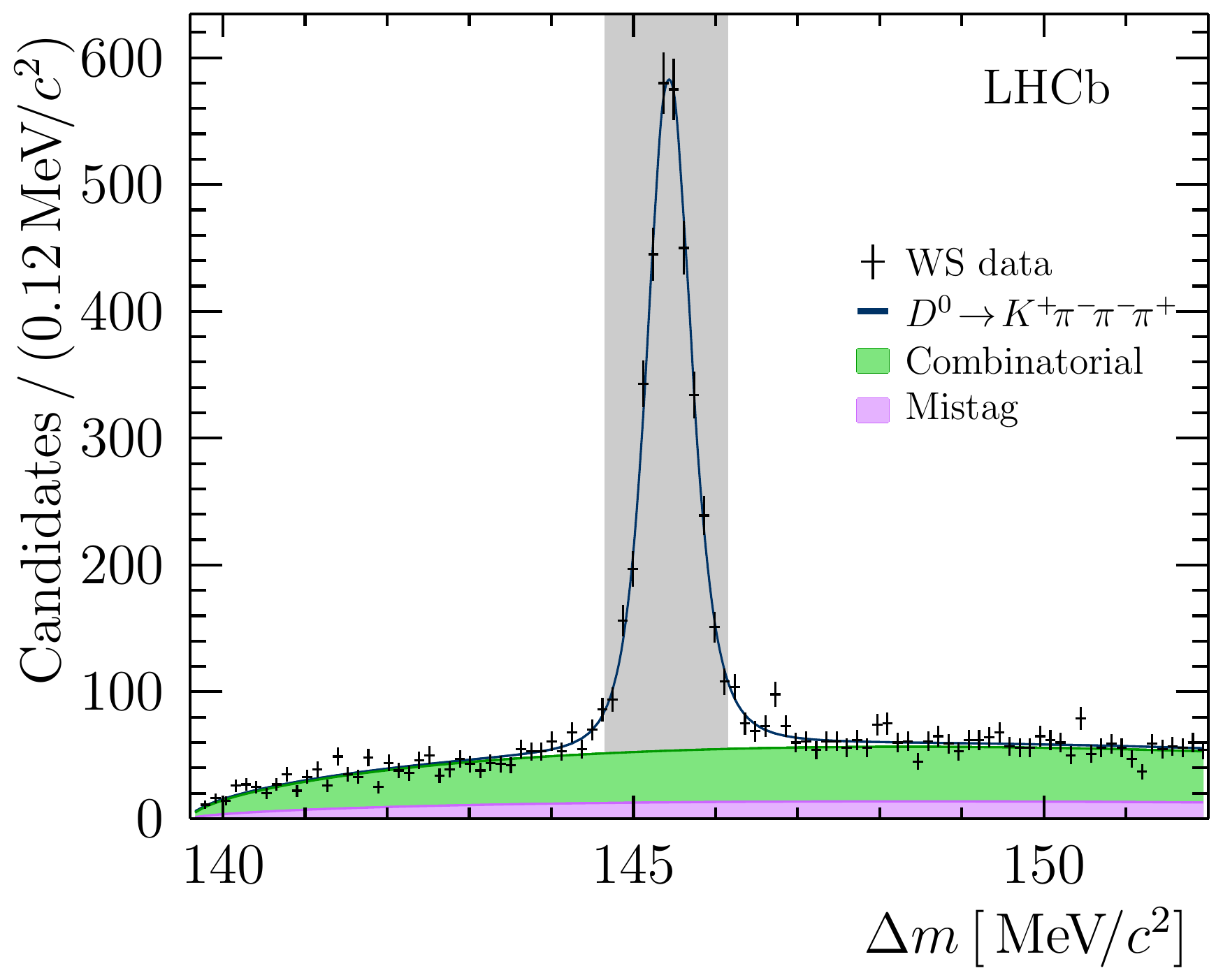}
  \caption{ 
     Invariant mass and mass difference distributions for RS (top) and WS (bottom) samples, shown with fit projections. 
    The signal region is indicated by the filled grey area, and 
    for each plot the mass window in the orthogonal projection is applied. 
  }
  \label{fig:InvariantMassFits}
\end{figure}

The signal and background yields in the signal region for each sample is determined by simultaneously fitting the two-dimensional $\Delta m$ vs. $m_{\PK\pi\pi\pi}$ distribution for both samples.
The $\Dz$, muon and slow pion candidates are constrained to originate from a common vertex in calculating the $\Dz$ and $\Dstarp$ masses. 
This requirement improves the resolution of the $\Delta m$ distribution by approximately a factor of two. 
The signal is modelled with a product of two Cruijff \cite{delAmoSanchez:2010ae} functions. The Cruijff shape parameters are shared between both samples. 
The combinatorial background is modelled by a first-order polynomial in $m_{\PK\pi\pi\pi}$, and by a threshold function in $\Delta m$, 
\begin{equation}
  \mathcal{P}(Q) \propto (1+pQ)\left(1+Q+pQ^2\right)^{a},
\end{equation}
where $Q=\Delta m - m_{\pi}$ and the parameters $p,a$ are determined by the fit. The background shape parameters, including those for the polynomial in $m_{\PK\pi\pi\pi}$, are allowed to differ between WS and RS samples. 
The mistag background component is a product of the signal shape in $m_{\PK\pi\pi\pi}$ and the combinatorial background shape in $\Delta m$.
The optimal requirement on the output of the \bdt classifier is selected by repeating the fit varying this requirement, and maximising the expected significance of the WS signal, which is defined as  
\begin{equation}
  S = \frac{\hat{N}_{\mathrm{sig}}}{\sqrt{\hat{N}_{\mathrm{sig}}+N_{\mathrm{bkg}}}},
\end{equation}
where $N_{\mathrm{bkg}}$ is the background yield in the signal region. 
The expected number of WS candidates, $\hat{N}_{\mathrm{sig}}$, is estimated by scaling the number of RS signal candidates in the signal region by the ratio of branching fractions. 
The yields of the various contributions for both samples are listed in Table \ref{tab:YieldTable}, and the $m_{\PK\pi\pi\pi}$ and $\Delta m$ distributions, with the fit projections superimposed, are shown in Fig.~\ref{fig:InvariantMassFits}.
The purities of the RS and WS samples after selection are found to be $99.6\%$ and $82.4\%$, respectively, with $4\%$ of WS candidates arising from mistagged decays. 
Studies of simulated data indicate that the selected sample has a relatively uniform acceptance across the phase space, with approximately 30\% reductions in acceptance near the edges of the kinematically allowed region. The samples also have a relatively uniform selection efficiency in decay time, being constant within $\pm10\%$ for lifetimes greater than one average lifetime of the \PD meson. 

\begin{table}
  \centering 
  \caption{Signal and background yields for both samples in the signal region, presented separately for each  year of data taking.}
  \label{tab:YieldTable}
  \begin{tabular}{l 
    >{\collectcell\num}r<{\endcollectcell} @{${}\pm{}$} >{\collectcell\num}l<{\endcollectcell}
    >{\collectcell\num}r<{\endcollectcell} @{${}\pm{}$} >{\collectcell\num}l<{\endcollectcell}
    >{\collectcell\num}r<{\endcollectcell} @{${}\pm{}$} >{\collectcell\num}l<{\endcollectcell}
    }
    \toprule
     & \multicolumn{6}{c}{Yield}\\
     \cmidrule{2-6}
      & \multicolumn{2}{c}{Signal} & \multicolumn{2}{c}{\kern-0.5em Combinatorial} & \multicolumn{2}{c}{Mistag}\\
      & \multicolumn{2}{c}{}       & \multicolumn{2}{c}{\kern-0.5em Background} & \multicolumn{2}{c}{Background}\\
    $\PD^{0}\rightarrow\PK^{-}\pi^{+}\pi^{+}\pi^{-}$ & \multicolumn{6}{c}{}\\
      \hline
      2011 & 266368 & 490 & 977 & 10 & \multicolumn{2}{c}{---}\\
      2012 & 624332 & 765 & 2475 & 19 & \multicolumn{2}{c}{---}\\
      Total & 890701 & 927 & 3452 & 24 & \multicolumn{2}{c}{---}\\
      \multicolumn{7}{c}{}\\
      $\PD^{0}\rightarrow\PK^{+}\pi^{-}\pi^{-}\pi^{+}$ & \multicolumn{6}{c}{}\\
      \hline
      2011 & 875 & 32 & 151 & 3 & 47 & 6\\
      2012 & 2154 & 51 & 340 & 5 & 108 & 9\\
      Total & 3028 & 61 & 491 & 7 & 155 & 11\\
      \bottomrule
  \end{tabular}

\end{table}

For the amplitude analysis, a kinematic fit is performed constraining the \Dz mass to its known value\cite{PDG2014}, which improves the resolution in the \Dz phase space. 
This also forces all candidates to lie inside the kinematically allowed region. Candidates are only accepted if this kinematic fit converges. 

\section{\label{sec:formalism}Formalism of amplitude model}        The amplitudes contributing to the decays $\Dz\rightarrow\Kmp\pipm\pipm\pimp$ are described in terms of a sequence of two-body states. 
It is assumed that once these two-body states are produced, rescattering against other particles can be neglected. 
Two-body processes are often referred to as {\it isobars} and this approximation as the {\it isobar model}.
Isobars can be described in terms of resonances, typically using the relativistic Breit-Wigner amplitude for narrow vector and tensor states. 
For scalar states, there typically are multiple broad overlapping resonances, in addition to significant nonresonant scattering amplitudes between the constituent particles of the state. 
Such states cannot be described in terms of Breit-Wigner amplitudes and instead the K-matrix formalism \cite{PhysRev.70.15,Chung:1995dx} is adopted, and will be denoted by $\left[\pip\pim\right]^{L=0}$ and $\left[\Kmp\pipm\right]^{L=0}$ throughout for $\pip\pim$ and $\Kmp\pipm$ S-waves, respectively.  

The following decay chains are considered: 

\begin{description}
 
  \item[Cascade decays] have the topology $\Dz\rightarrow X \left[ Y \left[P_1P_2\right]P_3 \right] P_4$ - the $\Dz$ meson decays into a stable pseudoscalar state $P_4$ and an unstable state $X$. 
    The unstable state then decays to three pseudoscalars $P_{1,2,3}$ via another intermediate unstable state ($Y$).
    There are three distinct possibilities for cascade decays. 
    The resonance $X$ can either have isospin $I=1/2$, and will therefore decay into the $\Kmp\pipm\pimp$ final state, or have isospin $I=1$ and therefore will decay into the $\pip\pim\pipm$ final state.
    In the $\Kmp\pipm\pimp$ case, the next state in the cascade $Y$ can either be in $\Kmp\pipm$ or $\pip\pim$, referred to as cases (1) and (2), respectively. 
    In the $\pip\pim\pipm$ case, there is only the $\pip\pim$ state, referred to as case (3). 
    \begin{enumerate}
      \item \makebox[3cm]{$\left[\Kmp\pipm\right]\pimp$\hfill} Example: $\Dz \rightarrow \KONE{1270}^{-} \left[ \Kstarb(892)^{0} \left[ \Km\pip \right] \pim \right] \pip$.
      \item \makebox[3cm]{$\Kmp\left[\pip\pim\right]$\hfill}  Example: $\Dz \rightarrow \KONE{1270}^{-} [ \rho(770)^{0} \left[ \pim\pip \right] \Km ] \pip$.
      \item \makebox[3cm]{$\pip\pim\pipm$\hfill}  Example: $\Dz \rightarrow a_1(1260)^{+} \left[ \rho(770)^{0} \left[ \pim\pip \right] \pip \right] \Km $.
    \end{enumerate}
    Two complex parameters can be used to describe cascade decays: the coupling between the $\Dz$ meson and the first isobar, and then the coupling between the first isobar and the second intermediate state. 
    One of the couplings between isobars can be fixed by convention, typically the dominant channel. 
    For example, for the $\AONE$ resonance,
    the couplings for subdominant decay chains such as $\AONE \to \left[\pip\pim\right]^{L=0}\pip$ are defined with respect to the dominant $\AONE \to \rho(770)^{0} \pip$ decay.
    
  \item[Quasi two-body decays] have the topology $\Dz\rightarrow X\left[ P_1P_2 \right] Y \left[P_3P_4\right]$ - the $\Dz$ meson decays into a pair of unstable states, which in turn each decay to a pair of stable pseudoscalar mesons.  
    The only possibility where $X,Y$ form resonances of conventional quark content is $X \left[\Km\pip\right] Y \left[ \pip\pim \right]$, with an example of a typical process being $\Dz\rightarrow \Kstarb(892)^{0}[\Km\pip]\rho(770)^{0}\left[\pip\pim\right]$. 
    The parameters to be determined describe the coupling between the $\Dz$ initial state and the quasi two-body state. In the above example, there are three different possible orbital configurations of the vector-vector system, and hence this component has three complex parameters. 
\end{description}

Decay chains are described using a product of dynamical functions for each isobar and a spin factor. 
The amplitude for each decay chain is explicitly made to respect Bose symmetry by summing over both possible permutations of same-sign pions. 
The total amplitude is then modelled as a coherent sum of these processes.
Spin factors are modelled using the Rarita-Schwinger formalism following the prescription in Ref.~\cite{Zou:2002ar}; the details of this formulation are included in Appendix~\ref{sec:SpinFormalism}.

Resonances are modelled with the relativistic Breit-Wigner function unless otherwise stated, which as a function of the invariant-mass squared, $s$, takes the form 
\begin{equation}
  \mathcal{T}(s) = \frac{\sqrt{k} B_L(q,0) }{ m_0^2 - s -im_0\Gamma(s) } ,
  \label{eq:BreitWigner}
\end{equation}
where the mass of the resonance is $m_0$ and $\Gamma(s)$ is the energy-dependent width.
The form factor for a decay in which the two decay products have relative orbital angular momentum $L$ is given by the normalised Blatt-Weisskopf function \cite{TheoreticalNuclearPhysics} $B_L(q,0)$, where $q$ is the three-momentum of either decay product in the rest frame of the resonance, and is normalised to unity at zero momentum transfer. 
The factor $k$ normalises the lineshape integrated over all values of $s$ if the Blatt-Weisskopf form-factor and energy dependence of the width are neglected, and is included to reduce correlations between the coupling to the channel and the mass and width of the resonance.

For a resonance that decays via a single channel to two stable particles, such as $\rho(770)^{0}\rightarrow\pip\pim$, the width is given by 
\begin{equation}
  \Gamma(s) = \frac{\Gamma_0 q m_0}{q_0 \sqrt{s}} \left(\frac{q}{q_0}\right)^{2L} B_L(q,q_0)^2,
  \label{eq:RunningWidth}
\end{equation}
where $\Gamma_0$ is the width at the resonance mass, and $q_0$ is the linear momentum of either decay product evaluated at the rest mass of the resonance.  
The energy-dependent width of a resonance that decays to a three-body final-state must account for the dynamics of the intermediate decay process, and follows that developed for the decay $\tau^{+}\rightarrow a_1(1260)^{+}\neutb$ by the CLEO collaboration in Ref.~\cite{PhysRevD.61.012002}.
The width of a resonance $R$ decaying into three bodies $abc$ can be expressed in terms of the spin-averaged matrix element of the decay $\mathcal{M}_{R\rightarrow abc}$ integrated over the phase space of the three-particle final state,
\begin{equation}
  \Gamma( s ) \propto \frac{1}{s} \int ds_{ab} ds_{bc} \left| \mathcal{M}_{R\rightarrow abc} \right|^2 ,
  \label{eq:TotalWidth}
\end{equation}
where the matrix element consists of a coherent sum over the intermediate states in the three-body system, described using the isobar model and using the fitted couplings between the resonance and the intermediate isobars.
In the example of the decay of the $\AONE$ resonance, these are predominately the couplings to the $\rho(770)^{0}\pip$ and $\left[\pip\pim\right]^{L=0}\pip$ intermediate states. 
The width is normalised such that $\Gamma(m_0^2) = \Gamma_0$. 
In the three-body case, exponential form-factors are used rather than normalised Blatt-Weisskopf functions, 
 \begin{equation}
   F(q) = e^{-r^2 q^2 / 2 },
   \label{eq:FormFactor}
 \end{equation}
 where $r$ characterises the radius of the decaying resonance.

The K-matrix formalism\cite{PhysRev.70.15} provides a convenient description of a two-particle scattering amplitude, which is particularly useful in parameterising S-wave systems.
This formulation can then be used in the description of multibody decays on the assumption that rescattering against the other particles in the decay can be neglected.
The K-matrix formalism is used in this analysis to describe the $\pip\pim$ and $\Kmp\pipm$ S-waves due to its relative success in parameterising the scalar contributions to three-body decays \cite{Aubert:2008bd,Pennington:2007se} of the \PD meson.

The $\pip\pim$ S-wave (isoscalar) amplitude is modelled using the K matrix from 
Ref.~\cite{Anisovich:2002ij, Aubert:2008bd},
which describes the amplitude in the mass range $280 \mevcc < \sqrt{s} < 1900 \mevcc$, 
considering the effects of five coupled channels, $\pi\pi$, $\PK\PK$, $\pi\pi\pi\pi$, $\Eta\Eta$, $\Etapr\Eta$, and five poles with masses which generate the resonances. 
The K matrix also includes polynomial terms that describe nonresonant scattering between hadrons. 
The coupling to each of these poles and the direct coupling to each of the five channels depend on the production mode, which is modelled using the production vector or P-vector approach, in which the amplitude is
\begin{equation}
  \mathcal{A}(s) = \left(I-i\hat{\rho} \hat{K}\right)^{-1} \hat{P}, 
  \label{eq:defFvector}
\end{equation}
where $\hat{\rho}$ is the two-body phase-space matrix. 
The complex-valued vector function, $\hat{P}$, has one component for each of the coupled channels, and describes the coupling between the initial state and either one of the poles or a direct coupling to one of these channels. 
The generic P-vector for the isoscalar K-matrix therefore has 10 complex parameters. 
An additional complexity in the four-body case is that there are several initial states that couple to the $\pip\pim$ S-wave, each of which has its own P vector. 
Several simplifying assumptions are therefore made to the P vector to avoid introducing an unreasonable number of degrees of freedom. 
The only direct production terms included in the P vector are to the $\pi\pi$ and $\PK\PK$ states,
 as the production of the $\pip\pim$ final state via a direct coupling to another channel all have similar structure below their corresponding production thresholds. 
The couplings to poles 3, 4 and 5 (where the numbering of the poles is defined in Ref.~\cite{Anisovich:2002ij}) are also fixed to zero, as production of these poles only has a small effect within the phase space. 
This choice reduces the number of free parameters per S-wave production mechanism to four complex numbers. The couplings to the poles are described by $\beta_0$ and $\beta_1$, while the direct couplings to each channel by $f_{\pi\pi}$ and $f_{\PK\PK}$. 
The production vectors used here should therefore be considered as a minimal simplified model.
For production of $\pip\pim$ S-wave states via resonances, such as the decay chain $\AONE\rightarrow[\pip\pim]^{L=0}\pip$, improved sensitivity to the structure of the $\pip\pim$ state can be achieved by studying a decay mode that produces the $\AONE$ with a larger phase space. 
In several cases, one or more of these couplings are found to be negligible for a given production mode, and therefore are fixed to zero. 

The $\Kmp\pipm$ S-wave is modelled using the K matrices from the analysis of $\Dp\rightarrow\Km\pip\pip$ by the FOCUS collaboration \cite{Pennington:2007se}. 
The $I=1/2$ K matrix considers two channels, $\PK\pi$ and $\PK\Etapr$, and a single pole which is responsible for generating the $K^*(1430)^{0}$ resonance.
Additionally, the K matrix includes polynomial terms that describe nonresonant scattering between the hadrons. 
The $\Kmp\pipm$ S-wave also contains a $I=3/2$ component. 
No poles or inelasticity are expected with this isospin, and therefore the associated amplitude can be modelled using a K matrix consisting of a single scalar term. 

The $I=1/2$ amplitudes are constructed in the Q-vector \cite{Chung:1995dx} approximation.
The P vector has the same pole structure as the K matrix, and therefore the approximation
\begin{equation}
  \hat{K}\hat{P} \approx \hat{\alpha}(s) 
\end{equation}
can be made, 
where $\hat{\alpha}(s)$ is a slowly varying complex vector. This is sometimes referred to as the Q-vector \cite{Chung:1995dx} approximation, and allows the insertion of $\hat{K}^{-1} \hat{K}$ into Eq.~\ref{eq:defFvector}, and the rephrasing of the $I=1/2$ decay amplitude, $\mathcal{A}_{1/2}$, in terms of the T-matrix elements from scattering: 
\begin{equation}
  \mathcal{A}_{1/2} = \alpha_{\PK\pi} \hat{T}_{11} + \alpha_{\PK\Etapr} \hat{T}_{12},
\end{equation}
where
\begin{equation}
  \hat{T} = \left( 1 - i\hat{\rho} \hat{K} \right)^{-1} \hat{K},
\end{equation}
which is the transition matrix associated with the $I=1/2$ scattering process. 
Given the relatively small energy range available to the $\Kmp\pipm$ system, it is reasonable to approximate $\hat{\alpha}(s)$ as a constant. 
Inclusion of polynomial terms in $\hat{\alpha}(s)$ is found not to improve the fit quality significantly.  
The coupling to the $\PK\Etapr$ channel, $\alpha_{\PK\Etapr}$, is defined with respect to the coupling to the $\PK\pi$ channel, $\alpha_{\PK\pi}$ in all production modes.
If the phase of $\alpha_{\PK\Etapr}$ is zero, the phase shift of the $I=1/2$ component matches that found in scattering experiments, which is the expected result if Watson's theorem \cite{PhysRev.88.1163} holds for these decays. 
Similar to the $\pip\pim$ S-wave, the components of $\hat{\alpha}$ and the coupling to the $I=3/2$ channel are allowed to differ between production modes.

\section{\label{sec:fitModel}Fit formalism and model construction} Independent fits are performed on the $\Dz\to\CF$ and $\Dz\to\DCS$ data sets, using an unbinned maximum likelihood procedure to determine the amplitude parameters. 
The formalism of the fit is described in Sects.~\ref{sec:Likelihood}--\ref{sec:FitFractions}, and the method for systematically selecting plausible models is discussed in Sect.~\ref{sec:ModelBuilding}.

\subsection{Likelihood}
\label{sec:Likelihood}
The probability density functions (PDFs) are functions of position in $\Dz$ decay phase-space, $\xp$, and are composed of the signal amplitude model and the two sources of background described in Sect.~\ref{sec:selection}:
\begin{equation}
  P(\xp) = \varepsilon(\xp) \phi(\xp) \left( 
  \frac{ Y_s }{\mathcal{N}_s} \left| \mathcal{M}(\xp) \right|^2 +
  \frac{ Y_c }{\mathcal{N}_c} \mathcal{P}_c(\xp)  +
  \frac{ Y_m }{\mathcal{N}_m} |\overline{\mathcal{M}}(\xp)|^2  \right) .
\end{equation}
The signal PDF is described by the function $\left|\mathcal{M}(\xp)\right|^2$, where $\mathcal{M}(\xp)$ is the total matrix element for the process, 
weighted by the four-body phase-space density $\phi(\xp)$, and the phase-space acceptance, $\varepsilon(\xp)$.
The mistag component involving $\overline{\mathcal{M}}(\xp)$, is only present in the WS sample, and is modelled using the RS signal PDF.
The combinatorial background is modelled by $\mathcal{P}_c(\xp)$, and is present in both samples. 
The normalisation of each component is given by the integral of the PDF over the phase space, $\mathcal{N}_{i}$, where $i=(c,s,m)$,  
weighted by the fractional yield, $Y_i$, determined in Sect.~\ref{sec:selection}.

The PDF that describes the combinatorial background in the WS sample is fixed to the results of a fit to the two sidebands of the $m_{\PK\pi\pi\pi}$ distribution, below $1844.5\mevcc$ and above $1888.5\mevcc$.
The components in this model are selected using the same algorithm to determine the resonant content of the signal modes, which is discussed in Sect.~\ref{sec:ModelBuilding}.
In this case, the PDF incoherently sums the different contributions and assumes no angular correlations between tracks. 
The contamination from combinatorial background in the RS sample is very low, and hence this contribution can safely be assumed to be distributed according to phase space, that is $\mathcal{P}_c(\xp) =1$.

The function to minimise is 
\begin{equation}
  \mathcal{L} = -2 \sum_{i \in \mathrm{data} } \log( P(\xp_i) ).
\end{equation}
As the efficiency variation across the phase space factorises in the PDF, these variations result in a constant shift in the likelihood everywhere except the normalisation integrals, and hence can be neglected in the minimisation procedure. 
Efficiency variations can then be included in the fit by performing all integrals using simulated events that have been propagated through the full \lhcb detector simulation and selection.
These events are referred to as the {\it integration sample}.
The values of the normalisation integrals are independent of the generator distribution of the integration sample, however the uncertainties on the integrals are minimised when integration events approximate the function being integrated, which is known as {\it importance sampling}.
Therefore, integration samples are generated using preliminary models that do not include efficiency effects. 

\subsection{Goodness of fit}
\label{sec:GoodnessOfFit}
The quality of fits is quantified by computing a $\chi^2$ metric.
Candidates are binned using an adaptive binning scheme. 
Five coordinates are selected, and the phase space is repeatedly divided in these coordinates such that each bin contains the same number of candidates, following the procedure described in Ref.~\cite{LHCb-PAPER-2015-057}.
The division is halted when each bin contains between 10 and 20 entries.
This procedure results in 32,768 approximately equally populated bins for the RS sample, and 256 for the WS sample. 
Five two- and three-body invariant mass-squared combinations are used as coordinates for the binning procedure, $s_{\pip\pim\pip}, s_{\Km\pip}, s_{\Km\pim}, s_{\pip\pim}$ and $ s_{\Km\pip\pim}$. 
The $\chi^2$ is defined as
\begin{equation}
  \chi^2 = \sum_{i \in \mathrm{bins}} \frac{  \left( N_{i} - \langle {N}_{i} \rangle \right)^2 } { N_{i} + \bar{\sigma}_{i}^{2} } ,
  \label{eq:defChi2}
\end{equation}
where $N_i$ is the observed number of candidates in bin $i$ and $\langle {N}_i \rangle$, the expected number of entries determined by reweighting the integration sample with the fitted PDF.
The statistical uncertainty from the limited size of the integration sample, $\bar{\sigma}_i$, is included in the definition of the $\chi^2$, and is estimated as 
\begin{equation}
  \bar{\sigma}_i^2 = \sum_{j \in \mathrm{bin}(i)} \omega_j^2 ,
  \label{eq:ErrorEstimate}
\end{equation}
where $\omega_j$ is the weight of integration event $j$.
The $\chi^2$ per degree of freedom is used as the metric to optimise the decay chains included in a model, using the model-building procedure described in Sect.~\ref{sec:ModelBuilding}.

\subsection{Fit fractions}
\label{sec:FitFractions}
The values of coupling parameters depend strongly on various choices of convention in the formalism.
Therefore, it is common to define the fractions in the data sample associated with each component of the amplitudes (fit fractions). 
In the limit of narrow resonances, the fit fractions are analogous to relative branching fractions. 
The fit fraction for component $i$ is  
\begin{equation}
  \mathcal{F}_{i} = \frac{ \int d \xp \left| \mathcal{M}_i (\xp) \right|^2 }{ 
  \int d \xp \left| \sum_{j} \mathcal{M}_j(\xp) \right|^2  }. 
\end{equation}
For cascade processes, the different secondary isobars contribute coherently to the fit fractions.
The {\it partial} fit fractions for each sub-process are then defined as the fit fraction with only the contributions from the parent isobar included in the denominator. 
\subsection{Model construction}

\label{sec:ModelBuilding}
The number of possible models that could be used to fit the amplitudes is extremely large due to the large number of possible decay chains ($\approx 100$). 
A full list of the components considered is included in Appendix~\ref{app:ListOfComponents}.

A model of ``reasonable'' complexity typically contains $\mathcal{O}(10)$ different decay chains.
Therefore, the number of possible models is extremely large, and only an infinitesimal fraction of these models can be tested. 
An algorithmic approach to model building is adopted, which begins with an initial model and attempts to iteratively improve the description by adding decay chains. 
For $\RS$ the initial model is that constructed by the Mark III collaboration \cite{PhysRevD.45.2196}, augmented by 
knowledge from other analyses, such as the additional decay channels of the $a_1(1260)^{+}$ found in the amplitude analysis of the decay ${\Dz\rightarrow \pip\pim\pip\pim}$ performed by the FOCUS collaboration \cite{PhysRevD.75.052003}. 
The two-body nonresonant terms in the Mark~III model are replaced with the relevant K matrices, and the four-body nonresonant term replaced with a quasi two-body scalar-scalar term $[\Km\pip]^{L=0}[\pip\pim]^{L=0}$, modelled using a product of K matrix amplitudes. 

For \WS, where no previous study exists, the initial model is obtained by inspection of the invariant-mass distributions. 
There are clear contributions from the $\Kstar(892)^{0}$ and $\rho(770)^{0}$ resonances, and therefore combined with the expectation that the 
vector-vector contributions should be similar between WS and RS, the quasi two-body mode $\Dz\rightarrow \Kstar(892)^{0}\rho(770)^{0}$ is included in all three allowed orbital states $L=(0,1,2)$.
The scalar-scalar contribution should also be comparable between WS and RS decay modes, and hence the quasi two-body term $\Dz\rightarrow [\Kp\pim]^{L=0} [\pip\pim]^{L=0}$ is also included. 

The steps of the model-building procedure are
\begin{enumerate}
  \item Take a model and a set of possible additional decay chains, initially the complete set discussed in Appendix.~\ref{app:ListOfComponents}.
    Perform a fit to the data using this model adding one of these decay chains.
  \item If adding this decay chain improves the $\chi^2$ per \dof by at least 0.02, then retain the model for further consideration.  
  \item On the first iteration, restrict the pool of decay chains that are added to the model to those 40 contributions that give the largest improvements to the fit. 
  \item Reiterate the model-building procedure, using the 15 models with the best fit quality from step 2 as starting points. Finish the procedure if no model has improved significantly.
\end{enumerate}
The model-building procedure therefore results in an ensemble of parametrisations of comparable fit quality.

\section{\label{sec:results}Fit results}                           This section presents fit results and systematic uncertainties, with the latter discussed first in Sect.~\ref{sec:Systematics}.
The model-building procedure outlined in Sect.~\ref{sec:ModelBuilding}
results in ensembles of parameterisations of comparable fit quality. 
The models discussed in this section, which are referred to as the baseline models, and are built to include all decay chains that are common to the majority of models that have a $\chi^2$ per \dof differing from the best-fitting models by less than $0.1$.
The results for these baseline models are shown and their features discussed in Sect.~\ref{sec:RSModel} and Sect.~\ref{sec:WSModel} for the RS decay and the WS decay, respectively. 
The general features of models in the ensembles are discussed in Sect.~\ref{sec:Ensemble}. 
In Sect.~\ref{sec:Coherence} the models are used to calculate the coherence factor of the decays, and an assessment is made of the stability of the predicted coherence factors, strong-phase differences and amplitude ratios with respect to the choice of WS model in regions of phase space. 

\subsection{Systematic uncertainties} 
\label{sec:Systematics}
Several sources of systematic uncertainty are considered. 
Experimental issues are discussed first, followed by uncertainties related to the model and the formalism. 

All parameters in the fit have a systematic uncertainty originating from the limited size of the integration sample used in the likelihood minimisation. 
This effect is reduced by importance sampling. 
The remaining uncertainty is estimated using a resampling technique. 
Half of the integration sample is randomly selected, and the fit performed using only this subsample. 
This procedure is repeated many times, and the systematic uncertainty from the finite integration statistics is taken to be $1/\sqrt{2}$ of the spread in fit parameters.

There is an additional systematic uncertainty due to the imperfect simulation, which affects the efficiency corrections. 
The RS data are divided into bins in the $\Dz$ transverse momentum, in which the efficiency corrections may be expected to vary, and the fit is performed independently in each bin. 
The results of these fits are combined in an uncertainty-weighted average, including the correlations between the different parameters, and the absolute difference between the parameters measured by this procedure and the usual fitting procedure is assigned as the systematic uncertainty. Additionally, the data is divided by data-taking year and software trigger category and independent fits performed using these subsamples. The fit results are found to be compatible within the assigned uncertainties between these samples, hence no additional systematic uncertainty is assigned.   

The uncertainty associated with the determination of the signal fraction and mistag fraction in each sample is measured by varying these fractions 
within the uncertainties found in the fit to the $m_{\KPI}$ \vs $\Delta{m}$ plane. 

Parameters that are fixed in the fit, such as the $\rho(770)^{0}$ mass and width, are randomly varied according to the uncertainties given in Ref. \cite{PDG2014}, 
and the corresponding spreads in fit results are assigned as the uncertainties.
It is assumed that input correlations between these parameters are negligible. 
When performing fits to the WS sample, several parameters, such as the mass, width and couplings of the $K_1(1270)^{\pm}$ resonance, are fixed to the values found in the RS fit.
The uncertainty on these parameters is propagated to the WS fit by randomly varying these parameters by their uncertainties.
The radii of several particles used in the Blatt-Weisskopf form factor are varied using the same procedure. 
The $\Dz$ radial parameter is varied by $\pm0.5\gev^{-1}c$. 

The uncertainty due to the background model in the WS fit is estimated using pseudo-experiments. 
A combination of simulated signal events generated with the final model and candidates from outside of the \Dz signal region is used to approximate the real data.
The composite dataset is then fitted using the signal model, and differences between the true and fitted values are taken as the systematic uncertainties on the background parametrisation. 

The choice of model is an additional source of systematic uncertainty. 
It is not meaningful to compare the coupling parameters between different parametrisations, as these are by definition the parameters of a given model. It is however useful to consider the impact the choice of parametrisation has on fit fractions and the fitted masses and widths. 
Therefore, the model choice is not included in the total systematic uncertainty, but considered separately in Sect.~\ref{sec:Ensemble}--\ref{sec:Coherence}.

The total systematic uncertainty is obtained by summing the components in quadrature. 
The total systematic uncertainty is significantly larger than the statistical uncertainty on the RS fit, with the largest contributions coming from the form factors that account for the finite size of the decaying mesons. 
For the WS fit, the total systematic uncertainty is comparable to the statistical uncertainty, with the largest contribution coming from the parametrisation of the combinatorial background.
A full breakdown of the different sources of systematic uncertainty for all parameters is given in Appendix~\ref{sec:SystematicBreakdown}.

\subsection{Results for the RS decay}

\label{sec:RSModel}

\begin{figure}[!htb]
  \centering
  \includegraphics[width=0.48\textwidth]{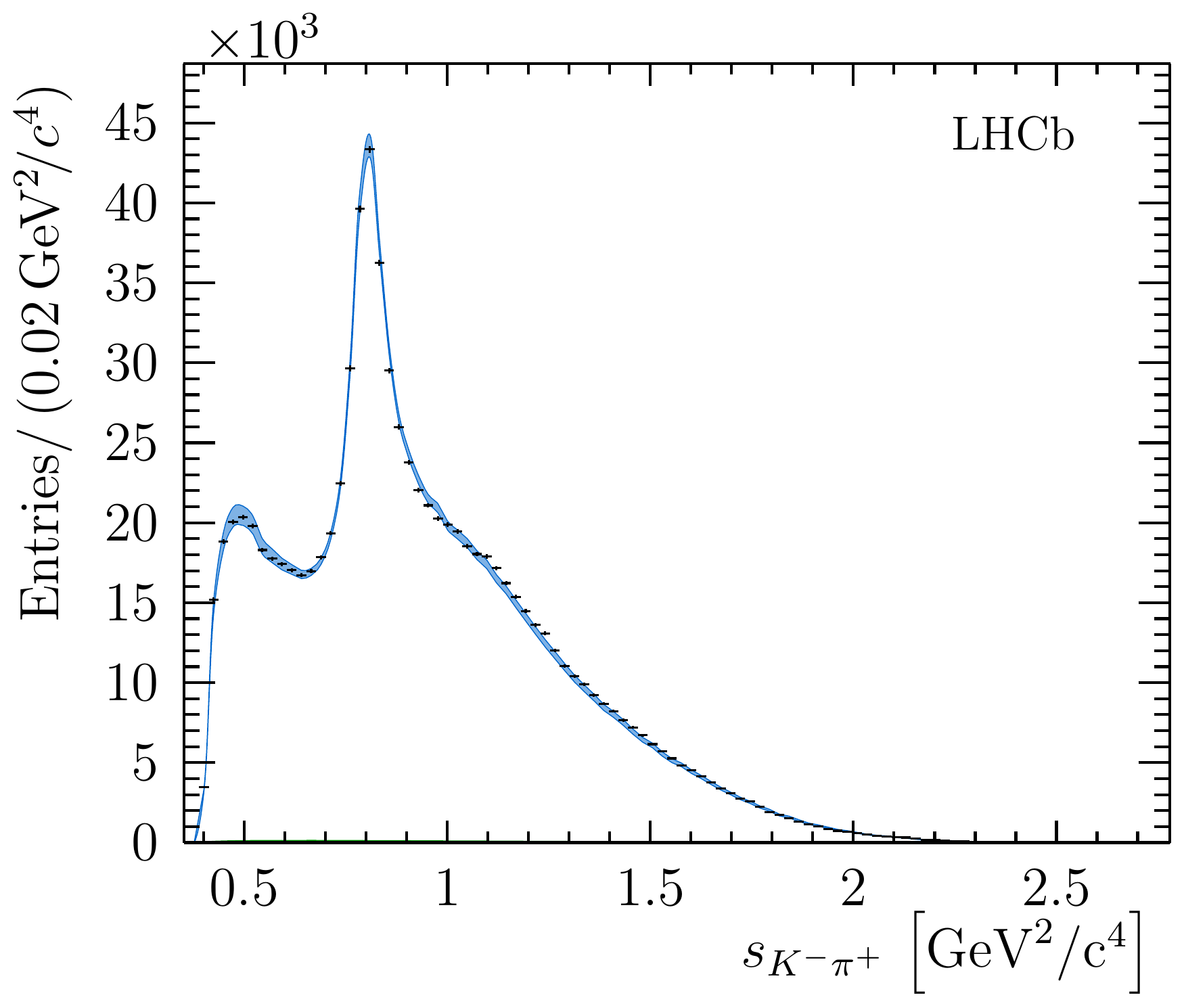} \includegraphics[width=0.48\textwidth]{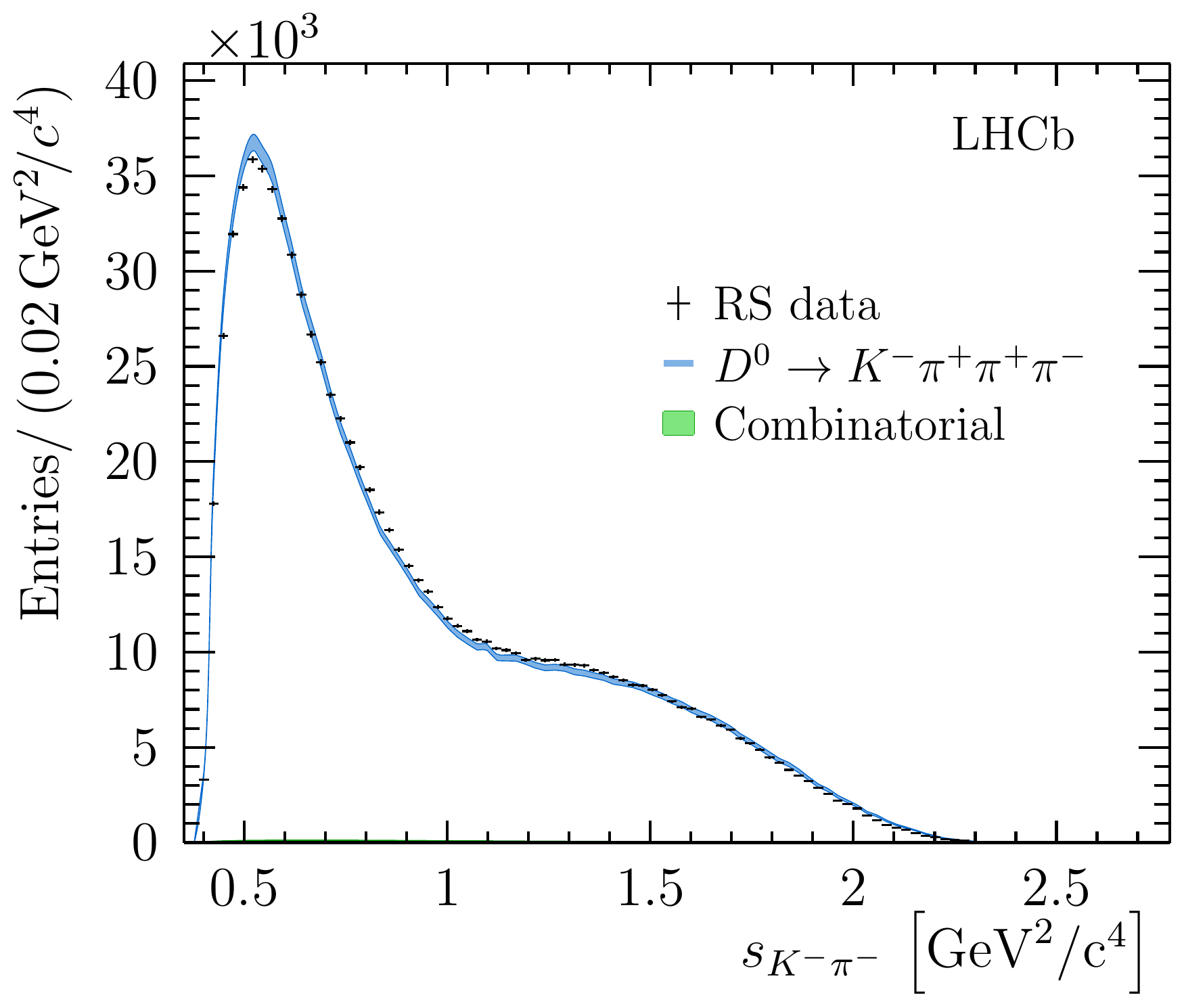}
  
  \includegraphics[width=0.48\textwidth]{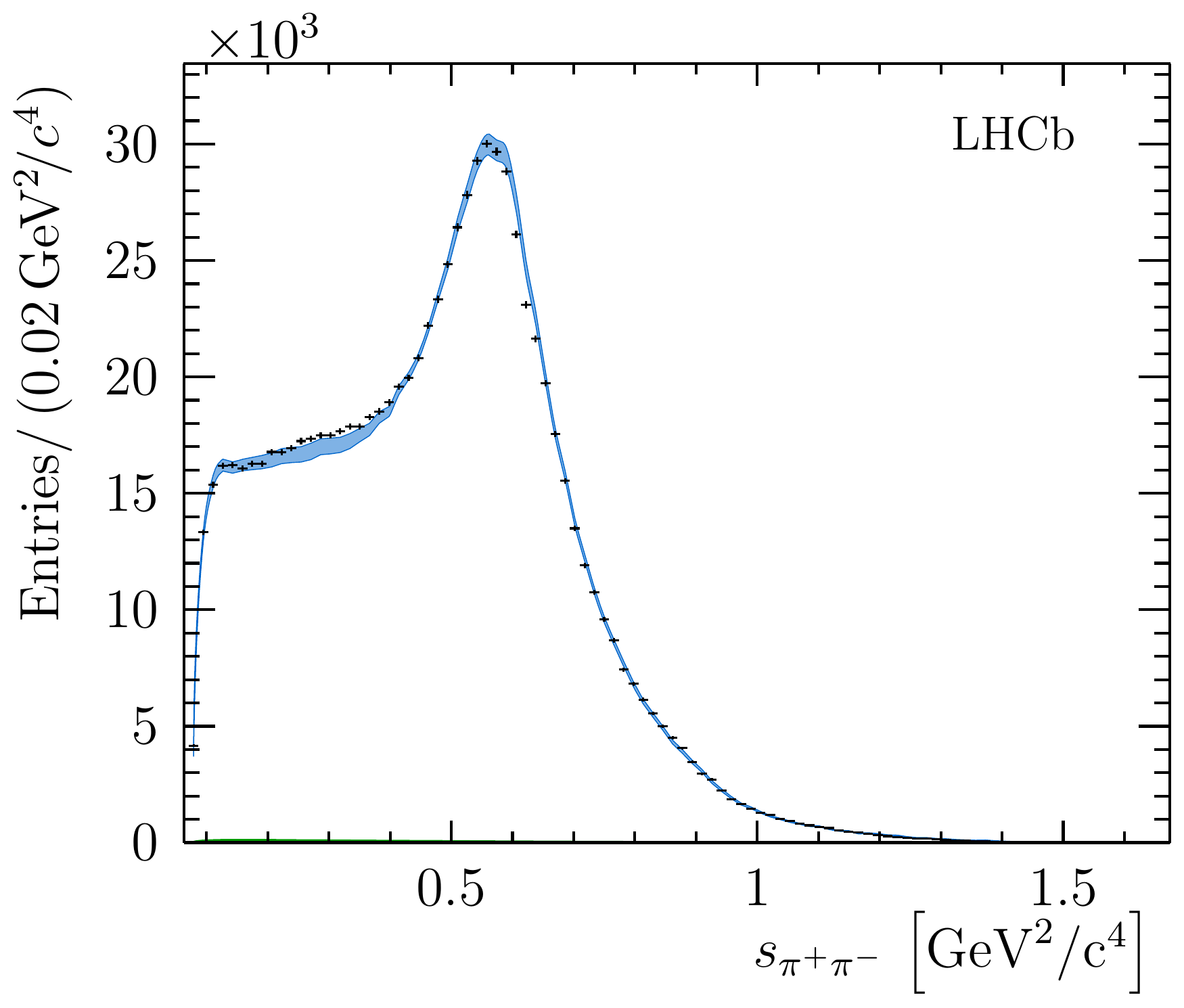} \includegraphics[width=0.48\textwidth]{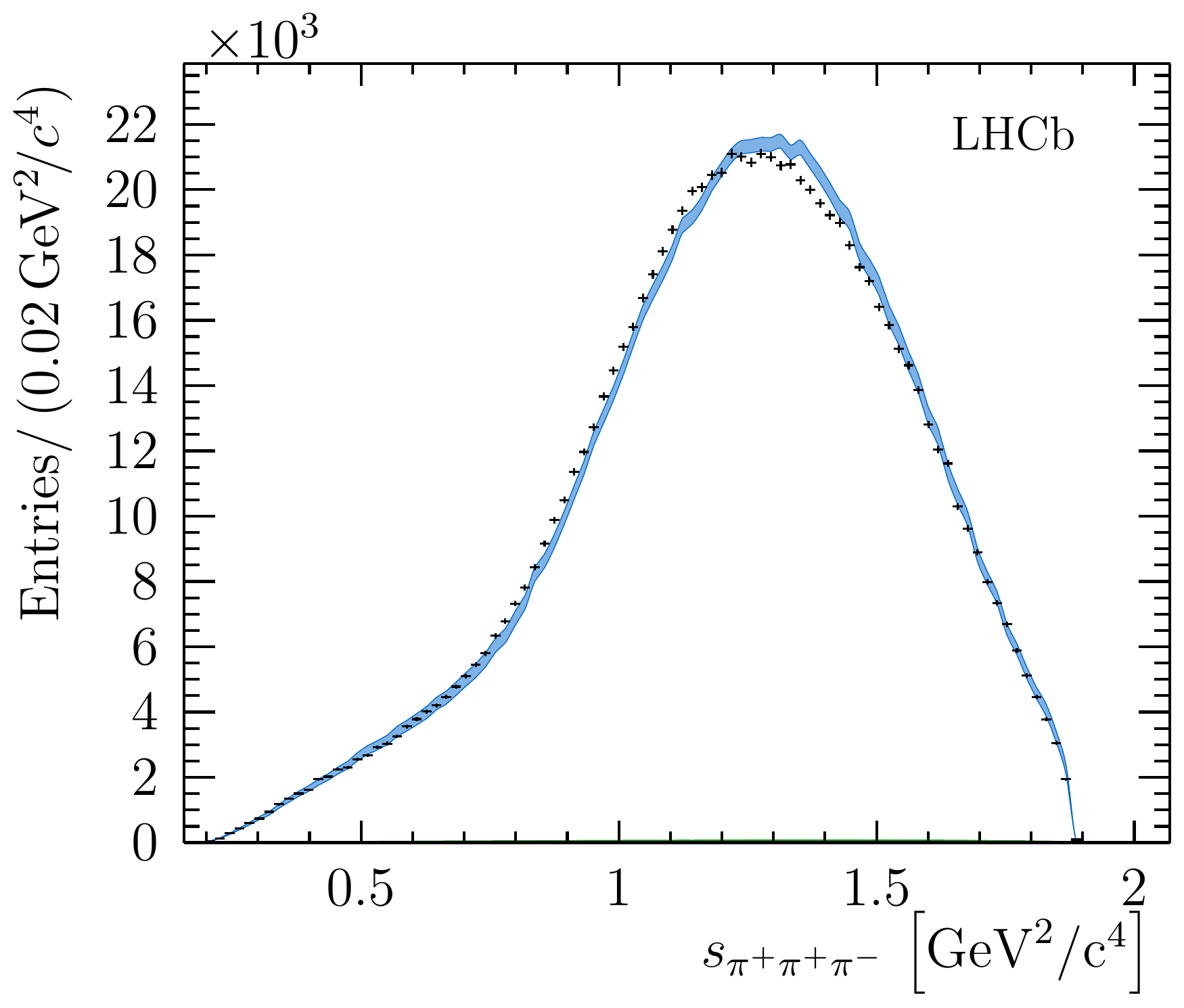}
  
  \includegraphics[width=0.48\textwidth]{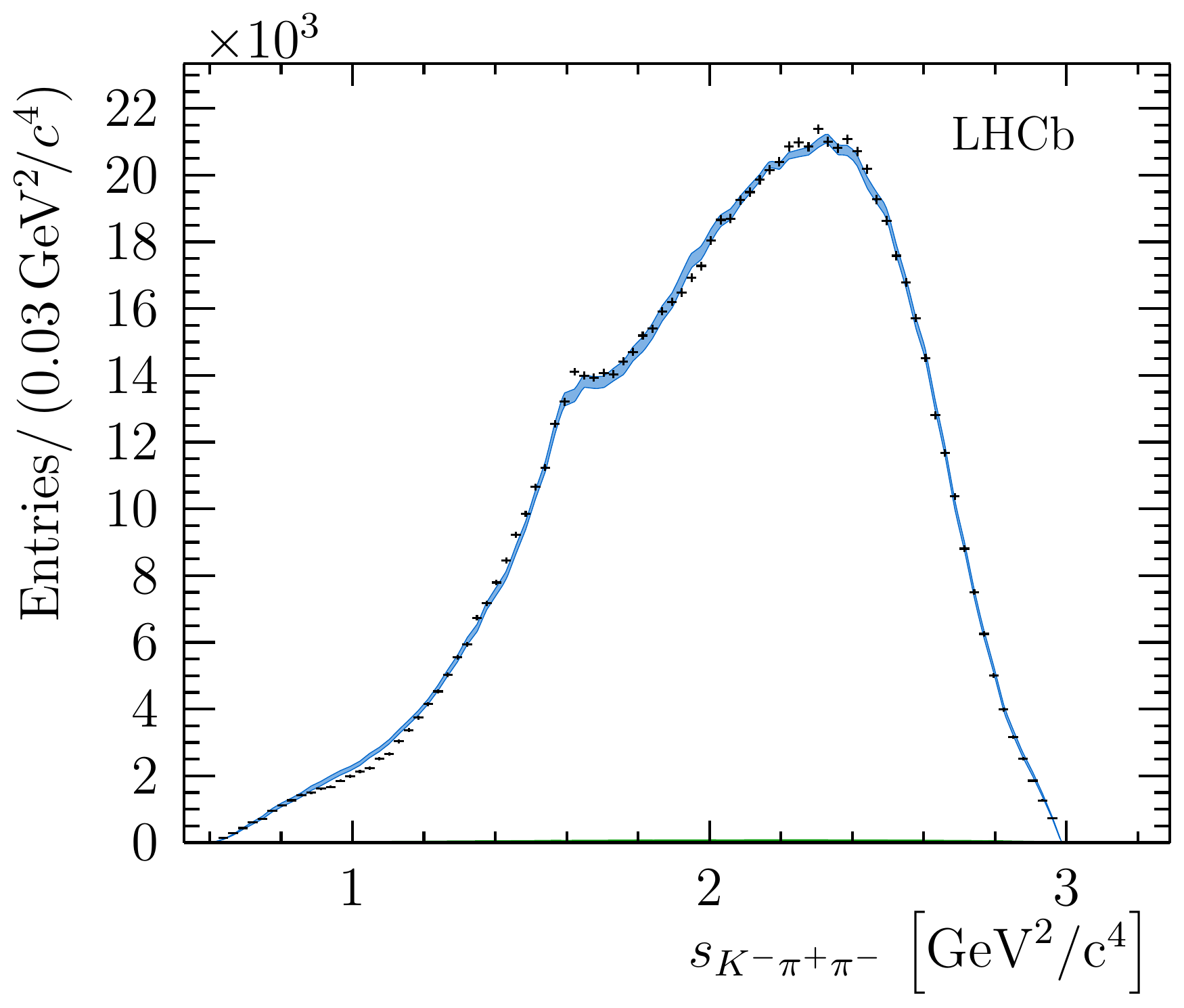} \includegraphics[width=0.48\textwidth]{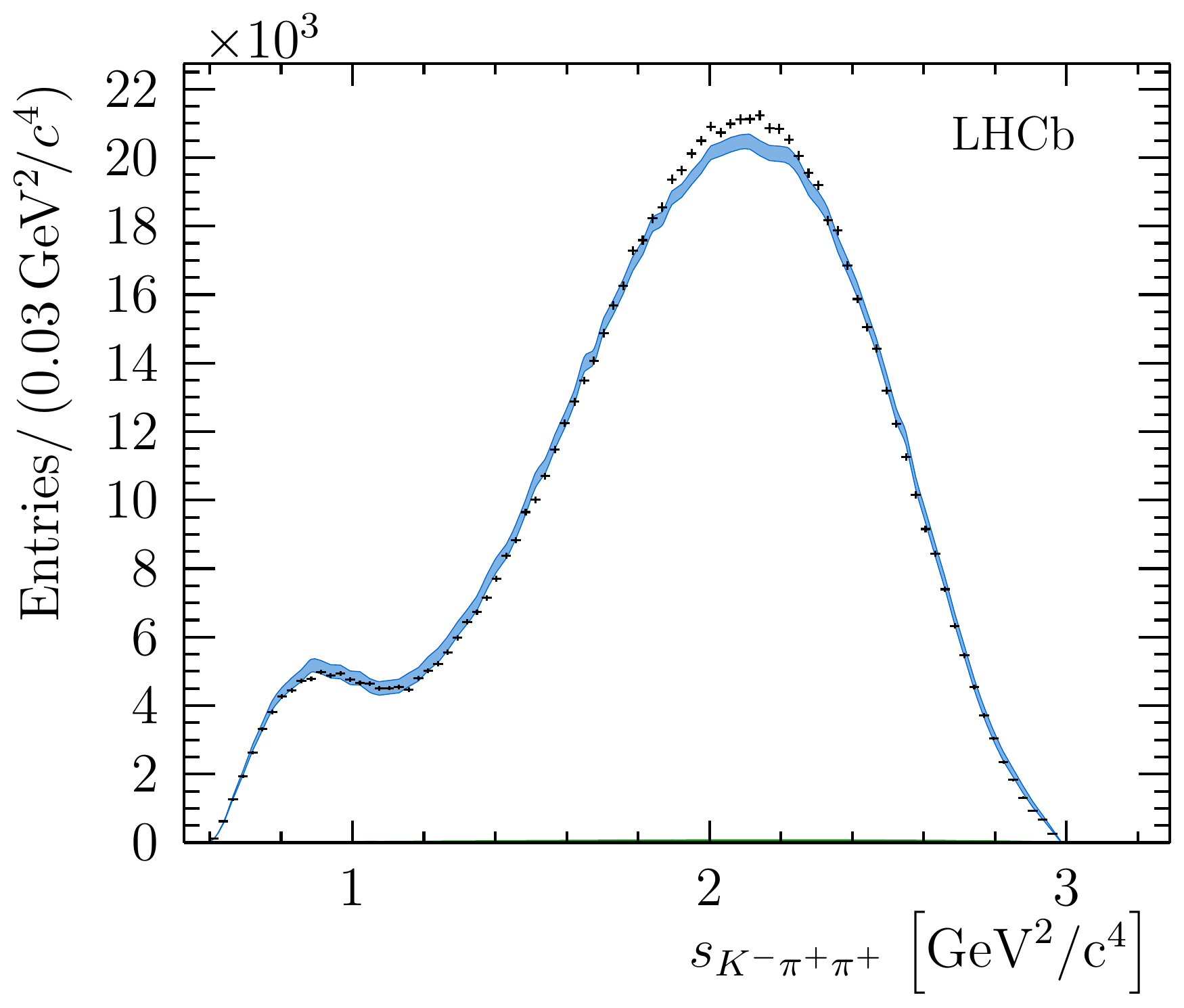}
  \caption{ Distributions for six invariant-mass observables in the RS decay \RS.
  Bands indicate the expectation from the model, with the width of the band indicating the total systematic uncertainty. The total background contribution, which is very low, is shown as a filled area.
  In figures that involve a single positively-charged pion, one of the two identical pions is selected randomly.  
  }
  \label{fig:RSPhase3}
\end{figure}

\begin{table}
  \centering  
  \caption{ Fit fractions and coupling parameters for the RS decay \RS. 
  For each parameter, the first uncertainty is statistical and the second systematic.
  Couplings $g$ are defined with respect to the coupling to the channel $\Dz\rightarrow[\Kstarb(892)^{0}\rho(770)^{0}]^{L=2}$.
  Also given are the $\chi^2$ and the number of degrees of freedom ($\nu$) from the fit and their ratio. }
  \scalebox{0.9}{\begin{tabular}{l 
>{\collectcell\num}r<{\endcollectcell} @{${}\pm{}$} >{\collectcell\num}l<{\endcollectcell} @{${}\pm{}$} >{\collectcell\num}l<{\endcollectcell}
>{\collectcell\num}r<{\endcollectcell} @{${}\pm{}$} >{\collectcell\num}l<{\endcollectcell} @{${}\pm{}$} >{\collectcell\num}l<{\endcollectcell}
>{\collectcell\num}r<{\endcollectcell} @{${}\pm{}$} >{\collectcell\num}l<{\endcollectcell} @{${}\pm{}$} >{\collectcell\num}l<{\endcollectcell}
}
\toprule
 & \multicolumn{3}{c}{Fit Fraction [\%]} & \multicolumn{3}{c}{$\left|g\right|$} & \multicolumn{3}{c}{$\mathrm{arg}(g) [^o]$}\\
\midrule
  $\left[\Kstarb(892)^{0}\rho(770)^{0}\right]^{L=0}$ & 7.34 & 0.08 & 0.47 & 0.196 & 0.001 & 0.015 & -22.4 & 0.4 & 1.6\\
$\left[\Kstarb(892)^{0}\rho(770)^{0}\right]^{L=1}$ & 6.03 & 0.05 & 0.25 & 0.362 & 0.002 & 0.010 & -102.9 & 0.4 & 1.7\\
$\left[\Kstarb(892)^{0}\rho(770)^{0}\right]^{L=2}$ & 8.47 & 0.09 & 0.67 & \multicolumn{3}{c}{} & \multicolumn{3}{c}{}\\
  $\left[\rho(1450)^{0}\Kstarb(892)^{0}\right]^{L=0}$ & 0.61 & 0.04 & 0.17 & 0.162 & 0.005 & 0.025 & -86.1 & 1.9 & 4.3\\
$\left[\rho(1450)^{0}\Kstarb(892)^{0}\right]^{L=1}$ & 1.98 & 0.03 & 0.33 & 0.643 & 0.006 & 0.058 & 97.3 & 0.5 & 2.8\\
$\left[\rho(1450)^{0}\Kstarb(892)^{0}\right]^{L=2}$ & 0.46 & 0.03 & 0.15 & 0.649 & 0.021 & 0.105 & -15.6 & 2.0 & 4.1\\
\midrule
$\rho(770)^{0}\left[K^{-}\pi^{+}\right]^{L=0}$ & 0.93 & 0.03 & 0.05 & 0.338 & 0.006 & 0.011 & 73.0 & 0.8 & 4.0\\
$\quad \alpha_{3/2}$ & \multicolumn{3}{c}{} & 1.073 & 0.008 & 0.021 & -130.9 & 0.5 & 1.8\\
$\Kstarb(892)^{0}\left[\pi^{+}\pi^{-}\right]^{L=0}$ & 2.35 & 0.09 & 0.33 & \multicolumn{3}{c}{} & \multicolumn{3}{c}{}\\
$\quad f_{\pi\pi}$ & \multicolumn{3}{c}{} & 0.261 & 0.005 & 0.024 & -149.0 & 0.9 & 2.7\\
$\quad \beta_1$ & \multicolumn{3}{c}{} & 0.305 & 0.011 & 0.046 & 65.6 & 1.5 & 4.0\\
\midrule
$a_{1}(1260)^{+}K^{-}$ & 38.07 & 0.24 & 1.38 & 0.813 & 0.006 & 0.025 & -149.2 & 0.5 & 3.1\\
$K_{1}(1270)^{-}\pi^{+}$ & 4.66 & 0.05 & 0.39 & 0.362 & 0.004 & 0.015 & 114.2 & 0.8 & 3.6\\
$K_{1}(1400)^{-}\left[\Kstarb(892)^{0}\pi^{-}\right]\pi^{+}$ & 1.15 & 0.04 & 0.20 & 0.127 & 0.002 & 0.011 & -169.8 & 1.1 & 5.9\\
\midrule
$K_{2}^{*}(1430)^{-}\left[\Kstarb(892)^{0}\pi^{-}\right]\pi^{+}$ & 0.46 & 0.01 & 0.03 & 0.302 & 0.004 & 0.011 & -77.7 & 0.7 & 2.1\\
\midrule
$K(1460)^{-}\pi^{+}$ & 3.75 & 0.10 & 0.37 & 0.122 & 0.002 & 0.012 & 172.7 & 2.2 & 8.2\\
\midrule
$\left[K^{-}\pi^{+}\right]^{L=0}\left[\pi^{+}\pi^{-}\right]^{L=0}$ & 22.04 & 0.28 & 2.09 & \multicolumn{3}{c}{} & \multicolumn{3}{c}{}\\
$\quad \alpha_{3/2}$ & \multicolumn{3}{c}{} & 0.870 & 0.010 & 0.030 & -149.2 & 0.7 & 3.5\\
$\quad \alpha_{K\eta^\prime}$ & \multicolumn{3}{c}{} & 2.614 & 0.141 & 0.281 & -19.1 & 2.4 & 12.0\\
$\quad \beta_1$ & \multicolumn{3}{c}{} & 0.554 & 0.009 & 0.053 & 35.3 & 0.7 & 1.6\\
$\quad f_{\pi\pi}$ & \multicolumn{3}{c}{} & 0.082 & 0.001 & 0.008 & -147.0 & 0.7 & 2.2\\
\midrule
Sum of Fit Fractions & 98.29 & 0.37 & 0.84 & \multicolumn{6}{c}{}\\
$\chi^2 / \nu$ & \multicolumn{3}{c}{$40483/32701 = 1.238 $} & \multicolumn{6}{c}{}\\
\bottomrule
\end{tabular}
}
  \label{tb:RSparams}
\end{table}

Invariant mass-squared projections for \RS are shown in Fig.~\ref{fig:RSPhase3} together with the expected distributions from the baseline model.
The coupling parameters, fit fractions and other quantities for this model are shown in Table~\ref{tb:RSparams}.
The $\chi^2$ per \dof for this model is calculated to be $40483/32701 = 1.238$, which indicates that although this is formally a poor fit, the model is providing a reasonable description of the data given the very large sample size. 
Three cascade contributions, from $a_1(1260)^+$, $\PK_1(1270)^-$ and $\PK(1460)^-$ resonances, 
are modelled using the three-body running-width treatment described in Sect.~\ref{sec:formalism}.
The masses and widths of these states are allowed to vary in the fit. 
The mass, width and coupling parameters for these resonances are presented in Tables \ref{tb:a1params},~\ref{tb:k1params} and \ref{tb:kparams}.
The values of these parameters are model dependent, in particular on the parametrisation of the running width described by Eq.~\ref{eq:TotalWidth} and of the form factors described by Eq.~\ref{eq:FormFactor}, and thus there is not a straightforward comparison with the values obtained by other experiments.

The largest contribution is found to come from the axial vector $a_1(1260)^{+}$, 
which is a result that was also obtained in the Mark III analysis \cite{PhysRevD.45.2196}.
This decay proceeds via the colour-favoured external $W$-emission diagram that is expected to dominate this final state.

There are also large contributions from the different orbital angular momentum configurations of the quasi two-body processes $\Dz\rightarrow\Kstarb(892)^{0}\rho(770)^{0}$, with a total contribution of around $20\%$. 
The polarisation structure of this component is not consistent with naive expectations, with the D wave being the dominant contribution and the overall hierarchy being $\mathrm{D}>\mathrm{S}>\mathrm{P}$. 
This result may be compared with that obtained for the study $\Dz\rightarrow\rho(770)^{0}\rho(770)^{0}$ in Ref.~\cite{dArgent:2017gzv}, where the D-wave polarisation of the amplitude was also found to be dominant. 

\begin{figure}
  \center \includegraphics[width=0.48\textwidth]{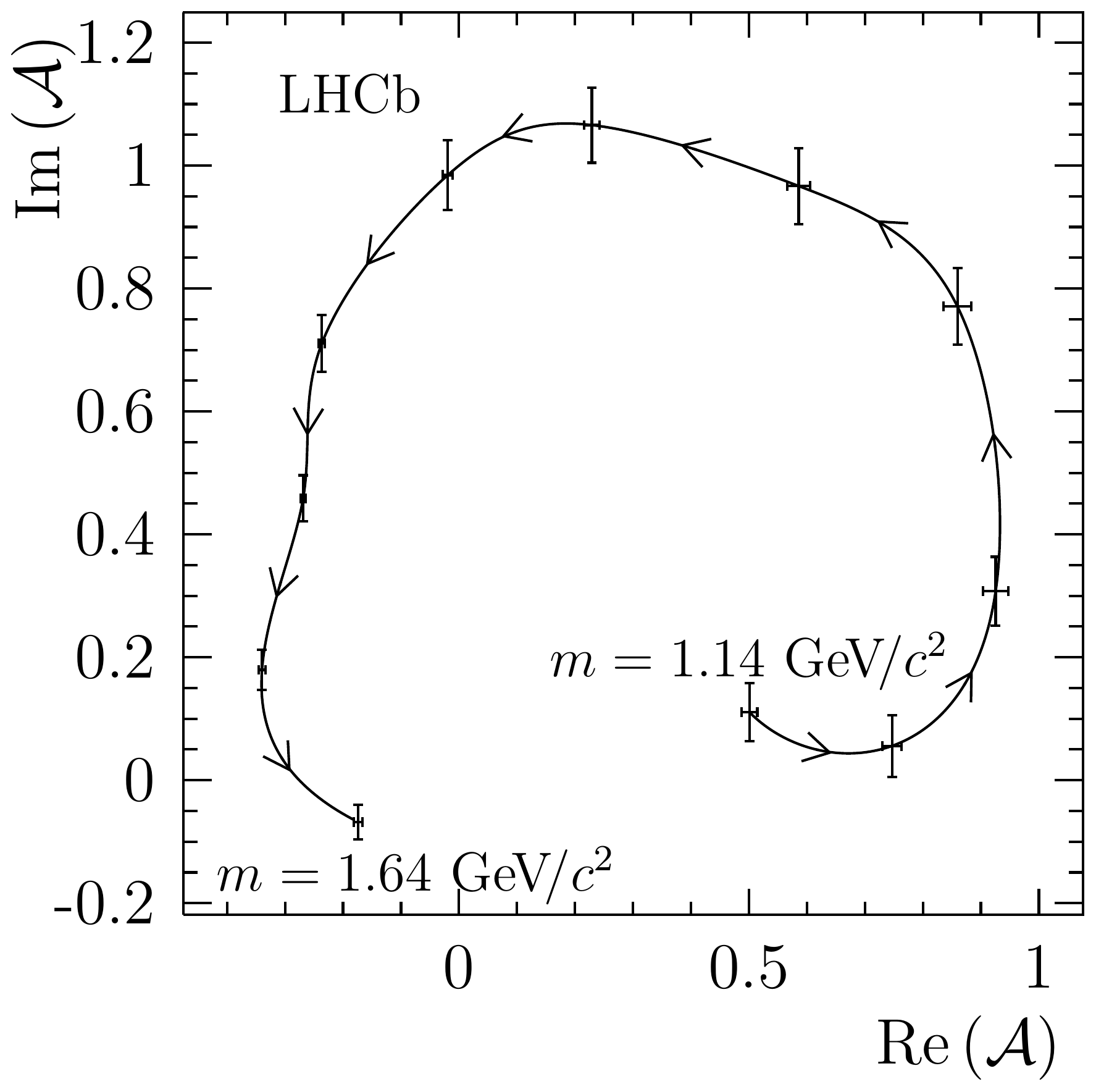} 
  \caption{Argand diagram for the model-independent partial-wave analysis (MIPWA) for the $\PK(1460)$ resonance. Points show the values of the amplitude that are determined by the fit, with only statistical uncertainties shown.}
  \label{fig:MIPWA}
\end{figure}

A significant contribution is found from the pseudoscalar state $\PK(1460)^{-}$. 
This resonance is a $2^{1}S_0$ excitation of the kaon \cite{PhysRevD.32.189}. 
Evidence for this state has been reported in partial-wave analyses
of the process $\Kpm\proton\rightarrow\Kpm\pip\pim\proton$ \cite{Daum:1981hb,Brandenburg:1976pg},
manifesting itself as a $0^{-}$ state with mass $\approx 1400\mevcc$ and width $\approx 250\mevcc$, coupling to the $\Kstarb(892)^{0}\pim$ and $[\pim\pip]^{L=0}\Km$ channels. 
The intermediate decays of the $\PK(1460)^{-}$ meson are found to be roughly consistent with previous studies, with approximately equal partial widths to $\Kstarb(892)^{0}\pim$ and $[\pip\pim]^{L=0}\Km$.
The resonant nature of this state is confirmed using a model-independent partial-wave analysis (MIPWA), following the method first used by the E791 collaboration~\cite{Meadows:2005ag,Aitala:2005yh}.
The relativistic Breit-Wigner lineshape is replaced by a parametrisation that treats the real and imaginary parts of the amplitude at 15 discrete positions in $s_{\Km\pip\pim}$ as independent pairs of free parameters to be determined by the fit. 
The amplitude is then modelled elsewhere by interpolating between these values using cubic splines \cite{doi:10.1137/0705007}.
The Argand diagram for this amplitude is shown in Fig.~\ref{fig:MIPWA}, with points indicating the values determined by the fit, and demonstrates the phase motion expected from a resonance.

Four-body weak decays contain amplitudes that are both even, such as $\PD\rightarrow [VV^{\prime}]^{L=0,2}$,
 where $V$ and $V^{\prime}$ are vector resonances, and odd, such as $\PD\rightarrow [VV^{\prime}]^{L=1}$, under parity transformations. 
Interference between these amplitudes can give rise to parity asymmetries which are different in $\Dz$ and $\Dzb$ decays. 
These asymmetries are the result of strong-phase differences, but can be mistaken for $\CP$ asymmetries \cite{Bigi:2001sg}.
Both sources of asymmetry can be studied by examining the distribution of the angle between the decay planes of the two quasi two-body systems, $\phi$,  which can be constructed from the three-momenta $\mathbf{p}$ of the decay products in the rest frame of the \Dz meson as  
\begin{align}
  \begin{split}
    \cos(\phi) &= \mathbf{\hat{n}}_{\Km\pi^{+}} \cdot \mathbf{\hat{n}}_{\pim\pi^{+}} \\
    \sin(\phi) &= \frac{\mathbf{p}_{\pi^{+}} \cdot \mathbf{\hat{n}}_{\Km\pi^{+}}} { \left| \mathbf{p}_{\pi^{+}} \times \mathbf{\hat{p}}_{\Km\pi^{+}} \right|} , 
  \end{split}
\end{align}
where $\mathbf{\hat{n}}_{ab}$ is the direction normal to the decay plane of a two-particle system $ab$,
\begin{equation}
  \mathbf{\hat{n}}_{ab} = \frac{ \mathbf{p}_a \times \mathbf{p}_b }{\left| \mathbf{p}_a \times \mathbf{p}_b \right| },
\end{equation}
and $\mathbf{\hat{p}}_{\Km\pip}$ is the direction of the combined momentum of the $\Km\pip$ system. 

\begin{figure}
  \centering  
  \includegraphics[width=0.48\textwidth]{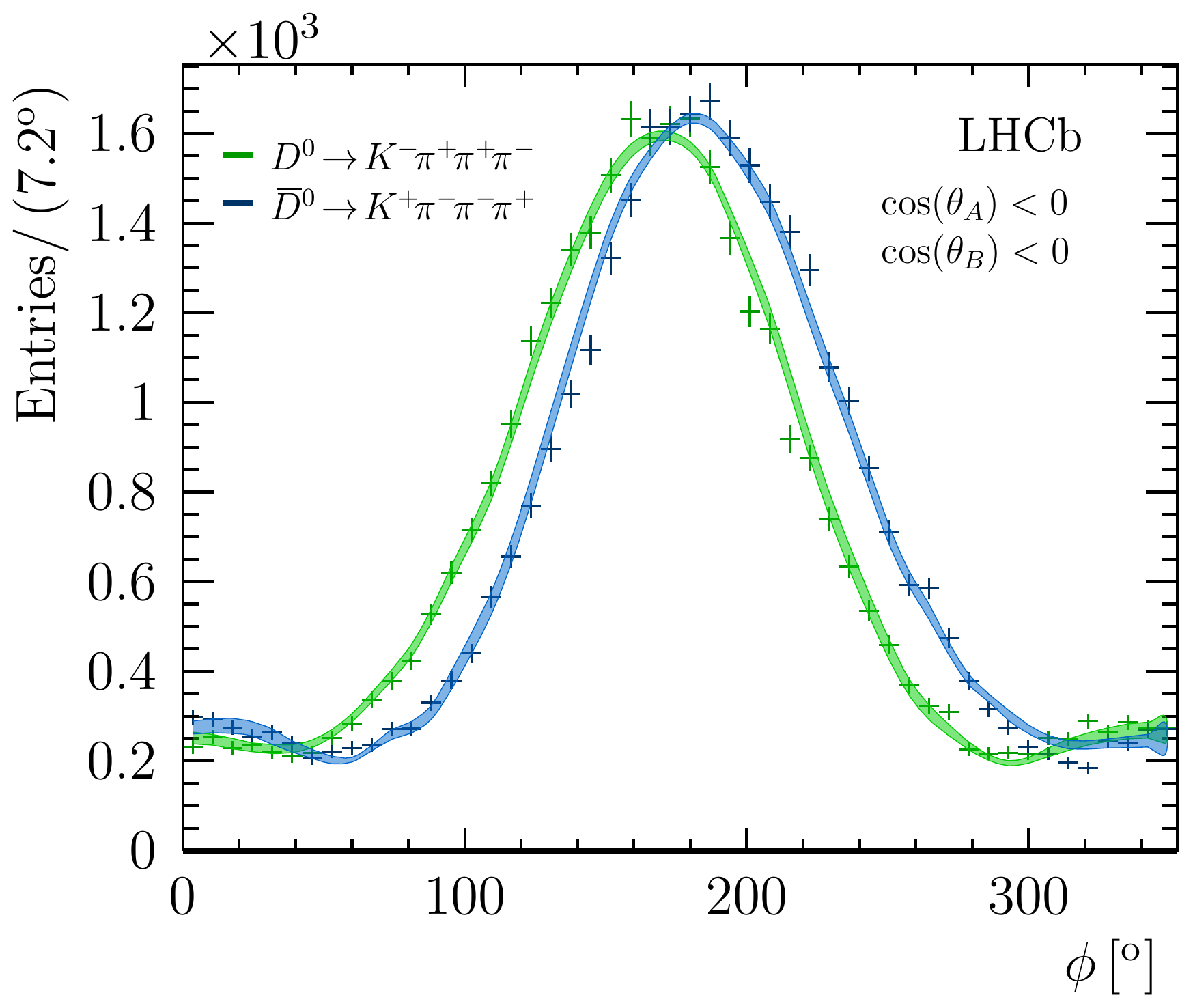}  \includegraphics[width=0.48\textwidth]{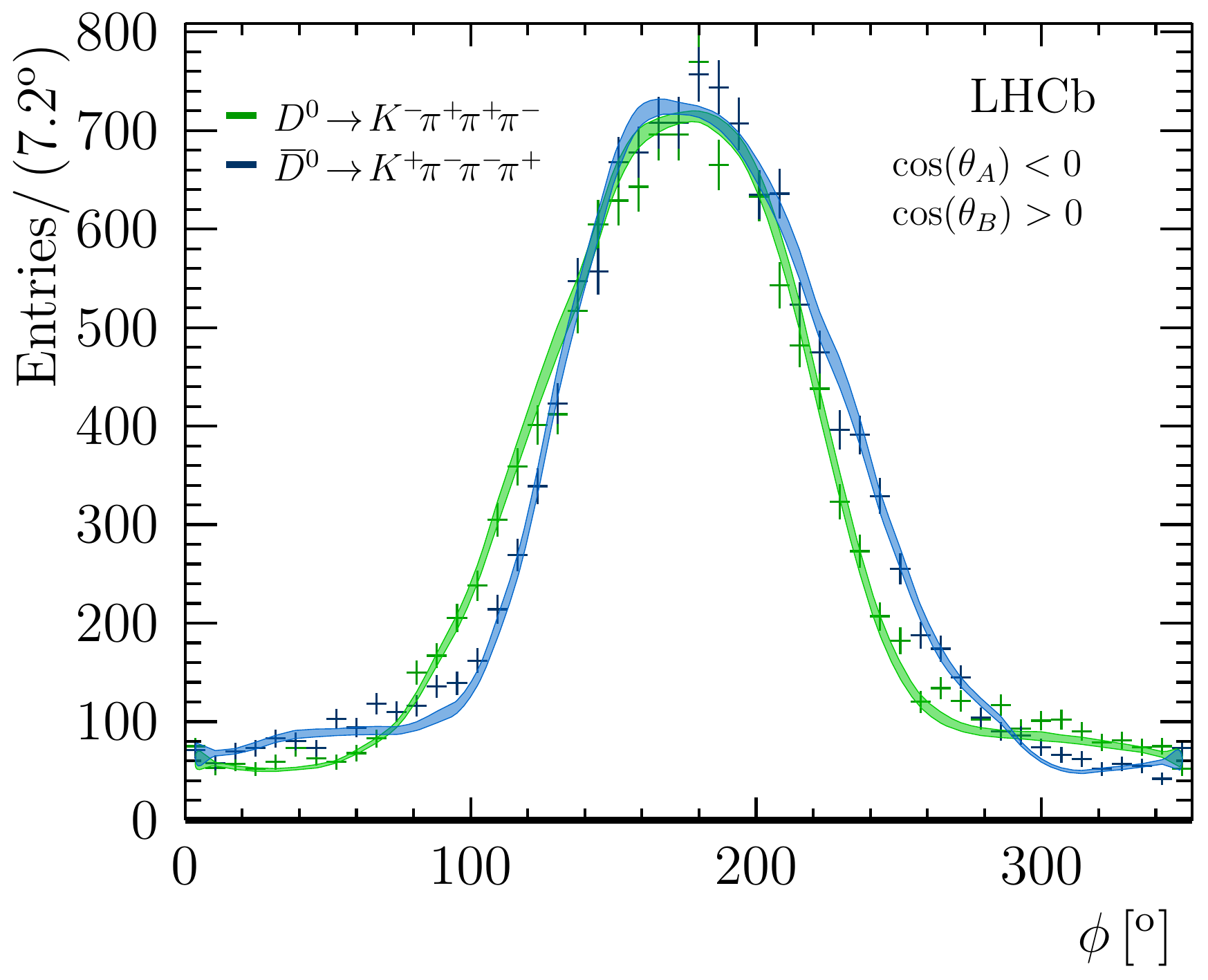}
  
  \includegraphics[width=0.48\textwidth]{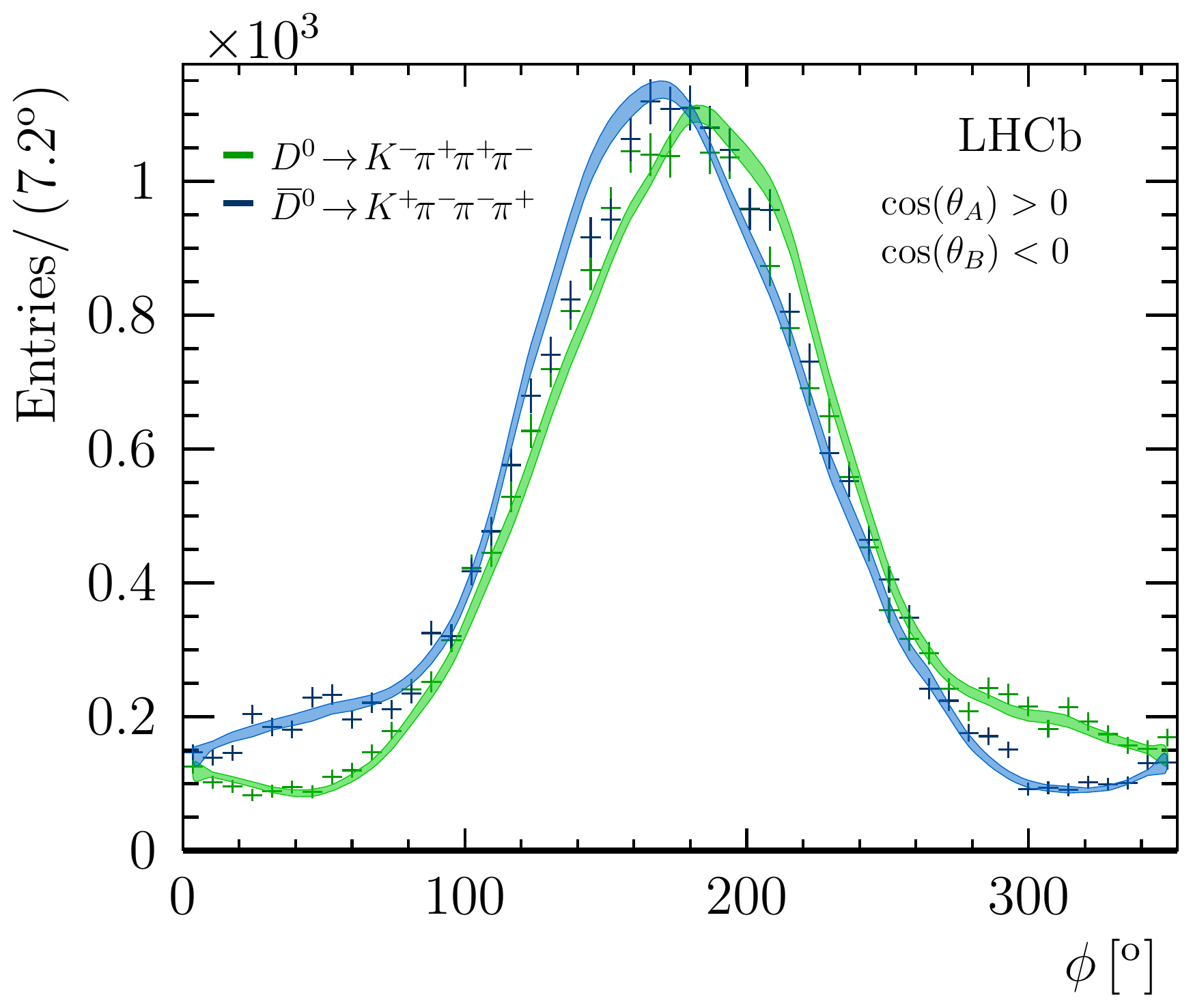}  \includegraphics[width=0.48\textwidth]{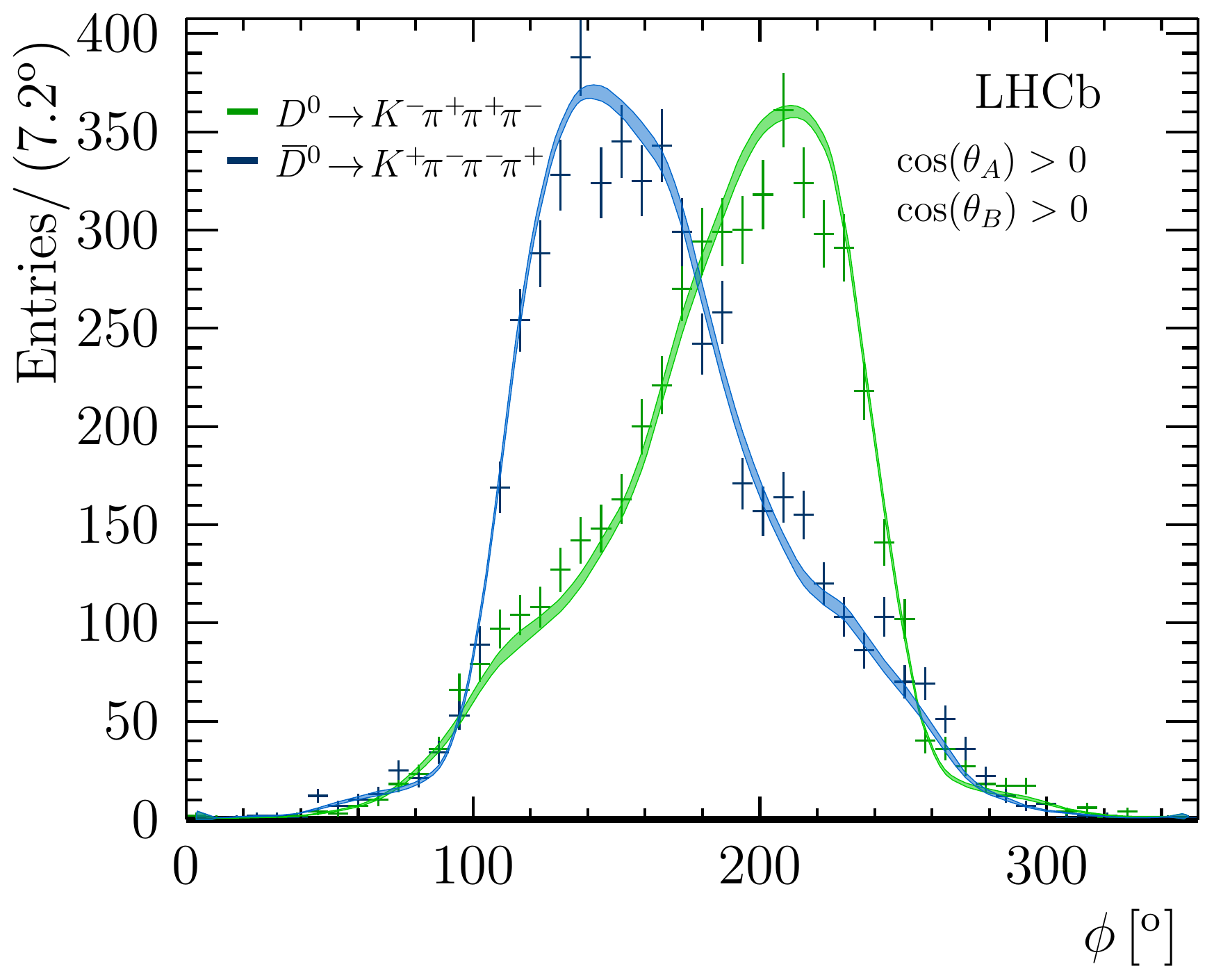}
  
  \caption{Parity violating distributions for the RS decay in the $\Kstarb(892)^{0}\rho(770)^{0}$ region defined by $\pm 35 \mev$($\pm 100 \mev$) mass windows about the nominal $\Kstarb(892)^{0}$ $(\rho(770)^{0})$ masses. Bands show the predictions of the fitted model including systematic uncertainties. }
  \label{fig:ParityViolation}
\end{figure}

The interference between $P$-even and $P$-odd amplitudes averages to zero when integrated over the entire phase space. 
Therefore, the angle $\phi$ is studied in regions of phase space.  
The region of the $\Kstarb(892)^{0}$ and $\rho(770)^{0}$ resonances is studied 
 as the largest $P$-odd amplitude is the decay $\Dz\rightarrow[\Kstarb(892)^{0}\rho(770)^{0}]^{L=1}$.
Selecting this region allows the identical pions to be distinguished, by one being part of the $\Kstarb(892)^{0}$-like system and the other in the $\rho(770)^{0}$-like system. 
The data in this region are shown in Fig.~\ref{fig:ParityViolation}, divided into quadrants of helicity angles, $\theta_A $ and $ \theta_B$, defined as the angle between the $\Km/\pim$ and the $\Dz$ in the rest frame of the $\Km\pip/\pim\pip$ system.
The distributions show clear asymmetries under reflection about $180\degrees$, indicating parity nonconservation. 
However, equal and opposite asymmetries are observed in the $\CP$-conjugate mode $\Dzb\rightarrow\Kp\pim\pim\pip$, indicating that these asymmetries originate from strong phases, rather than from $\CP$-violating effects. 
Bands show the expected asymmetries based on the amplitude model, which has been constructed according to the $\CP$-conserving hypothesis, and show reasonable agreement with the data.

\FloatBarrier
\begin{table}
  \centering
  \caption{Table of fit fractions and coupling parameters for the component involving the \AONE meson, 
  from the fit performed on the RS decay \RS. 
  The coupling parameters are defined with respect to the $\AONE\rightarrow\rho(770)^{0}\pip$ coupling. 
  For each parameter, the first uncertainty is statistical and the second systematic. }
  \scalebox{0.9}{\begin{tabular}{l 
>{\collectcell\num}r<{\endcollectcell} @{${}\pm{}$} >{\collectcell\num}l<{\endcollectcell} @{${}\pm{}$} >{\collectcell\num}l<{\endcollectcell}
>{\collectcell\num}r<{\endcollectcell} @{${}\pm{}$} >{\collectcell\num}l<{\endcollectcell} @{${}\pm{}$} >{\collectcell\num}l<{\endcollectcell}
>{\collectcell\num}r<{\endcollectcell} @{${}\pm{}$} >{\collectcell\num}l<{\endcollectcell} @{${}\pm{}$} >{\collectcell\num}l<{\endcollectcell}
}
\toprule
  \multicolumn{10}{l}
  { $a_1(1260)^{+}$\quad$m_0=1195.05\pm1.05\pm6.33\mevcc$; $\Gamma_0=422.01\pm2.10\pm12.72\mevcc$ } \\
  \rule{0pt}{3ex}    
  & \multicolumn{3}{c}{Partial Fractions [\%]} & \multicolumn{3}{c}{$\left|g\right|$} & \multicolumn{3}{c}{$\mathrm{arg}(g) [^o]$}\\
\midrule
$\quad \rho(770)^{0}\pi^{+}$ & 89.75 & 0.45 & 1.00 & \multicolumn{6}{c}{}\\
$\quad \left[\pi^{+}\pi^{-}\right]^{L=0}\pi^{+}$ & 2.42 & 0.06 & 0.12 & \multicolumn{6}{c}{}\\
$\quad \quad \beta_1$ & \multicolumn{3}{c}{} & 0.991 & 0.018 & 0.037 & -22.2 & 1.0 & 1.2\\
$\quad \quad \beta_0$ & \multicolumn{3}{c}{} & 0.291 & 0.007 & 0.017 & 165.8 & 1.3 & 3.1\\
$\quad \quad f_{\pi\pi}$ & \multicolumn{3}{c}{} & 0.117 & 0.002 & 0.007 & 170.5 & 1.2 & 2.2\\
$\quad \left[\rho(770)^{0}\pi^{+}\right]^{L=2}$ & 0.85 & 0.03 & 0.06 & 0.582 & 0.011 & 0.027 & -152.8 & 1.2 & 2.5\\
\bottomrule
\end{tabular}
}
  \label{tb:a1params}
\end{table}
\begin{table}
  \centering
  \caption{Table of fit fractions and coupling parameters for the component involving the $\KONE{1270}^{-}$ meson,
  from the fit performed on the RS decay \RS. 
  The coupling parameters are defined with respect to the $\KONE{1270}^{-}\rightarrow\rho(770)^{0}\Km$ coupling. 
  For each parameter, the first uncertainty is statistical and the second systematic. }
  \scalebox{0.9}{\begin{tabular}{l 
>{\collectcell\num}r<{\endcollectcell} @{${}\pm{}$} >{\collectcell\num}l<{\endcollectcell} @{${}\pm{}$} >{\collectcell\num}l<{\endcollectcell}
>{\collectcell\num}r<{\endcollectcell} @{${}\pm{}$} >{\collectcell\num}l<{\endcollectcell} @{${}\pm{}$} >{\collectcell\num}l<{\endcollectcell}
>{\collectcell\num}r<{\endcollectcell} @{${}\pm{}$} >{\collectcell\num}l<{\endcollectcell} @{${}\pm{}$} >{\collectcell\num}l<{\endcollectcell}
}
\toprule
  \multicolumn{10}{l}{ $K_1(1270)^{-}$\quad$m_0=1289.81\pm0.56\pm1.66 \mevcc$; $\Gamma_0=116.11\pm1.65\pm2.96\mevcc$}  \\
  \rule{0pt}{3ex}    
 & \multicolumn{3}{c}{Partial Fractions [\%]} & \multicolumn{3}{c}{$\left|g\right|$} & \multicolumn{3}{c}{$\mathrm{arg}(g) [^\mathrm{o}]$}\\
\midrule
$\quad \rho(770)^{0}\Km$ & 96.30 & 1.64 & 6.61 & \multicolumn{6}{c}{}\\
$\quad \rho(1450)^{0}\Km$ & 49.09 & 1.58 & 11.54 & 2.016 & 0.026 & 0.211 & -119.5 & 0.9 & 2.3\\
$\quad \Kstarb(892)^{0}\pi^{-}$ & 27.08 & 0.64 & 2.82 & 0.388 & 0.007 & 0.033 & -172.6 & 1.1 & 6.0\\
$\quad \left[\Km\pi^{+}\right]^{L=0}\pi^{-}$ & 22.90 & 0.72 & 1.89 & 0.554 & 0.010 & 0.037 & 53.2 & 1.1 & 1.9\\
$\quad \left[\Kstarb(892)^{0}\pi^{-}\right]^{L=2}$ & 3.47 & 0.17 & 0.31 & 0.769 & 0.021 & 0.048 & -19.3 & 1.6 & 6.7\\
$\quad \omega(782)\left[\pi^{+}\pi^{-}\right]\Km$ & 1.65 & 0.11 & 0.16 & 0.146 & 0.005 & 0.009 & 9.0 & 2.1 & 5.7\\
\bottomrule
\end{tabular}
}
  \label{tb:k1params}
\end{table}
\begin{table}
  \centering
  \caption{Table of fit fractions and coupling parameters for the component involving the \Kexc meson, 
  from the fit performed on the RS decay \RS.
  The coupling parameters are defined with respect to the $\Kexc\rightarrow\Kstarb(892)^{0}\pim$ coupling. 
  For each parameter, the first uncertainty is statistical and the second systematic. }
  \scalebox{0.9}{\begin{tabular}{l 
>{\collectcell\num}r<{\endcollectcell} @{${}\pm{}$} >{\collectcell\num}l<{\endcollectcell} @{${}\pm{}$} >{\collectcell\num}l<{\endcollectcell}
>{\collectcell\num}r<{\endcollectcell} @{${}\pm{}$} >{\collectcell\num}l<{\endcollectcell} @{${}\pm{}$} >{\collectcell\num}l<{\endcollectcell}
>{\collectcell\num}r<{\endcollectcell} @{${}\pm{}$} >{\collectcell\num}l<{\endcollectcell} @{${}\pm{}$} >{\collectcell\num}l<{\endcollectcell}
}
\toprule
  \multicolumn{10}{c}{ $K(1460)^{-}$\quad$m_0 = 1482.40\pm3.58\pm15.22\mevcc$ ; $\Gamma_0 = 335.60\pm6.20\pm8.65 \mevcc$ } \\
  \rule{0pt}{3ex}    
 & \multicolumn{3}{c}{Partial Fractions [\%]} & \multicolumn{3}{c}{$\left|g\right|$} & \multicolumn{3}{c}{$\mathrm{arg}(g) [^o]$}\\
\midrule
$\quad \Kstarb(892)^{0}\pi^{-}$ & 51.39 & 1.00 & 1.71 & \multicolumn{6}{c}{}\\
$\quad \left[\pi^{+}\pi^{-}\right]^{L=0}K^{-}$ & 31.23 & 0.83 & 1.78 & \multicolumn{6}{c}{}\\
$\quad \quad f_{KK}$ & \multicolumn{3}{c}{} & 1.819 & 0.059 & 0.189 & -80.8 & 2.2 & 6.6\\
$\quad \quad \beta_1$ & \multicolumn{3}{c}{} & 0.813 & 0.032 & 0.136 & 112.9 & 2.6 & 9.5\\
$\quad \quad \beta_0$ & \multicolumn{3}{c}{} & 0.315 & 0.010 & 0.022 & 46.7 & 1.9 & 3.0\\
\bottomrule
\end{tabular}
}
  \label{tb:kparams}
\end{table}
\FloatBarrier

\subsection{Results for the WS decay}

\label{sec:WSModel}
\begin{figure}[!htb]
  \centering
  \includegraphics[width=0.48\textwidth]{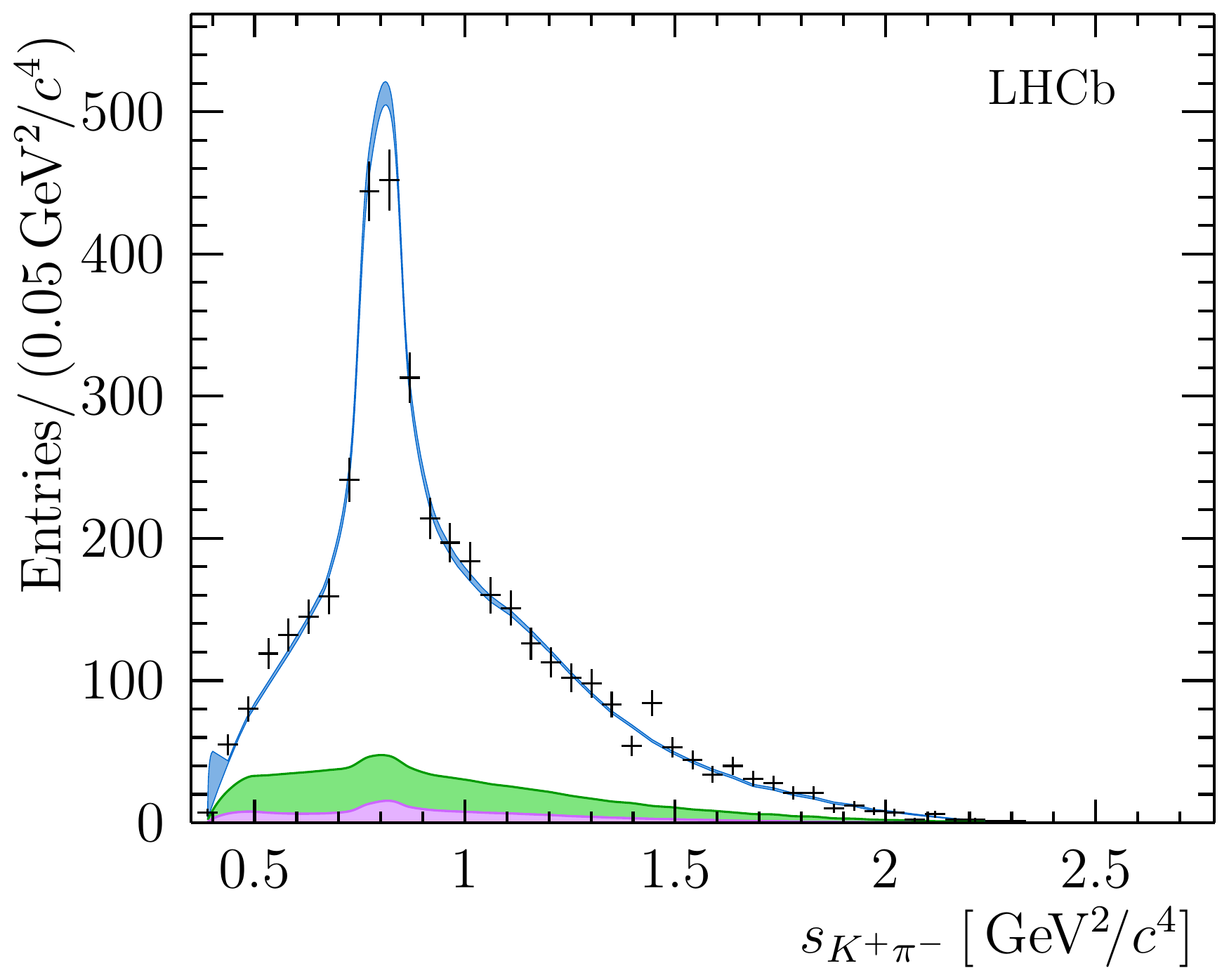} \includegraphics[width=0.48\textwidth]{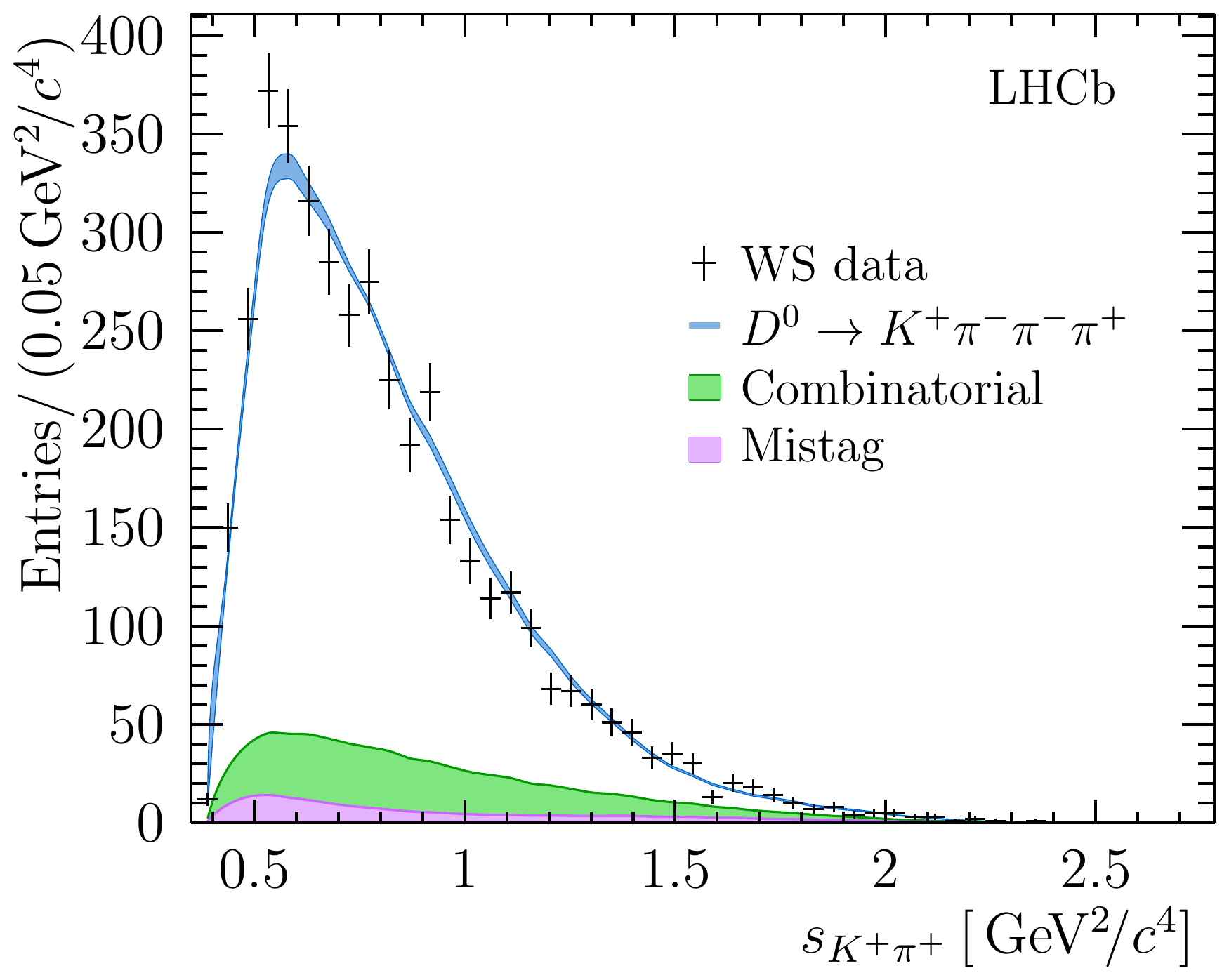}
  
  \includegraphics[width=0.48\textwidth]{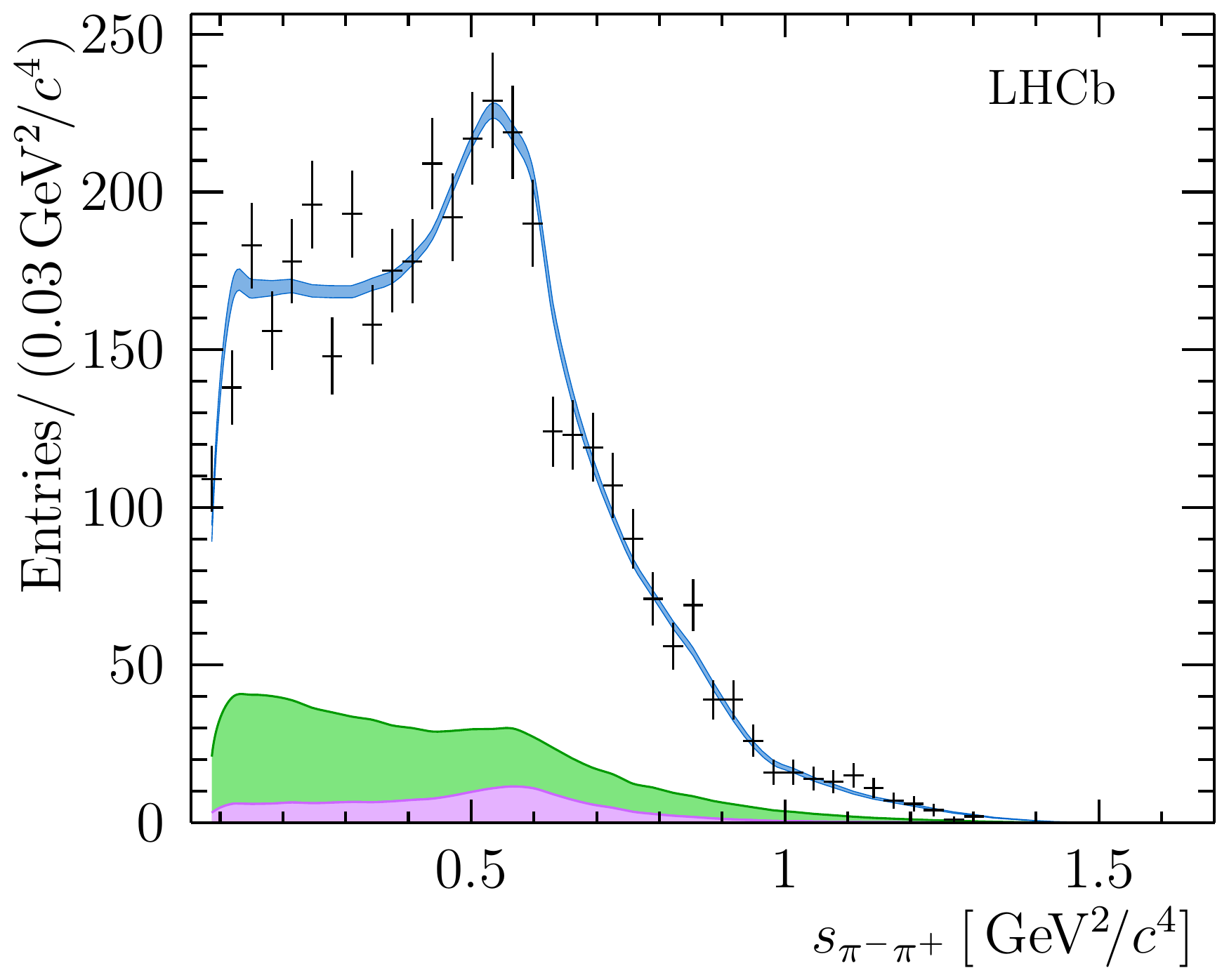} \includegraphics[width=0.48\textwidth]{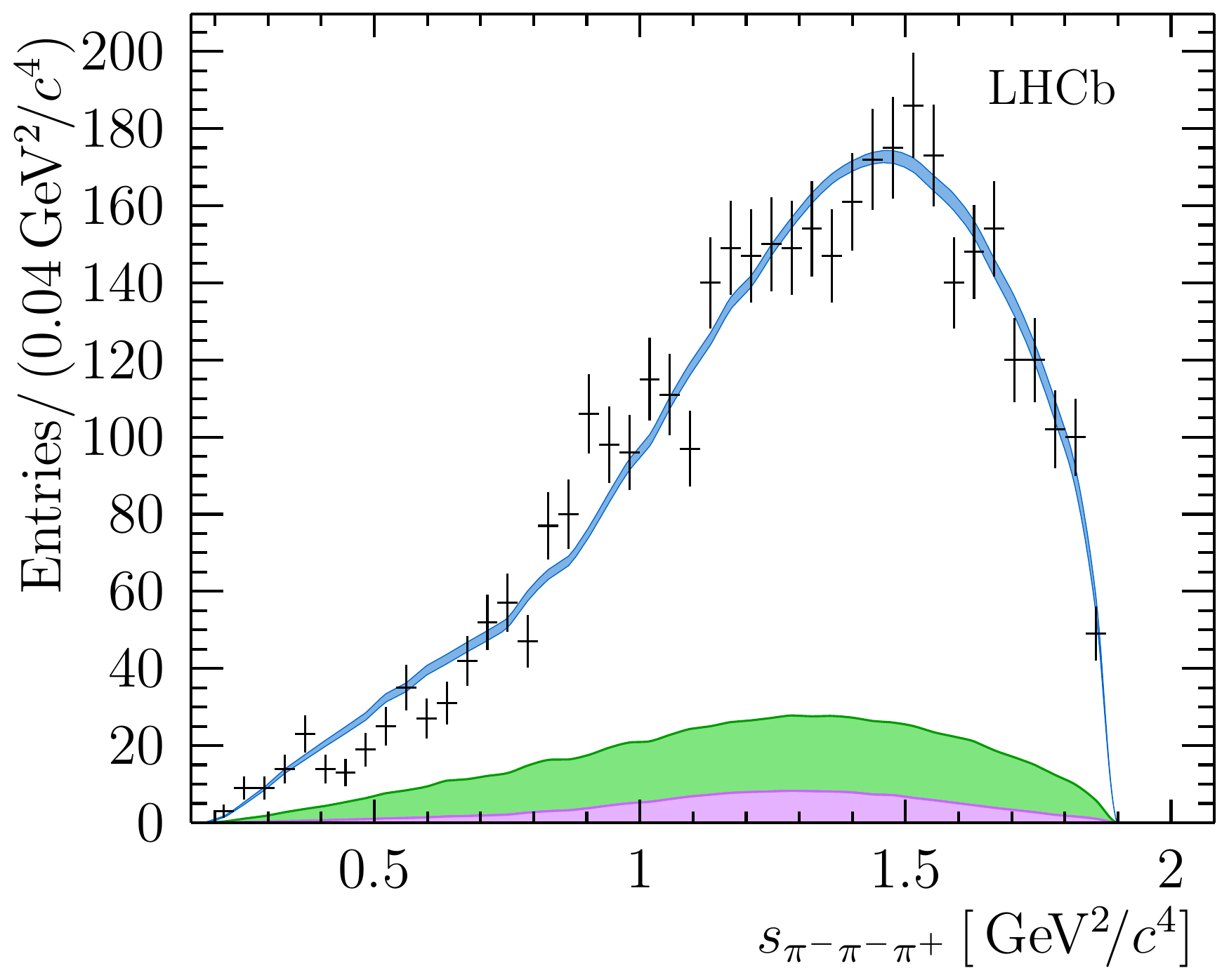}
  
  \includegraphics[width=0.48\textwidth]{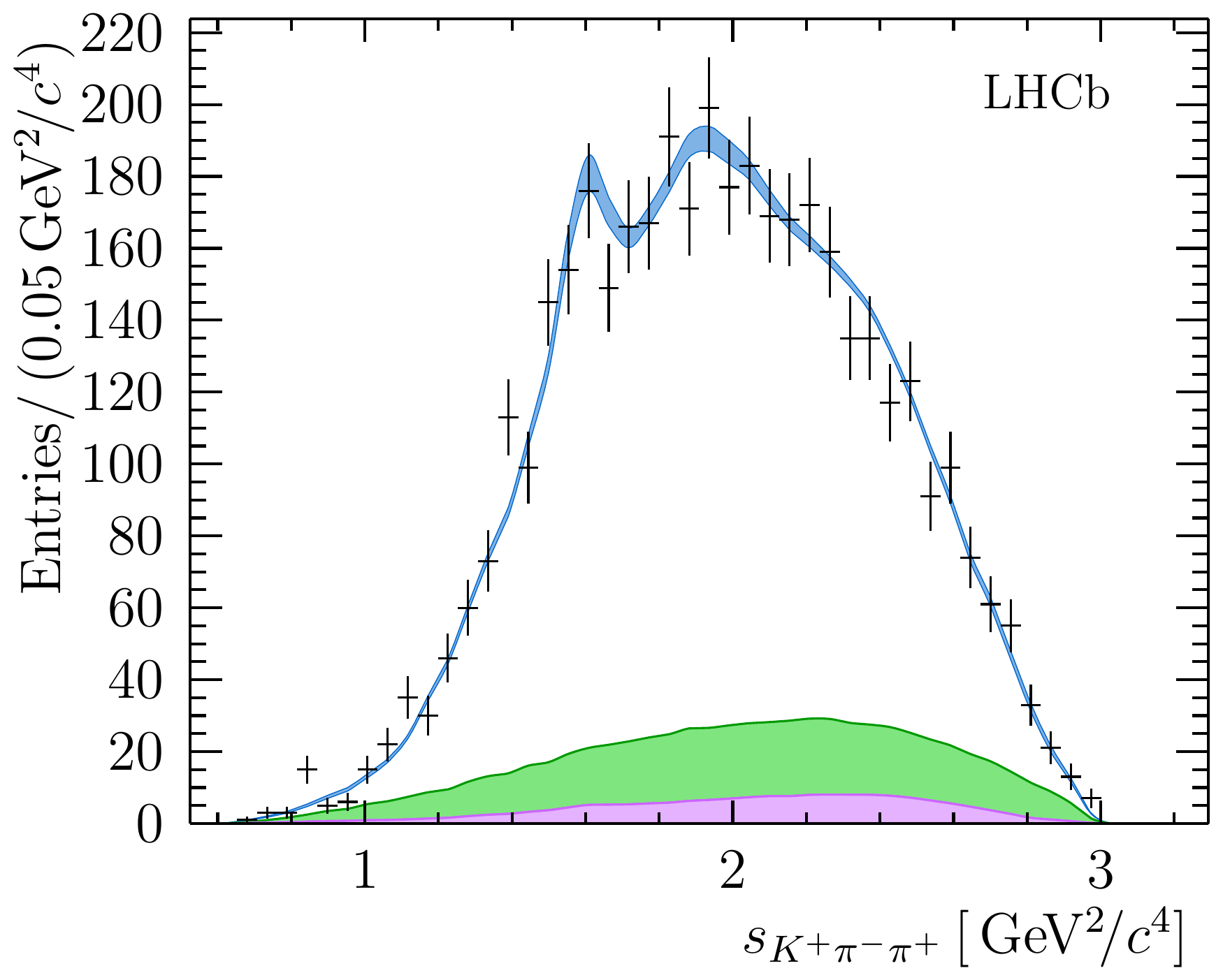} \includegraphics[width=0.48\textwidth]{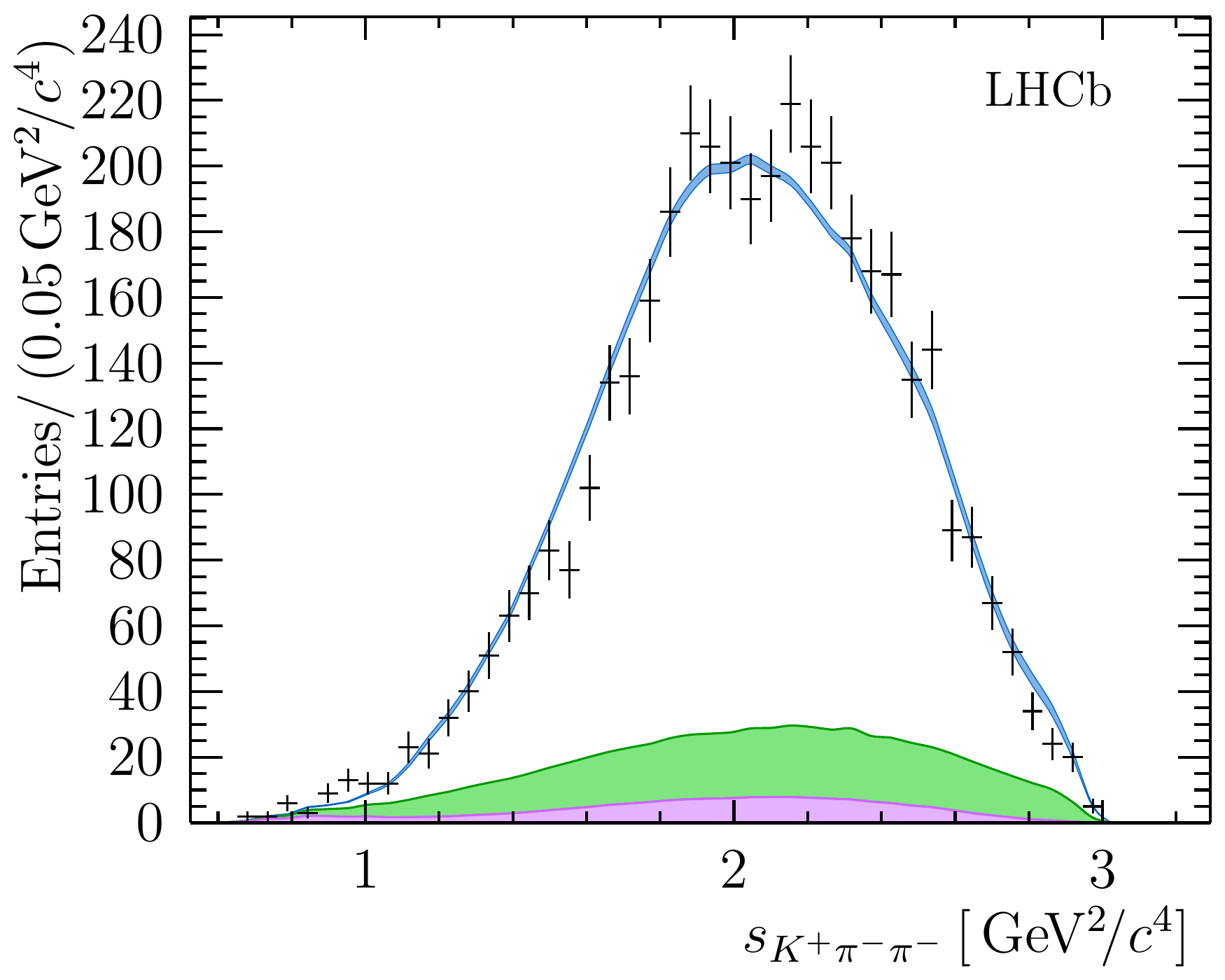}
  
  \caption{ \label{fig:WSPhase3} Distributions for six invariant-mass observables in the WS decay \WS.
  Bands indicate the expectation from the model, with the width of the band indicating the total systematic uncertainty. 
  The total background contribution is shown as a filled area, with the lower region indicating the expected contribution from mistagged $\Dzb\rightarrow\Kp\pim\pim\pip$ decays.
  In figures that involve a single negatively-charged pion, one of the two identical pions is selected randomly.}
\end{figure}

\begin{table}
  \caption{ \label{tb:WSparams} Fit fractions and coupling parameters for the WS decay \WS. 
  For each parameter, the first uncertainty is statistical and the second systematic.
  Couplings $g$ are defined with respect to the coupling to the decay $\Dz\rightarrow[\PK^{*}(892)^{0}\rho(770)^{0}]^{L=2}$. 
  Also given are the $\chi^2$ and the number of degrees of freedom ($\nu$) from the fit and their ratio. }
  \scalebox{0.9}{\begin{tabular}{l 
>{\collectcell\num}r<{\endcollectcell} @{${}\pm{}$} >{\collectcell\num}l<{\endcollectcell} @{${}\pm{}$} >{\collectcell\num}l<{\endcollectcell}
>{\collectcell\num}r<{\endcollectcell} @{${}\pm{}$} >{\collectcell\num}l<{\endcollectcell} @{${}\pm{}$} >{\collectcell\num}l<{\endcollectcell}
>{\collectcell\num}r<{\endcollectcell} @{${}\pm{}$} >{\collectcell\num}l<{\endcollectcell} @{${}\pm{}$} >{\collectcell\num}l<{\endcollectcell}
}
\toprule
 & \multicolumn{3}{c}{Fit Fraction [\%]} & \multicolumn{3}{c}{$\left|g\right|$} & \multicolumn{3}{c}{$\mathrm{arg}(g) [^o]$}\\
\midrule
  $\left[K^*(892)^{0}\rho(770)^{0}\right]^{L=0}$ & 9.62 & 1.58 & 1.03 & 0.205 & 0.019 & 0.010 & -8.5 & 4.7 & 4.4\\
$\left[K^*(892)^{0}\rho(770)^{0}\right]^{L=1}$ & 8.42 & 0.83 & 0.57 & 0.390 & 0.029 & 0.006 & -91.4 & 4.7 & 4.1\\
$\left[K^*(892)^{0}\rho(770)^{0}\right]^{L=2}$ & 10.19 & 1.03 & 0.79 & \multicolumn{3}{c}{} & \multicolumn{3}{c}{}\\
  $\left[\rho(1450)^{0}K^*(892)^{0}\right]^{L=0}$ & 8.16 & 1.24 & 1.69 & 0.541 & 0.042 & 0.055 & -21.8 & 6.5 & 5.5\\
\midrule
$K_{1}(1270)^{+}\pi^{-}$ & 18.15 & 1.11 & 2.30 & 0.653 & 0.040 & 0.058 & -110.7 & 5.1 & 4.9\\
$K_{1}(1400)^{+}\left[K^*(892)^{0}\pi^{+}\right]\pi^{-}$ & 26.55 & 1.97 & 2.13 & 0.560 & 0.037 & 0.031 & 29.8 & 4.2 & 4.6\\
\midrule
$\left[K^{+}\pi^{-}\right]^{L=0}\left[\pi^{+}\pi^{-}\right]^{L=0}$ & 20.90 & 1.30 & 1.50 & \multicolumn{3}{c}{} & \multicolumn{3}{c}{}\\
$\quad \alpha_{3/2}$ & \multicolumn{3}{c}{} & 0.686 & 0.043 & 0.022 & -149.4 & 4.3 & 2.9\\
$\quad \beta_1$ & \multicolumn{3}{c}{} & 0.438 & 0.044 & 0.030 & -132.4 & 6.5 & 3.0\\
$\quad f_{\pi\pi}$ & \multicolumn{3}{c}{} & 0.050 & 0.006 & 0.005 & 74.8 & 7.5 & 5.3\\
\midrule
Sum of Fit Fractions & 101.99 & 2.90 & 2.85 & \multicolumn{6}{c}{}\\
$\chi^2 / \nu$ & \multicolumn{3}{c}{$350/239 = 1.463 $} & \multicolumn{6}{c}{}\\
\bottomrule
\end{tabular}
}
\end{table}

Invariant mass-squared distributions for \WS are shown in Fig.~\ref{fig:WSPhase3}. 
Large contributions are clearly seen in $s_{\Kp\pim}$ from the $\Kstar(892)^{0}$ resonance. 
The fit fractions and amplitudes of the baseline model are given in Table~\ref{tb:WSparams}. 
The $\chi^2$ per \dof for the fit to the WS data is $350/243=1.463$.
If the true WS amplitude has a comparable structure to the RS amplitude, it contains several decay chains at the $\mathcal{O}(1\%)$ level that cannot be satisfactorily resolved given the small sample size, and hence the quality of the WS fit is degraded by the absence of these subdominant contributions.  

Dominant contributions are found from the axial kaons, $\PK_1(1270)^{+}$ and $\PK_1(1400)^{+}$, 
which are related to the same colour-favoured $W$-emission diagram that dominates the RS decay, where it manifests itself in the $\AONE\Km$ component. 
The contribution from the $\PK_1(1400)^{+}$ resonance is larger than that from the $\PK_1(1270)^{+}$ resonance.
It is instructive to consider this behaviour in terms of the quark states, $^1P_1$ and $^3P_1$, which mix almost equally to produce the mass eigenstates, 
\begin{align}
  \begin{split}
    \ket{ \PK_1(1400) } &= \cos(\theta_K) \ket{ ^3P_1} - \sin(\theta_K) \ket{ ^1P_1 }  \\
    \ket{ \PK_1(1270) } &= \sin(\theta_K) \ket{ ^3P_1} + \cos(\theta_K) \ket{ ^1P_1 },
  \end{split}
\end{align}
where $\theta_K$ is a mixing angle. 
The mixing is somewhat less than maximal, with Ref.~\cite{PhysRevD.47.1252} reporting a preferred value of $\theta_K = (33^{+6}_{-2}) ^{\mathrm{o}}$. 
In the WS decay, the axial kaons are produced via a weak current, which is decoupled from the $^1P_1$ state in the $\grpsuthree$ flavour-symmetry limit.
If the mixing were maximal, the mass eigenstates would be produced equally, 
but a smaller mixing angle results in a preference for the $\PK_1(1400)$, 
which is qualitatively consistent with the pattern seen in data. 
In the RS decay, the axial kaons are not produced by the external weak current, and hence there is no reason to expect either quark state to be preferred. 
The relatively small contribution from the $\PK_1(1400)$ is then understood as a consequence of approximately equal production of the quark states. 

The coupling and shape parameters of the $\PK_1(1270)^{+}$ resonance are fixed to the values measured in the RS nominal fit. 
A fit is also performed with these coupling parameters free to vary, and the parameters are found to be consistent with those measured in the RS decay.

A large contribution is found from $\Dz\rightarrow\rho(1450)^{0}\Kstar(892)^{0}$ decays in all models that describe the data well. 
This contribution resembles a quasi nonresonant component due to the large width of the $\rho(1450)^0$ resonance, 
and is likely to be an effective representation of several smaller decay chains involving the $\Kstar(892)^0$ resonance that cannot be resolved with the current sample size. 

\subsection{Alternative parametrisations}
\label{sec:Ensemble}

\begin{table}
  \centering 
  \caption{ \label{tb:AlternativeModels} 
  Decay chains taken into account in alternative parametrisations of the RS decay mode ${\Dz\rightarrow\Km\pip\pip\pim}$. 
  For each chain, the fraction of models in the ensemble that contain this decay, together with the associated average fit fraction, $\langle \mathcal{F} \rangle$, are shown.
  Components are not tabulated if they contribute to all models in the ensemble, or if they contribute to less than 5\% of the models. 
  }
  \begin{tabular}{l 
r 
r 
}
\toprule
Decay chain & Fraction & $\langle \mathcal{F} \rangle$\\
 & of models $[\%]$ & $[\%]$\\
\midrule
$K_{1}(1270)^{-}\left[\rho(1450)^{0}K^{-}\right]\pi^{+}$ & 68.9 & 1.61\\
$K_{1}(1400)^{-}\left[\rho(1450)^{0}K^{-}\right]\pi^{+}$ & 33.4 & 0.34\\
$a_{1}(1640)^{+}\left[\left[\pi^+\pi^-\right]^{L=0}\pi^{+}\right]K^{-}$ & 23.1 & 2.47\\
$K_{1}(1270)^{-}\left[\Kstarb(1680)^{0}\pi^{-}\right]\pi^{+}$ & 18.4 & 0.38\\
$K_{1}(1270)^{-}\left[\Kstarb(1410)^{0}\pi^{-}\right]\pi^{+}$ & 12.0 & 0.29\\
$K_{2}^{*}(1430)^{-}\left[\Kstarb(1410)^{0}\pi^{-}\right]\pi^{+}$ & 10.4 & 0.12\\
$K^{*}(1680)^{-}\left[\rho(770)^{0}K^{-}\right]\pi^{+}$ & 10.4 & 0.07\\
$K_{2}^{*}(1430)^{-}\left[\rho(1450)^{0}K^{-}\right]\pi^{+}$ & 10.4 & 0.10\\
$K_{2}^{*}(1430)^{-}\left[\Kstarb(1680)^{0}\pi^{-}\right]\pi^{+}$ & 10.4 & 0.13\\
$K_{1}(1400)^{-}\left[\rho(770)^{0}K^{-}\right]\pi^{+}$ & 10.4 & 0.44\\
$K_{1}(1400)^{-}\left[\Kstarb(1410)^{0}\pi^{-}\right]\pi^{+}$ & 10.4 & 0.11\\
  $K(1460)^{-}\left[ \Kbar^{*}_{2}(1430)^{0}\pi^{-}\right]\pi^{+}$ & 10.0 & 0.06\\
\bottomrule
\end{tabular}

\end{table}

\begin{table}
  \centering
  \caption{ \label{tb:FitFractionSystematicsk1270} Dependence of fit fractions (and partial fractions) on the  choice of the RS model. 
  This dependence is expressed as the mean value and the RMS of the values in the ensemble. 
  Also shown is the fit fractions of the baseline model presented in Sect.~\ref{sec:RSModel}. } 
  \begin{tabular}{l 
>{\collectcell\num}r<{\endcollectcell} @{${}\pm{}$} >{\collectcell\num}l<{\endcollectcell} @{${}\pm{}$} >{\collectcell\num}r<{\endcollectcell}
>{\collectcell\num}r<{\endcollectcell} @{${}\pm{}$} >{\collectcell\num}l<{\endcollectcell}
}
\toprule
 & \multicolumn{5}{c}{(Partial) Fraction [\%]}\\
  & \multicolumn{3}{c}{Baseline} & \multicolumn{2}{c}{Ensemble} \\
  & \multicolumn{3}{c}{} & \multicolumn{1}{r}{Mean} & \multicolumn{1}{l}{\!RMS} \\
\midrule
  $\left[\Kstarb(892)^{0}\rho(770)^{0}\right]^{L=0}$                                                 & 7.34  & 0.08 & 0.47 & 7.10  & 0.13\\
  $\left[\Kstarb(892)^{0}\rho(770)^{0}\right]^{L=1}$                              & 6.03  & 0.05 & 0.25 & 6.00  & 0.12\\
  $\left[\Kstarb(892)^{0}\rho(770)^{0}\right]^{L=2}$                              & 8.47  & 0.09 & 0.67 & 8.42  & 0.20\\
  $\left[\rho(1450)^{0}\Kstarb(892)^{0}\right]^{L=0}$                                                & 0.61  & 0.04 & 0.17 & 0.65  & 0.13\\
  $\left[\rho(1450)^{0}\Kstarb(892)^{0}\right]^{L=1}$                             & 1.98  & 0.03 & 0.33 & 1.91  & 0.06\\
  $\left[\rho(1450)^{0}\Kstarb(892)^{0}\right]^{L=2}$                             & 0.46  & 0.03 & 0.15 & 0.46  & 0.05\\
\midrule
  $\rho(770)^{0}\left[\Km\pip\right]^{L=0}$                              & 0.93  & 0.03 & 0.05 & 1.08  & 0.12\\
  $\Kstarb(892)^{0}\left[\pi^{+}\pi^{-}\right]^{L=0}$                             & 2.35  & 0.09 & 0.33 & 2.19  & 0.34\\
\midrule
  $a_{1}(1260)^{+}K^{-}$                                             & 38.07 & 0.24 & 1.38 & 38.06 & 2.08\\
  $\quad \rho(770)^{0}\pi^{+}$                                                         & 89.75 & 0.45 & 1.00 & 86.66 & 4.52\\
  $\quad \left[\pi^{+}\pi^{-}\right]^{L=0}\pi^{+}$                                     & 2.42  & 0.06 & 0.12 & 3.01 & 1.02\\
  $\quad \left[\rho(770)^{0}\pi^{+}\right]^{L=2}$                                      & 0.85  & 0.03 & 0.06 & 0.80 & 0.10\\
  $K_{1}(1270)^{-}\pi^{+}$                                                    & 4.66  & 0.05 & 0.39 & 4.74 & 0.24\\
  $\quad \rho(770)^{0}K^{-}$                                                  & 96.30 & 1.64 & 6.61 & 77.04 & 9.22\\
  $\quad \rho(1450)^{0}K^{-}$                                                 & 49.09 & 1.58 & 11.54& 34.13 & 8.19\\
  $\quad \omega(782)\left[\pi^{+}\pi^{-}\right]K^{-}$                         & 1.65  & 0.11 & 0.16 & 1.70 & 0.15\\
  $\quad \Kstarb(892)^{0}\pi^{-}$                                                 & 27.08 & 0.64 & 2.82 & 26.95 & 2.52\\
  $\quad \left[\Kstarb(892)^{0}\pi^{-}\right]^{L=2}$                              & 3.47  & 0.17 & 0.31 & 3.57 & 0.49\\
  $\quad \left[K^{-}\pi^{+}\right]\pi^{-}$                                    & 22.90 & 0.72 & 1.89  & 20.39 & 2.89\\
  $K_{1}(1400)^{-}\left[\Kstarb(892)^{0}\pi^{-}\right]\pi^{+}$           & 1.15  & 0.04 & 0.20 & 1.23 & 0.10\\
\midrule
  $K_{2}^{*}(1430)^{-}\left[\Kstarb(892)^{0}\pi^{-}\right]\pi^{+}$       & 0.46 & 0.01 & 0.03 & 0.44 & 0.04\\
\midrule
  $K(1460)^{-}\pi^{+}$                                                        & 3.75 & 0.10 & 0.37 & 3.63 & 0.27\\
  $\quad \Kstarb(892)^{0}\pi^{-}$                                                 & 51.39 & 1.00 & 1.71 & 53.18 & 1.52\\
  $\quad \left[\pi^{+}\pi^{-}\right]^{L=0}K^{-}$                              & 31.23 & 0.83 & 1.78 & 30.46 & 1.19\\
\midrule
  $\left[\Km\pi^{+}\right]^{L=0}\left[\pi^{+}\pi^{-}\right]^{L=0}$            & 22.04 & 0.28 & 2.09 & 21.87 & 1.51\\
\bottomrule
\end{tabular}

\end{table}

\begin{table}
  \centering
  \caption{ \label{tb:MassesAndWidths} Dependence of the fitted masses and widths on the final choice of the RS model. 
  This dependence is expressed as the mean value and the RMS of the values in the ensemble. 
  The values found for the baseline model presented in Sect.~\ref{sec:RSModel} are reported for comparison.} 
  \begin{tabular}{l 
>{\collectcell\num}r<{\endcollectcell} @{${}\pm{}$} >{\collectcell\num}l<{\endcollectcell} @{${}\pm{}$} >{\collectcell\num}r<{\endcollectcell}
>{\collectcell\num}r<{\endcollectcell} @{${}\pm{}$} >{\collectcell\num}r<{\endcollectcell}
}
\toprule
  & \multicolumn{3}{c}{Baseline} & \multicolumn{2}{c}{Ensemble} \\
\midrule
  $m(a_1(1260)^{+}) [\mevcc]$ & 1195.05 & 1.05 & 6.33 & 1196.85 & 6.21\\
  $\Gamma(a_1(1260)^{+}) [\mevcc]$   & 422.01 & 2.10 & 12.72 & 420.92 & 8.70\\
  $m(K_{1}(1270)^{-}) [\mevcc]$ & 1289.81 & 0.56 & 1.66  & 1287.77 & 3.97\\
  $\Gamma(    K_{1}(1270)^{-}) [\mevcc]$ & 116.11 & 1.65 & 2.96 & 114.27 & 7.57\\
  $m(K(1460)^{-}) [\mevcc]$     & 1482.40 & 3.58 & 15.22 & 1474.60 & 12.28\\
  $\Gamma(     K(1460)^{-}) [\mevcc]$    & 335.60 & 6.20 & 8.65 & 333.89 & 12.88\\
\bottomrule
\end{tabular}

\end{table}

The model-finding procedure outlined in Sect.~\ref{sec:ModelBuilding} results in ensembles of parametrisations of comparable quality and complexity. 
The decay chains included in the models discussed above are included in the majority of models of acceptable quality, 
with further variations made by addition of further small components. 
The fraction of models in this ensemble containing a given decay mode are shown in Table~\ref{tb:AlternativeModels} for the RS decay mode with the average fit fraction associated with each decay chain also tabulated. 
The ensemble of RS models consists of about 200 models with $\chi^2$ per \dof varying between 1.21 and 1.26. 
Many of the decay chains in the ensemble include resonances, such as the $\PK_1(1270)^{-}$, decaying via radially excited vector mesons, such as the $\rho(1450)^{0}$ and $\Kstar(1410)^{0}$ mesons. 
In particular, the decay $\PK_1(1270)^{-}\rightarrow \rho(1450)^{0}\Km$ is included in the models discussed in Sects.~\ref{sec:RSModel}~and~\ref{sec:WSModel} and is found in the majority of the models in the ensemble. 
This decay channel of the $\PK_1(1270)^{-}$ meson has a strong impact at low dipion masses due to the very large width of the $\rho(1450)^{0}$ resonance, of about 400\mevcc. 
Models excluding this component are presented as alternative parametrisations in Appendix~\ref{sec:AltModel} as this decay mode has not been studied extensively in other production mechanisms of the $\PK_1(1270)^{-}$ resonance, 
and the ensemble contains models without this decay chain of similar fit quality to the baseline model.
The situation can be clarified with independent measurements of the properties of these resonances. 
The $a_1(1640)^{+}$ resonance is also found in many models in the ensemble, and is likely to be present at some level despite only the low-mass tail of this resonance impacting the phase space. 
This resonance strongly interferes with the dominant $a_1(1260)^{+}$ component and, 
 as the parameters of this resonance are poorly known, improved external inputs will be required to correctly constrain this component.

The coupling parameters cannot strictly be compared between different models, as in many cases these coupling parameters have a different interpretation depending on the choice of the model.
However, it is instructive to consider how the fit fractions vary depending on the choice of model, which is shown in Table~\ref{tb:FitFractionSystematicsk1270}.
It is also useful to consider how the choice of model impacts upon the fitted masses and widths, which is shown in Table~\ref{tb:MassesAndWidths}. 
The values for the model described in Sect.~\ref{sec:RSModel} are also shown, which has compatible values with the ensemble. 
The variation with respect to the choice of model is characterised by the RMS of the parameters in the ensemble, and is of a comparable size to the combined systematic uncertainty from other sources on these parameters.

\begin{table}
  \centering 
  \caption{ \label{tb:AlternativeModels_WS} 
  Decay chains taken into account in alternative parametrisations of the WS decay mode ${\Dz\rightarrow\Kp\pim\pim\pip}$. 
  For each chain, the fraction of models in the ensemble that contain this decay, together with the associated average fit fraction, $\langle \mathcal{F} \rangle$, are shown. 
  Components are not tabulated if they contribute to all models in the ensemble, or if they contribute to less than 5\% of the models. 
  }
  \begin{tabular}{l
r
r
}
\toprule
Decay Chain & Fraction & $\langle \mathcal{F} \rangle$\\
 & of models $[\%]$ & $[\%]$\\
\midrule
$K_{1}(1270)^{+}\left[\rho(770)^{0}K^{+}\right]^{L=2}\pi^{-}$ & 47.2 & 1.21\\
$K^{*}(1680)^{+}\left[K^{*}(1680)^{0}\pi^{+}\right]\pi^{-}$ & 38.0 & 2.89\\
$K^{*}(1680)^{+}\left[\rho(770)^{0}K^{+}\right]\pi^{-}$ & 33.3 & 2.58\\
$a_{1}(1640)^{-}\left[\left[\pi^+\pi^-\right]^{L=0}\pi^{-}\right]K^{+}$ & 27.8 & 3.24\\
$K^{*}(1680)^{+}\left[\rho(1450)^{0}K^{+}\right]\pi^{-}$ & 22.2 & 2.53\\
$K_{1}(1270)^{+}\left[K^*(1410)^{0}\pi^{+}\right]^{L=2}\pi^{-}$ & 22.2 & 0.60\\
$K_{1}(1270)^{+}\left[\left[\pi^+\pi^-\right]^{L=0}K^{+}\right]\pi^{-}$ & 21.3 & 0.26\\
$K^{*}(1680)^{+}\left[K^*(1410)^{0}\pi^{+}\right]\pi^{-}$ & 17.6 & 1.98\\
$\rho(770)^{0}\left[K^{+}\pi^{-}\right]^{L=0}$ & 17.6 & 3.49\\
$K^{*}(1680)^{+}\left[K_{2}^{*}(1430)^{0}\pi^{+}\right]\pi^{-}$ & 16.7 & 0.82\\
$K_{1}(1400)^{+}\left[\left[\pi^+\pi^-\right]^{L=0}K^{+}\right]\pi^{-}$ & 13.0 & 0.29\\
$K_{2}^{*}(1430)^{0}\left[K^{+}\pi^{-}\right]\rho(770)^{0}$ & 13.0 & 0.35\\
$K^*(1410)^{0}\rho(770)^{0}$ & 10.2 & 3.50\\
\bottomrule
\end{tabular}

\end{table}

The $\WS$ ensemble consists of 108 models, all of which have a $\chi^2$ per \dof of less than $1.45$, with the best models in the ensemble having a $\chi^2$ per \dof of about $1.35$. 
The fraction of models in this ensemble containing a given decay mode are shown in Table~\ref{tb:AlternativeModels_WS}. 
In particular, there should be percent-level contributions from some of the decay chains present in the $\RS$ mode, such as $\Dz\to a_1(1260)^{-}\Kp$ and $\Dz\to\PK^{*}(892)^{0}\left[\pip\pim\right]^{L=0}$.
In addition to the marginal decays of the $\PK_1(1270)^{+}$ present in the $\WS$ ensemble, the models suggest contributions from the $\PK^*(1680)$, which resembles a nonresonant component due to its large width and position on the edge of the phase space. As is the case for the large $\Dz\rightarrow \PK^*(892)^{0}\rho(1450)$ component, this contribution is likely to be mimicking several smaller decay channels that cannot be resolved with the current sample size.

\FloatBarrier

\FloatBarrier
\subsection{Coherence factor}
\label{sec:Coherence}

The coherence factor $R_{\PK3\pi}$ and average strong-phase difference $\delta_{\PK3\pi}$ are measures of the phase-space-averaged interference properties between suppressed and favoured amplitudes, and are defined as \cite{Atwood:2003mj} 
\begin{equation}
  R_{\PK3\pi} e^{-i \delta_{\PK3\pi}} = \frac{ \int d\xp \mathcal{A} \left(\Dz\rightarrow\Kp\pim\pim\pip\right) \mathcal{A}^{*}\left(\Dzb\rightarrow\Kp\pim\pim\pip\right) }{
    \sqrt{ \int d\xp\left| \mathcal{A}\left(\Dz\rightarrow \Kp\pim\pim\pip \right) \right|^2 
           \int d\xp\left| \mathcal{A}\left(\Dzb\rightarrow\Kp\pim\pim\pip\right) \right|^2 } },
\label{eq:CoherenceFactor}
\end{equation}
  where $\mathcal{A}(\Dz\rightarrow\Kp3\pi)$ is the amplitude of the suppressed decay and $\mathcal{A}(\Dzb\rightarrow\Kp3\pi)$ is the favoured amplitude for $\Dzb$ decays. 
  Additionally, it is useful to define the average ratio of amplitudes as
\begin{equation}
  r_{\PK3\pi} = \sqrt{ \frac{ \int d\xp \left| \mathcal{A}\left({\Dz\rightarrow\Kp\pim\pim\pip}\right) \right|^2 }{\int d\xp \left| \mathcal{A}\left({\Dzb\rightarrow\Kp\pim\pim\pip}\right) \right|^2} }.
\end{equation}
  Knowledge of these parameters is necessary when making use of this decay in $\Bm\rightarrow\PD\Km$
  transitions for measuring the \CP-violating phase $\gamma$ \cite{Atwood:2003mj}, 
  and can also be exploited for charm mixing studies. 
  Observables with direct sensitivity to the coherence factor and related parameters have been measured in $e^{+}e^{-}$ collisions at the $\psi(3770)$ resonance with \cleo-c data \cite{Evans:2016}, and through charm mixing at \lhcb \cite{LHCb-PAPER-2015-057}. 
  A global analysis of these results \cite{Evans:2016} yields
  \begin{align*}
    \begin{split}
      R_{\PK3\pi}        &= 0.43^{+0.17}_{-0.13}  \\
      \delta_{\PK3\pi} &= (128^{+28}_{-17})^{\mathrm{o}} \\
      r_{K3\pi}      &= (5.49\pm0.06)\times 10^{-2} .
    \end{split}
  \end{align*}
The baseline models presented in Sect.~\ref{sec:results} can be used to calculate the model-derived coherence factor
\begin{equation*}
  R_{\PK3\pi}^{\mathrm{mod}} = 0.458 \pm 0.010 \pm 0.012 \pm 0.020,
\end{equation*}
where the first uncertainty is statistical, the second systematic, and the third the uncertainty from the choice of WS model. This uncertainty is assigned by taking the spread in values from an ensemble of alternative models from the model-building algorithm, 
requiring that models have a $\chi^2$ per \dof of less than 1.5, 
and that all unconstrained components in the fit have a significance of $> 2\sigma$.
This result is in good agreement with the direct measurement in Ref.~\cite{Evans:2016}. 
This analysis has no sensitivity to $\delta_{\PK3\pi}$ and $r_{\PK3\pi}$ as each amplitude contains an arbitrary independent amplitude and phase. 

The stability of the local phase description can also be verified by evaluating the model-derived coherence factor and associated parameters in different regions of phase space.  
This is equivalent to changing the definition of Eq.~\ref{eq:CoherenceFactor} 
such that integrals are performed over a limited region of phase space. 
In this case, it is also possible to determine the local values of $\delta_{\PK3\pi}$ and $r_{\PK3\pi}$ relative to the phase-space averaged values.
Therefore, overall normalisation factors are fixed such that the central values of the direct measurement are correctly reproduced.

In order to define these regions, the phase space is divided into hypercubes using the algorithm described in Sect.~\ref{sec:GoodnessOfFit}. 
The division is done such that the hypercubes cannot be smaller in any dimension than $50\mevcc$. 
The hypercubes are grouped into bins of average phase difference between the two amplitudes in the bin, using the amplitude models described in Sect.~\ref{sec:results}.
The range $[-180\degree,180\degree]$ in phase difference between the two decay modes is split into eight bins.
The division of this range is done such that each bin is expected to have an approximately equal population of WS events within the bin. 
The coherence factors, average strong phases and amplitude ratios and their RMS spread arising from the choice of WS model are summarised in Table \ref{tb:CoherenceTable}.
Good stability with respect to the choice of model is observed, which is a consequence of the dominant features of the amplitude being common for all models, and gives confidence to using the models presented in this paper to define regions of interest 
for future binned measurements of $\gamma$ or studies of charm mixing. 
The relatively high coherence factor in some regions of phase-space demonstrates the potential 
improvements in sensitivity to measurements of $\CP$-violating observables.

\begin{table}
  \centering
  \caption{Coherence factor and average strong-phase differences in regions of phase space. The spread of coherence factors, average strong-phase difference and ratio of amplitudes from choice of WS model characterised with the RMS of the distribution.}
  \label{tb:CoherenceTable}
  \begin{tabular}{l 
>{\collectcell\num}r<{\endcollectcell} @{${}\pm{}$} >{\collectcell\num}l<{\endcollectcell}
>{\collectcell\num}r<{\endcollectcell} @{${}\pm{}$} >{\collectcell\num}l<{\endcollectcell}
>{\collectcell\num}r<{\endcollectcell} @{${}\pm{}$} >{\collectcell\num}l<{\endcollectcell}
}
\toprule
Bin & \multicolumn{2}{c}{$R_{\mathrm{K}3\pi}$} & \multicolumn{2}{c}{$\delta_{\mathrm{K}3\pi} [^\mathrm{o}]$} & \multicolumn{2}{c}{$r_{\mathrm{K}3\pi} \times 10^{-2}$}\\
\midrule
1 & 0.701 & 0.017 & 169 & 3 & 5.287 & 0.034\\
2 & 0.691 & 0.016 & 151 & 1 & 5.679 & 0.032\\
3 & 0.726 & 0.010 & 133 & 1 & 6.051 & 0.032\\
4 & 0.742 & 0.008 & 117 & 1 & 6.083 & 0.030\\
5 & 0.783 & 0.005 & 102 & 2 & 5.886 & 0.031\\
6 & 0.764 & 0.007 & 84 & 3 & 5.727 & 0.033\\
7 & 0.424 & 0.013 & 26 & 3 & 5.390 & 0.061\\
8 & 0.473 & 0.030 & -149 & 7 & 4.467 & 0.065\\
\bottomrule
\end{tabular}

\end{table}

\FloatBarrier

\section{\label{sec:conclusions}Conclusions}                       The four-body decay modes $\Dz\rightarrow\Kmp\pipm\pipm\pimp$ have been studied using high-purity time-integrated samples obtained from doubly tagged $\Bb\rightarrow\Dstarp(2010)[\Dz\pip]\mu X$ decays. 
For the RS decay mode \RS, the analysis is performed with a sample around sixty times larger than that exploited in any previous analysis of this decay.
For the WS mode \WS, the resonance substructure is studied for the first time with $\approx 3000$ signal candidates. 

Both amplitude models are found to have large contributions from axial resonances, the decays $\Dz\rightarrow a_1(1260)^{+}\Km$ and $\Dz\rightarrow \PK_1(1270/1400)^{+}\pim$ for \RS and \WS, respectively. 
This is consistent with the general picture that $W$-emission topologies are crucial in describing these decays. 
Interference between the parity-even and parity-odd amplitudes causes large local parity violations, 
which are shown to be reasonably well modelled in the RS decay. 
A significant contribution from the pseudoscalar resonance $\PK(1460)^{-}$ is identified, which is validated using the model-independent partial waves method.  

The coherence factor is calculated using the models, and is found to be consistent with direct measurements. 
It is found that the calculated value is relatively stable with respect to the parametrisation of subdominant amplitudes in the WS model.
These models therefore provide a valuable input to future binned measurements of the $\CP$-violating parameter $\gamma$ and charm-mixing studies.

\section*{Acknowledgements}
%
%
\noindent We express our gratitude to our colleagues in the CERN
accelerator departments for the excellent performance of the LHC. We
thank the technical and administrative staff at the LHCb
institutes. We acknowledge support from CERN and from the national
agencies: CAPES, CNPq, FAPERJ and FINEP (Brazil); MOST and NSFC
(China); CNRS/IN2P3 (France); BMBF, DFG and MPG (Germany); INFN
(Italy); NWO (The Netherlands); MNiSW and NCN (Poland); MEN/IFA
(Romania); MinES and FASO (Russia); MinECo (Spain); SNSF and SER
(Switzerland); NASU (Ukraine); STFC (United Kingdom); NSF (USA).  We
acknowledge the computing resources that are provided by CERN, IN2P3
(France), KIT and DESY (Germany), INFN (Italy), SURF (The
Netherlands), PIC (Spain), GridPP (United Kingdom), RRCKI and Yandex
LLC (Russia), CSCS (Switzerland), IFIN-HH (Romania), CBPF (Brazil),
PL-GRID (Poland) and OSC (USA). We are indebted to the communities
behind the multiple open-source software packages on which we depend.
Individual groups or members have received support from AvH Foundation
(Germany), EPLANET, Marie Sk\l{}odowska-Curie Actions and ERC
(European Union), ANR, Labex P2IO and OCEVU, and R\'{e}gion
Auvergne-Rh\^{o}ne-Alpes (France), RFBR, RSF and Yandex LLC (Russia),
GVA, XuntaGal and GENCAT (Spain), Herchel Smith Fund, the Royal
Society, the English-Speaking Union and the Leverhulme Trust (United
Kingdom).

\begin{appendices}
\section{\label{sec:SpinFormalism}Spin formalism}                                             The effects of spin and orbital angular momentum are calculated using the Rarita-Schwinger formalism, following a similar prescription to that described in Ref.~\cite{Zou:2002ar}.
Spin-matrix elements for quasi two-body processes are constructed in terms of a series of polarisation and pure orbital angular momentum tensors.
Consider the decay of particle $a$ that has integer spin $J$,
into particles $b$ and $c$, which have integer spin $s_b$, $s_c$, respectively. 
All three particles have an associated polarisation tensor, $\epsilon^{(a,b,c)}$, which is of rank equal to the spin of the particle. 
The decay products $b,c$ will also in general have a relative orbital angular momentum $l$, which is expressed in terms of the pure orbital angular momentum tensor, $L_{\mu ... \nu}$, which is of rank $l$.
The matrix element for the decay is 
\begin{equation}
  \mathcal{M}_{a\rightarrow bc} = \epsilon_{\mu_a ... \nu_a}^{(a)*} \epsilon_{\mu_b ... \nu_b }^{(b)} \epsilon_{\mu_c ... \nu_c }^{(c)} L_{ \mu_l ... \nu_l }^{(l)} G^{ \mu_a ... \nu_a \mu_b ... \nu_b \mu_c ... \nu_c \mu_l ... \nu_l } ,
  \label{eq:MatrixElementDefinition}
\end{equation}
where the tensor $G^{...}$ combines the polarisation and pure orbital angular momentum tensor to produce a scalar object. 
This tensor is constructed from combinations of the metric tensor $g_{\mu\nu}$ and the Levi-Civita tensor contracted with the four-momenta of the decaying particle,
$\varepsilon_{\mu\nu\alpha\beta}P^{\mu}$.
The second of these tensors is used only if $J-(l-s_b-s_c)$ is odd, and ensures that matrix elements have the correct properties under parity transformations.
The matrix element can also be written by defining the current, $\mathcal{J}$, of the decaying particle:
\begin{equation}
  \mathcal{M}_{a\to bc} = \epsilon_{\underline{\mu}}^{(a)*}\mathcal{J}^{(a)\underline{\mu}} ,
  \label{eq:CurrentDefinition}
\end{equation}
where the $\underline{\mu}$ represents a set of Lorentz indices $\mu ... \nu$, a shorthand which will be used throughout this section.  
The isobar model factorises an $N$-body decay into a sequence of two-body processes. 
Each of these quasi two-body decays can be described with a single spin matrix element, and hence the total matrix element is the product of $N-1$ matrix elements:
\begin{equation}
  \mathcal{M} = \prod_{i=0}^{N-1} \mathcal{M}_{a_i \rightarrow b_ic_i}.
  \label{eq:MatrixElementCascade}
\end{equation}
For example, 
consider the quasi two-body decay $P\to X [ab] Y[cd]$. The matrix element for this decay is
\begin{equation}
  \mathcal{M} = \sum_i \sum_j \mathcal{M}_{P \to X_iY_j} \mathcal{M}_{X_i\to ab } \mathcal{M}_{Y_j\to cd},
\end{equation}
where the sums are over the possible polarisations of the intermediate states. 

It is preferable to build a generic formulation of the total matrix element for arbitrary topologies, spins and angular momenta, rather than performing an explicit computation for each possible process.
A generic approach to computing matrix elements is to introduce a generalised ``current'' associated with a decaying particle that has absorbed the matrix elements of its decay products. 
This current can be written in terms of the currents of its decay products as  
\begin{equation}
  \mathcal{J}^{\underline{\mu}} = 
  L^{ (l) }_{\underline{\beta}} 
  G^{ \underline{\mu \nu \alpha \beta} }  
  \times \left( \mathcal{S}^{1}_{\underline{\nu\gamma}}  \mathcal{J}_1^{\underline{\gamma}} \right)
  \times \left( \mathcal{S}^{2}_{\underline{\alpha\eta}} \mathcal{J}_2^{\underline{\eta}} \right),
  \label{eq:MatrixElementExpansion}
\end{equation}
where $\mathcal{S}^{1,2}_{\underline{\mu}}$ is the spin-projection operator of decay products (1,2), 
which has been used to sum intermediate polarisation tensors, using the definition  
\begin{equation}
  \sum_i \epsilon_{i\underline{a}} {\epsilon_{i\underline{b}}}^* = \mathcal{S}_{\underline{ab}} .
  \label{eq:ProjectionDefinition}
\end{equation}
The first few projection operations, which are sufficient for describing charm decays, are
\begin{align}
  \begin{split}
    \mathcal{S}_{\mu\nu}(P)    \quad  & = \quad -g_{\mu\nu} + \frac{ P_{\mu}P_{\nu} }{P^2}  \\
    \mathcal{S}_{\mu\nu\alpha\beta}(P) \quad &= \quad \frac{1}{2}\left( \mathcal{S}_{\mu\alpha}S_{\nu\beta} + \mathcal{S}_{\mu\beta}\mathcal{S}_{\nu\alpha} \right) - \frac{1}{3}\mathcal{S}_{\mu\nu}\mathcal{S}_{\alpha\beta}.
    \label{eq:ProjectionOperators}
  \end{split}
\end{align}
This operator projects out the component of a tensor that is orthogonal to the four-momentum of a particle, 
 and has rank $2J$ for a particle of spin $J$.
The orbital angular momentum tensors are also constructed from the spin projection operators and the relative momentum of the decay products, $Q_a$ \cite{Zou:2002ar}, and are written as
\begin{align}
  \begin{split}
    L_\mu &= - \mathcal{S}_{\mu\nu}(P_a) Q_a^\nu \\ 
    L_{\mu\nu} &= \mathcal{S}_{\mu\nu\alpha\beta}(P_a) Q_a^\alpha Q_a^\beta.
    \label{eq:OrbitalAngularMomentumTensors}
  \end{split}
\end{align}

The matrix element for a generic cascade of particle decays can then be calculated recursively. 
In the case of the decay of a spinless particle, the matrix element for the total decay process is identical  to the current of the decaying particle. 
The generalised current is therefore merely a convenient device for organising the computation of spin matrix elements, 
but is not in general associated with the propagation of angular momentum.
It is also useful to define the spin-projected currents, $\mathcal{S}_{\underline{\mu\nu}} J^{\underline{\nu}}$, which will be written as $S,V^{\mu},T^{\mu\nu}$ for (pseudo)scalar, (pseudo)vector and (pseudo)tensor states, respectively.  

The rules for how the different spin-projected currents are written in terms of each other is given in Table~\ref{tb:CurrentRules}, where these relations are derived by considering the symmetries of Lorentz indices and the parity properties of the matrix element. 
All of the coupling structures necessary to describe $P\rightarrow 4 P$ are uniquely determined by these constraints, although this property does not hold in general. 
This allows complicated spin configurations to be calculated in terms of a simple and consistent set of rules.
The rules are written with consistent dependencies to clarify their derivations, and in some cases simplified forms are also given. 
These simplifications typically rely on the symmetry properties of the Levi-Civita tensor and the relationship $\mathcal{S}^{\underline{ab}}\mathcal{S}_{\underline{bc}} = \mathcal{S}^{\underline{a}}_{\underline{c}}$, which is the defining characteristic of a projection operator.

\begin{table}
  \caption{Rules for calculating the current associated with a given decay chain in terms of the currents of the decay products.
    Where relevant, the spin projection operator $\mathcal{S}$ and the orbital angular momentum operators $L$ are those for the decaying particle. 
  }
  \label{tb:CurrentRules}
  \renewcommand{\arraystretch}{1.5}%
  \centering 
  \scalebox{0.9}{
  \begin{tabular}{lll}
    \toprule
    Topology & Current & Simplified current \\
    \midrule
    $ \displaystyle S\rightarrow [ S_{1} S_{2} ] $ & $\displaystyle S_{1} S_{2} $ & \\
    $ \displaystyle S\rightarrow  [ V S_{1} ]^{L=1} $ & $ \displaystyle L^{\mu} V_{\mu} S_1 $ &\\
    $ \displaystyle S\rightarrow  [ V_{1} V_{2} ]^{L=0} $ & $ \displaystyle g_{\mu\nu} V_1^{\mu} V_2^{\nu}  $ &\\
    $ \displaystyle S\rightarrow  [ V_{1} V_{2} ]^{L=1}$ & $ \displaystyle \varepsilon_{\mu\nu\alpha\beta}{P_S^\mu} L^{\nu} V_1^{\alpha} V_2^{\beta} $ &
    $ \displaystyle \varepsilon_{\mu\nu\alpha\beta} P_S^\mu Q_S^\nu V_1^\alpha V_2^\beta $ \\
    $ \displaystyle S\rightarrow  [ V_{1} V_{2} ]^{L=2}$ & $ \displaystyle L_{\mu\nu} V_1^{\mu} V_2^{\nu} $ & \\
    $ \displaystyle S\rightarrow [ T S_1 ]^{L=2}$ & $ \displaystyle L_{\mu\nu} T^{\mu\nu} S_1 $ & \\
    $ \displaystyle S\rightarrow [ T V ]^{L=1}$ & $ \displaystyle L^{\mu} T_{\mu\nu} V^{\nu} $ & \\
    $ \displaystyle S\rightarrow [ T V ]^{L=2}$ & $ \displaystyle L^{\mu\nu} \varepsilon_{\nu\alpha\beta\gamma}  P_{S}^{\alpha} T^{\beta}_{\mu} V^{\gamma} $ &
    $ \displaystyle \varepsilon_{\nu\alpha\beta\gamma} P_S^\alpha Q_S^\nu  L_{\mu} T^{\beta\mu} V^\gamma $ \\
    $ \displaystyle S\rightarrow [ T_1 T_2 ]^{L=0}$ & $ \displaystyle T_{1}^{\mu\nu} T_{2\mu\nu} $ & \\    
    \midrule
    $ \displaystyle V_\mu \rightarrow [ S_{1} S_{2} ]^{L=1} $ & $ \displaystyle \mathcal{S}_{\mu\nu} L^{\nu} S_1 S_2 $ & $ \displaystyle L_{\mu} S_1 S_2 $ \\
    $ \displaystyle V_\mu \rightarrow [ V_{1} S]^{L=0} $ & $ \displaystyle \mathcal{S}_{\mu\nu} V_1^{\nu} S $ & \\
    $ \displaystyle V_\mu \rightarrow [ V_{1} S ]^{L=1} $ & $ \displaystyle \mathcal{S}_{\mu\nu} \varepsilon^{\nu\alpha\beta\gamma}{P_{V\alpha}} L_{\beta} V_{1\gamma} S $ &
    $ \displaystyle - \varepsilon_{\mu\alpha\beta\gamma} P_{V}^\alpha Q_{V}^\beta V_{1}^{\gamma} S $ \\
    $ \displaystyle V_\mu \rightarrow [ V_{1} S ]^{L=2} $ & $ \displaystyle \mathcal{S}_{\mu\nu} L^{\nu\alpha} V_{1\alpha} S $ & $\displaystyle L_{\mu\alpha} V_{1}^{\alpha} S  $ \\

    $ \displaystyle V_\mu \rightarrow [ T S ]^{L=1} $ & $ \displaystyle \mathcal{S}_{\mu\nu} L_{\alpha} T^{\nu\alpha} $ & \\
    $ \displaystyle V_\mu \rightarrow [ T S ]^{L=2} $ & $ \displaystyle \mathcal{S}_{\mu\nu} \varepsilon^{\nu\alpha\beta\gamma}{P_{V\alpha}} L_{\beta}^{\eta} T_{\gamma\eta} S $ & $\displaystyle - \varepsilon_{\mu\alpha\beta\gamma} P_V^\alpha Q_V^\beta T^{\gamma\eta} L_{\eta} $  \\
    $ \displaystyle V_\mu \rightarrow [ T V_1 ]^{L=0} $ & $ \displaystyle \mathcal{S}_{\mu\nu} T^{\nu\alpha} V_{1\alpha} $ \\

    \midrule
    $\displaystyle T_{\mu\nu} \rightarrow [ S_{1} S_{2} ]^{L=2} $ & $\displaystyle \mathcal{S}_{\mu\nu\alpha\beta} L^{\alpha\beta} S_{1} S_{2} $ & $L_{\mu\nu} S_1 S_2 $  \\

    $\displaystyle T_{\mu\nu} \rightarrow [ V S ]^{L=1} $ & $\displaystyle \mathcal{S}_{\mu\nu\alpha\beta} L_1^{\alpha} V^{\beta} S $ & 
    $\displaystyle \left( \frac{1}{2} \left( L_{\mu} S_{\nu\beta} + S_{\mu\beta} L_{\nu} \right) - \frac{1}{3}S_{\mu\nu }L_{\beta} \right) V^{\beta} $ \\

    $\displaystyle T_{\mu\nu} \rightarrow [ V S ]^{L=2} $ &
    $\displaystyle \mathcal{S}_{\mu\nu\alpha\beta} \varepsilon^{\alpha\gamma\eta\lambda} P_{T\gamma} L_{2\eta}^{\beta} V^{\lambda} S $ &
    $\displaystyle - \frac{1}{2}\left( \varepsilon_{\mu\gamma\eta\lambda} L_{\nu} + \varepsilon_{\nu\gamma\eta\lambda} L_{\mu} \right) P_T^\gamma Q_T^\eta V^\lambda $ \\

    $\displaystyle T_{\mu\nu} \rightarrow [ T_1 S ] $ & $\displaystyle \mathcal{S}_{\mu\nu\alpha\beta} T_1^{\alpha\beta}$ & \\

    \bottomrule
  \end{tabular}

  }
\end{table}

\section{\label{app:ListOfComponents}List of decay chains}                                    The list of possible decay chains is built from what is allowed by the relevant conservation laws. 
  Approximately one hundred different decay chains modes are included as possible contributions to the model. 
  Certain cascade decays already have well known sub-branching ratios. 
  For example, although the $\PK_1(1400)$ decays almost exclusively via the $\Kstar(892)$, 
  the various decays of the $\PK_1(1400)$ are treated separately without assumption about their branching ratios. 
  The different components can be split into the same groups as in Sect.~\ref{sec:formalism}: 

  \begin{itemize}

    \item $\Dz \rightarrow Y_{\pi\pi}\left[\pi\pi\right] Y_{K\pi}\left[K\pi\right]$, where $Y_{\pi\pi}$ is one of the following states: 
      $\rho(770)^{0}$, $\rho(1450)^{0}$, $f_2(1270)$ or $[\pip\pim]^{L=0}$, and $Y_{K\pi}$ is one of the following: 
      $\ensuremath{\PK}^{*}(892)^{0}$, $\ensuremath{\PK}^{*}(1410)^{0}$, 
      $\ensuremath{\PK}^{*}(1680)^{0}$,
      $\ensuremath{\PK}_{2}^{*}(1430)^{0}$ or
      $[\Kmp\pipm]^{L=0}$.

      The $[\pip\pim]^{L=0}$ and $[\Kmp\pipm]^{L=0}$ contributions are modelled using K matrices. 
      In cases with a scalar contribution and a radial recurrence of a vector state, such as $\rho(1450)^{0}[\Kmp\pipm]^{L=0}$, 
      the K matrix is fixed to be the same as the first vector, i.e. the K-matrix parameters of $\rho(770)^{0}[\Kmp\pipm]^{L=0}$.
      For vector-vector and vector-tensor contributions, the different possible polarisation states are included together in the model building. 
      The contributions from the radial excitations of the kaon are only included as a possibility when included with the $\pip\pim$ S-wave,
      as the other decay chains involving this resonance, for example the decay $\PK^*(1410)^{0}\rho(770)^{0}$, tend to have large interference terms, which requires fine tuning with other amplitudes and hence are considered to be unphysical. 
    \item $\Dz \rightarrow X_{\pi\pi\pi} \left[ Y_{\pi\pi} \left[\pi\pi\right] \pi \right] K$,
      where $X_{\pi\pi\pi}$ is one of the following states: 
      $a_{1}(1260)^{\pm}$, 
      $a_1(1640)^\pm$,
      $\pi(1300)^\pm$ or 
      $a_{2}(1320)^\pm$ . 

    \item $\Dz \rightarrow X_{K\pi\pi} \left[ Y_{K\pi}  \left[K\pi  \right] \pi  \right] \pi$, 
      $\Dz \rightarrow X_{K\pi\pi} \left[ Y_{\pi\pi}\left[\pi\pi\right] K \right] \pi$,
      where $X_{K\pi\pi}$ is one of the following states:
      $\ensuremath{\PK}_{1}(1270)^\pm$, 
      $\ensuremath{\PK}_{1}(1400)^\pm$, 
      $\ensuremath{\PK}^{*}(1410)^\pm$, 
      $\ensuremath{\PK}^{*}(1680)^\pm$, 
      $\ensuremath{\PK}_{2}^{*}(1430)^\pm$ or
      $\ensuremath{\PK}(1460)^\pm$. 

  \end{itemize}

  All of these states are considered under all possible orbital configurations that obey the respective conservation laws. 

\section{\label{sec:SystematicBreakdown}Systematic uncertainties}                             \FloatBarrier

\begin{table}[!htb]
  \centering
  \caption{ \label{tb:SysLegend} Legend for systematic uncertainties, including whether this sources of uncertainty is considered on the RS/WS decay mode.}
  \begin{tabular}{clcc}
\toprule 
  & Description & RS & WS \\
\midrule  
  I & Efficiency variations & \checkmark & \\
  II & Simulation statistics & \checkmark & \checkmark \\
  III & Masses and widths & \checkmark & \checkmark \\
  IV & Form factor radii & \checkmark & \checkmark \\
  V & Background fraction & \checkmark & \checkmark \\
  VI & Background parameterisation & & \checkmark \\
 VII & RS parameters & & \checkmark \\
\bottomrule
\end{tabular}

\end{table} 

The various contributions assigned for different systematic uncertainties are summarised in this appendix by a series of tables. 
The legend for these is given in Table~\ref{tb:SysLegend}, including which sources of uncertainty are considered on each decay mode. 
The breakdown of systematic uncertainties for the RS decay \RS for coupling parameters, fit fractions and other parameters are given in Tables~\ref{tb:SysSummary} and \ref{tb:Fractions} for the quasi two-body decay chains and cascade decay chains, respectively. 
The systematic uncertainties for the WS mode \WS are given in Table~\ref{tb:WS_SysSummary} for both coupling parameters and the fit fractions. 

\begin{table}
  \centering 
  \caption{ \label{tb:SysSummary} Systematic uncertainties on the RS decay coupling parameters and fit fractions for quasi two-body decay chains.}
  \scalebox{0.75}{\begin{tabular}{l 
l 
>{\collectcell\num}r<{\endcollectcell} @{${}\pm{}$} >{\collectcell\num}l<{\endcollectcell} @{${}\pm{}$} >{\collectcell\num}l<{\endcollectcell}
l 
l 
l 
l 
l 
}
\toprule
 &  & \multicolumn{3}{c}{} & \multicolumn{1}{c}{I} & \multicolumn{1}{c}{II} & \multicolumn{1}{c}{III} & \multicolumn{1}{c}{IV} & \multicolumn{1}{c}{V}\\
\midrule
$\Kstarb(892)^{0}\rho(770)^{0}$ & $\mathcal{F}$ & 7.340 & 0.084 & 0.637 & 0.426 & 0.050 & 0.063 & 0.466 & 0.025\\
 & $\left| g \right|$ & 0.196 & 0.001 & 0.015 & 0.000 & 0.001 & 0.001 & 0.015 & 0.000\\
 & $arg(g) [^o]$ & -22.363 & 0.361 & 1.644 & 1.309 & 0.239 & 0.119 & 0.955 & 0.075\\
$\left[\Kstarb(892)^{0}\rho(770)^{0}\right]^{L=1}$ & $\mathcal{F}$ & 6.031 & 0.049 & 0.436 & 0.358 & 0.029 & 0.061 & 0.239 & 0.006\\
 & $\left| g \right|$ & 0.362 & 0.002 & 0.010 & 0.002 & 0.001 & 0.002 & 0.009 & 0.000\\
 & $arg(g) [^o]$ & -102.907 & 0.380 & 1.667 & 1.431 & 0.224 & 0.321 & 0.760 & 0.025\\
$\left[\Kstarb(892)^{0}\rho(770)^{0}\right]^{L=2}$ & $\mathcal{F}$ & 8.475 & 0.086 & 0.826 & 0.492 & 0.051 & 0.059 & 0.659 & 0.023\\
$\rho(1450)^{0}\Kstarb(892)^{0}$ & $\mathcal{F}$ & 0.608 & 0.040 & 0.165 & 0.061 & 0.032 & 0.134 & 0.065 & 0.019\\
 & $\left| g \right|$ & 0.162 & 0.005 & 0.025 & 0.007 & 0.004 & 0.018 & 0.015 & 0.003\\
 & $arg(g) [^o]$ & -86.122 & 1.852 & 4.345 & 1.933 & 1.570 & 2.485 & 2.152 & 1.368\\
$\left[\rho(1450)^{0}\Kstarb(892)^{0}\right]^{L=1}$ & $\mathcal{F}$ & 1.975 & 0.029 & 0.351 & 0.115 & 0.017 & 0.315 & 0.103 & 0.003\\
 & $\left| g \right|$ & 0.643 & 0.006 & 0.058 & 0.001 & 0.003 & 0.050 & 0.029 & 0.001\\
 & $arg(g) [^o]$ & 97.304 & 0.516 & 2.770 & 2.249 & 0.288 & 1.341 & 0.854 & 0.031\\
$\left[\rho(1450)^{0}\Kstarb(892)^{0}\right]^{L=2}$ & $\mathcal{F}$ & 0.455 & 0.028 & 0.163 & 0.078 & 0.016 & 0.090 & 0.110 & 0.004\\
 & $\left| g \right|$ & 0.649 & 0.021 & 0.105 & 0.052 & 0.011 & 0.063 & 0.065 & 0.003\\
 & $arg(g) [^o]$ & -15.564 & 1.960 & 4.109 & 1.208 & 1.323 & 2.631 & 2.484 & 0.762\\
\midrule
$\rho(770)^{0}\left[K^{-}\pi^{+}\right]^{L=0}$ & $\mathcal{F}$ & 0.926 & 0.032 & 0.083 & 0.069 & 0.019 & 0.016 & 0.039 & 0.006\\
 & $\left| g \right|$ & 0.338 & 0.006 & 0.011 & 0.000 & 0.004 & 0.002 & 0.010 & 0.002\\
 & $arg(g) [^o]$ & 73.048 & 0.795 & 3.951 & 3.567 & 0.469 & 0.481 & 1.549 & 0.185\\
$\quad \alpha_{3/2}$ & $\left| g \right|$ & 1.073 & 0.008 & 0.021 & 0.018 & 0.005 & 0.005 & 0.009 & 0.003\\
 & $arg(g) [^o]$ & -130.856 & 0.457 & 1.786 & 1.679 & 0.282 & 0.274 & 0.435 & 0.155\\
$\Kstarb(892)^{0}\left[\pi^{+}\pi^{-}\right]^{L=0}$ & $\mathcal{F}$ & 2.347 & 0.089 & 0.557 & 0.483 & 0.079 & 0.148 & 0.206 & 0.076\\
$\quad f_{\pi\pi}$ & $\left| g \right|$ & 0.261 & 0.005 & 0.024 & 0.022 & 0.004 & 0.006 & 0.007 & 0.003\\
 & $arg(g) [^o]$ & -149.023 & 0.943 & 2.696 & 2.275 & 0.540 & 1.176 & 0.617 & 0.196\\
$\quad \beta_1$ & $\left| g \right|$ & 0.305 & 0.011 & 0.046 & 0.040 & 0.010 & 0.013 & 0.013 & 0.007\\
 & $arg(g) [^o]$ & 65.554 & 1.534 & 4.004 & 3.017 & 0.857 & 2.322 & 0.771 & 0.455\\
\midrule
$\left[K^{-}\pi^{+}\right]^{L=0}\left[\pi^{+}\pi^{-}\right]^{L=0}$ & $\mathcal{F}$ & 22.044 & 0.282 & 4.137 & 3.631 & 0.268 & 0.213 & 1.945 & 0.188\\
$\quad \alpha_{3/2}$ & $\left| g \right|$ & 0.870 & 0.010 & 0.030 & 0.029 & 0.005 & 0.003 & 0.004 & 0.002\\
 & $arg(g) [^o]$ & -149.187 & 0.712 & 3.503 & 3.467 & 0.350 & 0.250 & 0.194 & 0.157\\
$\quad \alpha_{K\eta^\prime}$ & $\left| g \right|$ & 2.614 & 0.141 & 0.281 & 0.263 & 0.063 & 0.041 & 0.062 & 0.018\\
 & $arg(g) [^o]$ & -19.073 & 2.414 & 11.979 & 11.775 & 1.507 & 1.151 & 0.816 & 0.755\\
$\quad \beta_1$ & $\left| g \right|$ & 0.554 & 0.009 & 0.053 & 0.019 & 0.005 & 0.004 & 0.050 & 0.002\\
 & $arg(g) [^o]$ & 35.310 & 0.662 & 1.627 & 0.969 & 0.439 & 0.588 & 1.069 & 0.168\\
$\quad f_{\pi\pi}$ & $\left| g \right|$ & 0.082 & 0.001 & 0.008 & 0.004 & 0.001 & 0.001 & 0.007 & 0.000\\
 & $arg(g) [^o]$ & -146.991 & 0.718 & 2.248 & 1.849 & 0.463 & 0.593 & 1.003 & 0.252\\
\bottomrule
\end{tabular}
}
\end{table}

\begin{table}
  \centering 
  \caption{ \label{tb:Fractions} Systematic uncertainties on the RS decay coupling parameters, fit fractions and masses and widths of resonances for cascade topology decay chains. }
  \scalebox{0.75}{\begin{tabular}{l 
l 
>{\collectcell\num}r<{\endcollectcell} @{${}\pm{}$} >{\collectcell\num}l<{\endcollectcell} @{${}\pm{}$} >{\collectcell\num}l<{\endcollectcell}
l 
l 
l 
l 
l 
}
\toprule
 &  & \multicolumn{3}{c}{} & \multicolumn{1}{c}{I} & \multicolumn{1}{c}{II} & \multicolumn{1}{c}{III} & \multicolumn{1}{c}{IV} & \multicolumn{1}{c}{V}\\
\midrule
$a_{1}(1260)^{+}\mathrm{K}^{-}$ & $\mathcal{F}$ & 38.073 & 0.245 & 2.594 & 2.198 & 0.155 & 0.171 & 1.356 & 0.053\\
 & $\left| g \right|$ & 0.813 & 0.006 & 0.025 & 0.002 & 0.003 & 0.004 & 0.024 & 0.001\\
 & $\mathrm{arg}(g) [^\mathrm{o}]$ & -149.155 & 0.453 & 3.132 & 2.628 & 0.321 & 0.531 & 1.579 & 0.162\\
$\quad \rho(770)^{0}\pi^{+}$ & $\mathcal{F}$ & 89.745 & 0.452 & 1.498 & 1.116 & 0.298 & 0.596 & 0.720 & 0.192\\
$\quad \left[\pi^{+}\pi^{-}\right]^{L=0}\pi^{+}$ & $\mathcal{F}$ & 2.420 & 0.060 & 0.202 & 0.165 & 0.043 & 0.037 & 0.102 & 0.010\\
$\quad \quad \beta_1$ & $\left| g \right|$ & 0.991 & 0.018 & 0.037 & 0.005 & 0.015 & 0.012 & 0.031 & 0.006\\
 & $\mathrm{arg}(g) [^\mathrm{o}]$ & -22.185 & 1.044 & 1.195 & 0.769 & 0.597 & 0.393 & 0.545 & 0.169\\
$\quad \quad \beta_0$ & $\left| g \right|$ & 0.291 & 0.007 & 0.017 & 0.012 & 0.006 & 0.003 & 0.010 & 0.001\\
 & $\mathrm{arg}(g) [^\mathrm{o}]$ & 165.819 & 1.325 & 3.076 & 2.155 & 0.802 & 0.819 & 1.845 & 0.318\\
$\quad \quad f_{\pi\pi}$ & $\left| g \right|$ & 0.117 & 0.002 & 0.007 & 0.001 & 0.002 & 0.002 & 0.007 & 0.001\\
 & $\mathrm{arg}(g) [^\mathrm{o}]$ & 170.501 & 1.235 & 2.243 & 0.151 & 0.765 & 0.960 & 1.722 & 0.731\\
$\quad \left[\rho(770)^{0}\pi^{+}\right]^{L=2}$ & $\mathcal{F}$ & 0.850 & 0.032 & 0.077 & 0.058 & 0.021 & 0.023 & 0.040 & 0.007\\
 & $\left| g \right|$ & 0.582 & 0.011 & 0.027 & 0.020 & 0.007 & 0.008 & 0.015 & 0.002\\
 & $\mathrm{arg}(g) [^\mathrm{o}]$ & -152.829 & 1.195 & 2.512 & 1.691 & 0.710 & 0.755 & 1.520 & 0.258\\
  $a_1(1260)^{+}$ & $m_0 \left[\mevcc\right]$ & 1195.050 & 1.045 & 6.333 & 3.187 & 0.784 & 0.497 & 5.371 & 0.493\\
 & $\Gamma_0 \left[\mevcc\right]$ & 422.013 & 2.096 & 12.723 & 2.638 & 1.335 & 0.723 & 12.341 & 0.549\\
$K_{1}(1270)^{-}\pi^{+}$ & $\mathcal{F}$ & 4.664 & 0.053 & 0.624 & 0.485 & 0.037 & 0.285 & 0.268 & 0.012\\
 & $\left| g \right|$ & 0.362 & 0.004 & 0.015 & 0.013 & 0.002 & 0.002 & 0.008 & 0.001\\
 & $\mathrm{arg}(g) [^\mathrm{o}]$ & 114.207 & 0.760 & 3.612 & 3.320 & 0.526 & 0.441 & 1.227 & 0.219\\
$\quad \rho(770)^{0}\mathrm{K}^{-}$ & $\mathcal{F}$ & 96.301 & 1.644 & 8.237 & 5.523 & 1.082 & 5.624 & 2.110 & 0.286\\
$\quad \rho(1450)^{0}\mathrm{K}^{-}$ & $\mathcal{F}$ & 49.089 & 1.580 & 13.727 & 7.467 & 1.062 & 11.159 & 2.611 & 0.452\\
 & $\left| g \right|$ & 2.016 & 0.026 & 0.211 & 0.108 & 0.017 & 0.172 & 0.053 & 0.007\\
 & $\mathrm{arg}(g) [^\mathrm{o}]$ & -119.504 & 0.856 & 2.333 & 1.597 & 0.489 & 1.102 & 1.190 & 0.146\\
$\quad \Kstarb(892)^{0}\pi^{-}$ & $\mathcal{F}$ & 27.082 & 0.639 & 4.039 & 2.943 & 0.410 & 2.525 & 1.046 & 0.097\\
 & $\left| g \right|$ & 0.388 & 0.007 & 0.033 & 0.025 & 0.004 & 0.017 & 0.011 & 0.001\\
 & $\mathrm{arg}(g) [^\mathrm{o}]$ & -172.577 & 1.087 & 5.957 & 5.653 & 0.712 & 1.482 & 0.876 & 0.255\\
$\quad [\mathrm{K}^{-}\pi^{+}]^{L=0}\pi^{-}$ & $\mathcal{F}$ & 22.899 & 0.722 & 3.091 & 2.483 & 0.457 & 1.490 & 0.973 & 0.119\\
 & $\left| g \right|$ & 0.554 & 0.010 & 0.037 & 0.033 & 0.007 & 0.005 & 0.015 & 0.001\\
 & $\mathrm{arg}(g) [^\mathrm{o}]$ & 53.170 & 1.068 & 1.920 & 1.564 & 0.659 & 0.401 & 0.735 & 0.323\\
$\quad \left[\Kstarb(892)^{0}\pi^{-}\right]^{L=2}$ & $\mathcal{F}$ & 3.465 & 0.168 & 0.469 & 0.362 & 0.117 & 0.204 & 0.176 & 0.043\\
 & $\left| g \right|$ & 0.769 & 0.021 & 0.048 & 0.035 & 0.014 & 0.011 & 0.027 & 0.004\\
 & $\mathrm{arg}(g) [^\mathrm{o}]$ & -19.286 & 1.616 & 6.657 & 6.463 & 1.013 & 0.914 & 0.800 & 0.207\\
$\quad \omega(782)\left[\pi^{+}\pi^{-}\right]\mathrm{K}^{-}$ & $\mathcal{F}$ & 1.649 & 0.109 & 0.228 & 0.161 & 0.083 & 0.120 & 0.069 & 0.007\\
 & $\left| g \right|$ & 0.146 & 0.005 & 0.009 & 0.006 & 0.004 & 0.002 & 0.004 & 0.000\\
 & $\mathrm{arg}(g) [^\mathrm{o}]$ & 9.041 & 2.114 & 5.673 & 5.401 & 1.402 & 0.587 & 0.826 & 0.126\\
  $K_1(1270)^{-}$ & $m_0 \left[\mevcc\right]$ & 1289.810 & 0.558 & 1.656 & 1.197 & 0.436 & 0.244 & 1.010 & 0.198\\
  & $\Gamma_0 \left[\mevcc\right]$ & 116.114 & 1.649 & 2.963 & 1.289 & 1.221 & 0.981 & 2.090 & 0.545\\
$K_{1}(1400)^{-}\left[\Kstarb(892)^{0}\pi^{-}\right]\pi^{+}$ & $\mathcal{F}$ & 1.147 & 0.038 & 0.205 & 0.079 & 0.022 & 0.181 & 0.049 & 0.003\\
 & $\left| g \right|$ & 0.127 & 0.002 & 0.011 & 0.002 & 0.001 & 0.010 & 0.005 & 0.000\\
 & $\mathrm{arg}(g) [^\mathrm{o}]$ & -169.822 & 1.102 & 5.879 & 2.052 & 0.687 & 5.343 & 1.124 & 0.270\\
\midrule
$K_{2}^{*}(1430)^{-}\left[\Kstarb(892)^{0}\pi^{-}\right]\pi^{+}$ & $\mathcal{F}$ & 0.458 & 0.011 & 0.041 & 0.031 & 0.007 & 0.010 & 0.024 & 0.001\\
 & $\left| g \right|$ & 0.302 & 0.004 & 0.011 & 0.005 & 0.002 & 0.003 & 0.009 & 0.000\\
 & $\mathrm{arg}(g) [^\mathrm{o}]$ & -77.690 & 0.732 & 2.051 & 0.898 & 0.409 & 1.174 & 1.360 & 0.051\\
\midrule
$K(1460)^{-}\pi^{+}$ & $\mathcal{F}$ & 3.749 & 0.095 & 0.803 & 0.717 & 0.066 & 0.076 & 0.341 & 0.064\\
 & $\left| g \right|$ & 0.122 & 0.002 & 0.012 & 0.002 & 0.001 & 0.002 & 0.012 & 0.001\\
 & $\mathrm{arg}(g) [^\mathrm{o}]$ & 172.675 & 2.227 & 8.208 & 6.826 & 2.235 & 2.413 & 2.619 & 1.761\\
$\quad \Kstarb(892)^{0}\pi^{-}$ & $\mathcal{F}$ & 51.387 & 0.996 & 9.581 & 9.490 & 0.529 & 0.629 & 0.974 & 0.333\\
$\quad \left[\pi^{+}\pi^{-}\right]^{L=0}\mathrm{K}^{-}$ & $\mathcal{F}$ & 31.228 & 0.833 & 11.085 & 11.021 & 0.454 & 0.414 & 0.989 & 0.247\\
$\quad \quad f_{\mathrm{K}\mathrm{K}}$ & $\left| g \right|$ & 1.819 & 0.059 & 0.189 & 0.180 & 0.027 & 0.030 & 0.036 & 0.025\\
 & $\mathrm{arg}(g) [^\mathrm{o}]$ & -80.790 & 2.225 & 6.563 & 5.820 & 1.617 & 1.740 & 1.361 & 1.305\\
$\quad \quad \beta_1$ & $\left| g \right|$ & 0.813 & 0.032 & 0.136 & 0.132 & 0.016 & 0.018 & 0.018 & 0.015\\
 & $\mathrm{arg}(g) [^\mathrm{o}]$ & 112.871 & 2.555 & 9.487 & 8.636 & 2.025 & 2.241 & 1.817 & 1.730\\
$\quad \quad \beta_0$ & $\left| g \right|$ & 0.315 & 0.010 & 0.022 & 0.019 & 0.005 & 0.005 & 0.009 & 0.002\\
 & $\mathrm{arg}(g) [^\mathrm{o}]$ & 46.734 & 1.946 & 2.952 & 1.110 & 1.576 & 1.416 & 1.121 & 1.318\\
$K(1460)^{-}$ & $m_0 \left[\mevcc\right]$ & 1482.400 & 3.576 & 15.216 & 13.873 & 3.466 & 3.216 & 3.611 & 1.916\\
 & $\Gamma_0 \left[\mevcc\right]$&  335.595 & 6.196 & 8.651 & 1.524 & 4.234 & 2.017 & 5.901 & 3.962\\
\bottomrule
\end{tabular}
}
\end{table}

\begin{table}
  \centering 
  \caption{ \label{tb:WS_SysSummary} Systematic uncertainties on the WS decay coupling parameters and fit fractions.}
  \scalebox{0.75}{\begin{tabular}{l 
l 
>{\collectcell\num}r<{\endcollectcell} @{${}\pm{}$} >{\collectcell\num}l<{\endcollectcell} @{${}\pm{}$} >{\collectcell\num}l<{\endcollectcell}
l 
l 
l 
l 
l 
l 
}
\toprule
 &  & \multicolumn{3}{c}{} & \multicolumn{1}{c}{II} & \multicolumn{1}{c}{III} & \multicolumn{1}{c}{IV} & \multicolumn{1}{c}{V} & \multicolumn{1}{c}{VI} & \multicolumn{1}{c}{VII}\\
\midrule
\multirow{2}{*}{$K^*(892)^{0}\rho(770)^{0}$} & $\left| g \right|$ & 0.205 & 0.019 & 0.010 & 0.002 & 0.006 & 0.003 & 0.001 & 0.005 & 0.006\\
 & $\mathrm{arg}(g) [^o]$ & -8.502 & 4.662 & 4.439 & 0.433 & 1.272 & 0.112 & 0.148 & 4.150 & 0.799\\
 & $\mathcal{F}$ & 9.617 & 1.584 & 1.028 & 0.134 & 0.436 & 0.344 & 0.069 & 0.567 & 0.637\\
  
\multirow{3}{*}{$\left[K^*(892)^{0}\rho(770)^{0}\right]^{L=1}$} & $\left| g \right|$ & 0.390 & 0.029 & 0.006 & 0.002 & 0.003 & 0.000 & 0.001 & 0.004 & 0.003\\
 & $\mathrm{arg}(g) [^o]$ & -91.359 & 4.728 & 4.132 & 0.406 & 0.827 & 0.128 & 0.101 & 3.951 & 0.766\\
 & $\mathcal{F}$ & 8.424 & 0.827 & 0.573 & 0.069 & 0.091 & 0.210 & 0.020 & 0.458 & 0.249\\

 $\left[K^*(892)^{0}\rho(770)^{0}\right]^{L=2}$ & $\mathcal{F}$ & 10.191 & 1.028 & 0.789 & 0.089 & 0.130 & 0.255 & 0.018 & 0.658 & 0.314\\
  \multirow{3}{*}{$\rho(1450)^{0}K^*(892)^{0}$} & $\left| g \right|$ & 0.541 & 0.042 & 0.055 & 0.004 & 0.043 & 0.018 & 0.001 & 0.024 & 0.016\\
 & $\mathrm{arg}(g) [^o]$ & -21.798 & 6.536 & 5.483 & 0.573 & 4.532 & 0.547 & 0.254 & 0.254 & 2.960\\
  & $\mathcal{F}$ & 8.162 & 1.242 & 1.686 & 0.107 & 1.381 & 0.474 & 0.031 & 0.718 & 0.428\\
\midrule
\multirow{2}{*}{$K_{1}(1270)^{+}\pi^{-}$} & $\left| g \right|$ & 0.653 & 0.040 & 0.058 & 0.004 & 0.017 & 0.009 & 0.001 & 0.049 & 0.024\\
 & $\mathrm{arg}(g) [^o]$ & -110.715 & 5.054 & 4.854 & 0.481 & 1.484 & 0.219 & 0.056 & 4.236 & 1.770\\
 & $\mathcal{F}$ & 18.147 & 1.114 & 2.301 & 0.104 & 0.800 & 0.423 & 0.021 & 1.788 & 1.125\\
\multirow{2}{*}{$K_{1}(1400)^{+}\left[K^*(892)^{0}\pi^{+}\right]\pi^{-}$} & $\left| g \right|$ & 0.560 & 0.037 & 0.031 & 0.003 & 0.020 & 0.011 & 0.001 & 0.018 & 0.010\\
 & $\mathrm{arg}(g) [^o]$ & 29.769 & 4.220 & 4.565 & 0.396 & 4.055 & 0.211 & 0.060 & 1.638 & 1.227\\
 & $\mathcal{F}$ & 26.549 & 1.973 & 2.128 & 0.190 & 1.715 & 0.469 & 0.046 & 0.940 & 0.667\\
\midrule
  $\left[K^{+}\pi^{-}\right]^{L=0}\left[\pi^{+}\pi^{-}\right]^{L=0}$ & $\mathcal{F}$ & 20.901 & 1.295 & 1.500 & 0.129 & 0.328 & 0.565 & 0.134 & 1.246 & 0.486\\
\multirow{2}{*}{$\quad \alpha_{3/2}$} & $\left| g \right|$ & 0.686 & 0.043 & 0.022 & 0.004 & 0.007 & 0.002 & 0.002 & 0.019 & 0.007\\
 & $\mathrm{arg}(g) [^o]$ & -149.399 & 4.260 & 2.946 & 0.502 & 0.277 & 0.181 & 0.082 & 2.809 & 0.651\\
\multirow{2}{*}{$\quad \beta_1$} & $\left| g \right|$ & 0.438 & 0.044 & 0.030 & 0.004 & 0.006 & 0.010 & 0.001 & 0.026 & 0.010\\
 & $\mathrm{arg}(g) [^o]$ & -132.424 & 6.507 & 2.972 & 0.618 & 1.109 & 0.357 & 0.200 & 2.382 & 1.174\\
\multirow{2}{*}{$\quad f_{\pi\pi}$} & $\left| g \right|$ & 0.050 & 0.006 & 0.005 & 0.001 & 0.001 & 0.001 & 0.000 & 0.004 & 0.002\\
 & $\mathrm{arg}(g) [^o]$ & 74.821 & 7.528 & 5.282 & 0.695 & 0.745 & 0.149 & 0.472 & 5.050 & 1.058\\
\bottomrule
\end{tabular}
}
\end{table}

\FloatBarrier

\section{\label{sec:Interference}Interference fractions}                                      The interference fraction between decay chains $a$ and $b$ is 
\begin{equation}
  I(a,b) = \frac{ \mathrm{Re}\left( \int d \xp \mathcal{M}_a (\xp) \mathcal{M}_b(\xp)^{*} \right) }{ \int d\xp \left| 
  \sum_{j}\mathcal{M}_j(\xp) \right|^2 },
\end{equation}
where the sum over ${j}$ is over all of the decay chains. For cascade processes, the different secondary isobars contribute coherently to the interference fractions. 
The interference fractions are presented in Tables~\ref{tb:RSIF}~and~\ref{tb:WSIF} for RS and WS decay modes, respectively. 
For each decay mode, the largest interference fractions are between the axial vector decay chain, and the lowest orbital angular momentum vector-vector decay chain. 

\begin{table}
  \caption{\label{tb:RSIF} Interference fractions for the RS mode \RS, only shown for fractions $>0.5\%$. For each fraction, the first uncertainty is statistical and the second systematic. }
  \begin{tabular}{l l 
>{\collectcell\num}r<{\endcollectcell} @{${}\pm{}$} >{\collectcell\num}l<{\endcollectcell} @{${}\pm{}$} >{\collectcell\num}l<{\endcollectcell}
}
\toprule
  Decay chain $a$ & Decay chain $b$ & \multicolumn{3}{c}{Interference Fraction $\left[\%\right]$} \\
\midrule
$\Kstarb(892)^{0}\rho(770)^{0}  $&$ {a}_{1}(1260)^{+}{K}^{-}$ & 5.74  & 0.03  & 0.1 \\
$\left[\Kstarb(892)^{0}\rho(770)^{0}\right]^{L=2} $&$ \Kstarb(892)^{0}\rho(770)^{0}$  & -2.59 & 0.02  & 0.07  \\
$\left[{K}^{+}\pi^{-}\right]^{L=0}\left[\pi^{+}\pi^{-}\right]^{L=0} $&$ {a}_{1}(1260)^{+}{K}^{-}$ & 2.4 & 0.03  & 0.14  \\
${a}_{1}(1260)^{+}{K}^{-} $&$ \rho(770)^{0}\left[{K}^{-}\pi^{+}\right]^{L=0}$ & 2.14  & 0.07  & 0.26  \\
$\left[\Kstarb(892)^{0}\rho(770)^{0}\right]^{L=2} $&$ {a}_{1}(1260)^{+}{K}^{-}$ & -1.76 & 0.01  & 0.08  \\
$\left[\Kstarb(892)^{0}\rho(770)^{0}\right]^{L=1} $&$ \left[\rho(1450)^{0}\Kstarb(892)^{0}\right]^{L=1}$  & -1.55 & 0.02  & 0.18  \\
${K}_{1}(1270)^{-}\pi^{+} $&$ \Kstarb(892)^{0}\rho(770)^{0}$  & -1.05 & 0.02  & 0.14  \\
${K}_{1}(1400)^{-}\left[\Kstarb(892)^{0}\pi^{-}\right]\pi^{+} $&$ \Kstarb(892)^{0}\rho(770)^{0}$ & 0.96  & 0.02  & 0.1 \\
$\Kstarb(892)^{0}\rho(770)^{0}  $&$ \rho(1450)^{0}\Kstarb(892)^{0}$ & -0.83 & 0.05  & 0.11  \\
$\left[\Kstarb(892)^{0}\rho(770)^{0}\right]^{L=2} $&$ \left[\rho(1450)^{0}\Kstarb(892)^{0}\right]^{L=2}$  & 0.81  & 0.04  & 0.13  \\
${K}(1460)^{-}\pi^{+} $&$ \Kstarb(892)^{0}\left[\pi^{+}\pi^{-}\right]^{L=0}$  & 0.78  & 0.03  & 0.1 \\
$\Kstarb(892)^{0}\rho(770)^{0}  $&$ \left[\Km\pip\right]^{L=0}\left[\pi^{+}\pi^{-}\right]^{L=0}$ & 0.73  & 0.01  & 0.03  \\
$\Kstarb(892)^{0}\left[\pi^{+}\pi^{-}\right]^{L=0}  $&$ {a}_{1}(1260)^{+}{K}^{-}$ & -0.68 & 0.01  & 0.07  \\
${K}_{1}(1270)^{-}\pi^{+} $&$ {K}_{1}(1400)^{-}\left[\Kstarb(892)^{0}\pi^{-}\right]\pi^{+}$ & -0.67 & 0.02  & 0.12  \\
${K}(1460)^{-}\pi^{+} $&$ \Kstarb(892)^{0}\rho(770)^{0}$  & -0.66 & 0.02  & 0.05  \\
${a}_{1}(1260)^{+}{K}^{-} $&$ \rho(1450)^{0}\Kstarb(892)^{0}$ & -0.63 & 0.02  & 0.08  \\
$\left[\Kstarb(892)^{0}\rho(770)^{0}\right]^{L=2} $&$ {K}(1460)^{-}\pi^{+}$ & -0.6  & 0.02  & 0.07  \\
${K}(1460)^{-}\pi^{+} $&$ \rho(1450)^{0}\Kstarb(892)^{0}$ & 0.51  & 0.01  & 0.06  \\
\bottomrule
\end{tabular}

\end{table}

\begin{table}
  \caption{\label{tb:WSIF} Interference fractions for the WS mode \WS, only shown for fractions $>0.5\%$. For each fraction, the first uncertainty is statistical and the second systematic. }
  
\begin{tabular}{l l 
>{\collectcell\num}r<{\endcollectcell} @{${}\pm{}$} >{\collectcell\num}l<{\endcollectcell} @{${}\pm{}$} >{\collectcell\num}l<{\endcollectcell}
}
\toprule
  Decay chain $a$ & Decay chain $b$ & \multicolumn{3}{c}{Interference Fraction $\left[\%\right]$} \\
\midrule
  $K_{1}(1400)^{+}\left[K^*(892)^{0}\pi^{+}\right]\pi^{-} $ & $ K^{*}(892)^{0}\rho(770)^{0}$                                       &  5.09&  0.49 &  0.56\\ 
  $\left[K^*(892)^{0}\rho(770)^{0}\right]^{L=2} $&$ K^*(892)^{0}\rho(770)^{0}$                                         & -3.48&  0.36 &  0.26\\ 
  $K_{1}(1270)^{+}\pi^{-} $&$ \rho(1450)^{0}K^*(892)^{0}$                                                                          & -2.17&  0.24 &  0.37\\ 
  $K_{1}(1400)^{+}\left[K^*(892)^{0}\pi^{+}\right]\pi^{-} $&$ \rho(1450)^{0}K^*(892)^{0}$                                          & -1.78&  0.88 &  0.63\\ 
  $\rho(1450)^{0}K^*(892)^{0} $&$ K^{*}(892)^{0}\rho(770)^{0}$                                                                    &  1.59&  0.69 &  0.77\\ 
  $\left[K^*(892)^{0}\rho(770)^{0}\right]^{L=2} $&$ \rho(1450)^{0}K^*(892)^{0}$                                                    & -1.49&  0.29 &  0.30\\ 
  $\left[K^*(892)^{0}\rho(770)^{0}\right]^{L=2} $&$ K_{1}(1400)^{+}\left[K^*(892)^{0}\pi^{+}\right]\pi^{-}$                        & -1.36&  0.13 &  0.12\\ 
  $K^*(892)^{0}\rho(770)^{0} $ & $ \left[K^{+}\pi^{-}\right]^{L=0}\left[\pi^{+}\pi^{-}\right]^{L=0}$                                 &  1.14&  0.13 &  0.11\\ 
  $K_{1}(1400)^{+}\left[K^*(892)^{0}\pi^{+}\right]\pi^{-} $ &$ \left[K^{+}\pi^{-}\right]^{L=0}\left[\pi^{+}\pi^{-}\right]^{L=0}$    &  1.03&  0.10 &  0.10\\ 
  $K_{1}(1270)^{+}\pi^{-} $&$ K_{1}(1400)^{+}\left[K^*(892)^{0}\pi^{+}\right]\pi^{-}$                                              &  0.82&  0.51 &  0.79\\ 
  $\rho(1450)^{0}K^*(892)^{0} $&$ \left[K^{+}\pi^{-}\right]^{L=0}\left[\pi^{+}\pi^{-}\right]^{L=0}$                                & -0.65&  0.11 &  0.09\\ 
  $K_{1}(1270)^{+}\pi^{-} $ & $ K^*(892)^{0}\rho(770)^{0}$                                                                           &  0.65&  0.29 &  0.33\\ 
\bottomrule
\end{tabular}

\end{table}

\section{\label{sec:AltModel}Models excluding \boldmath$K_1(1270)^{-} \to \rho(1450)^{0}\Km$} \begin{figure}
  \centering
  \includegraphics[width=0.48\textwidth]{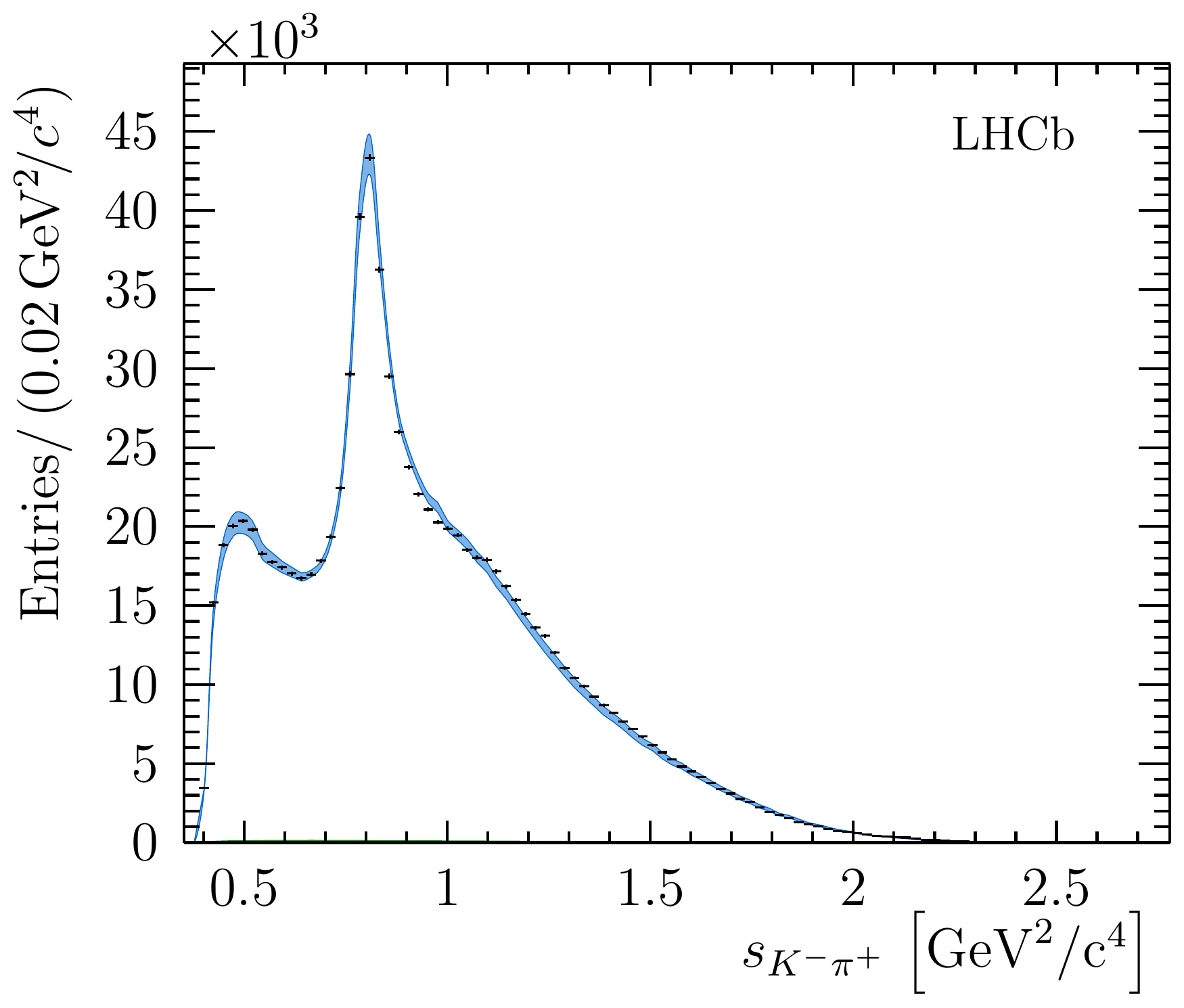} \includegraphics[width=0.48\textwidth]{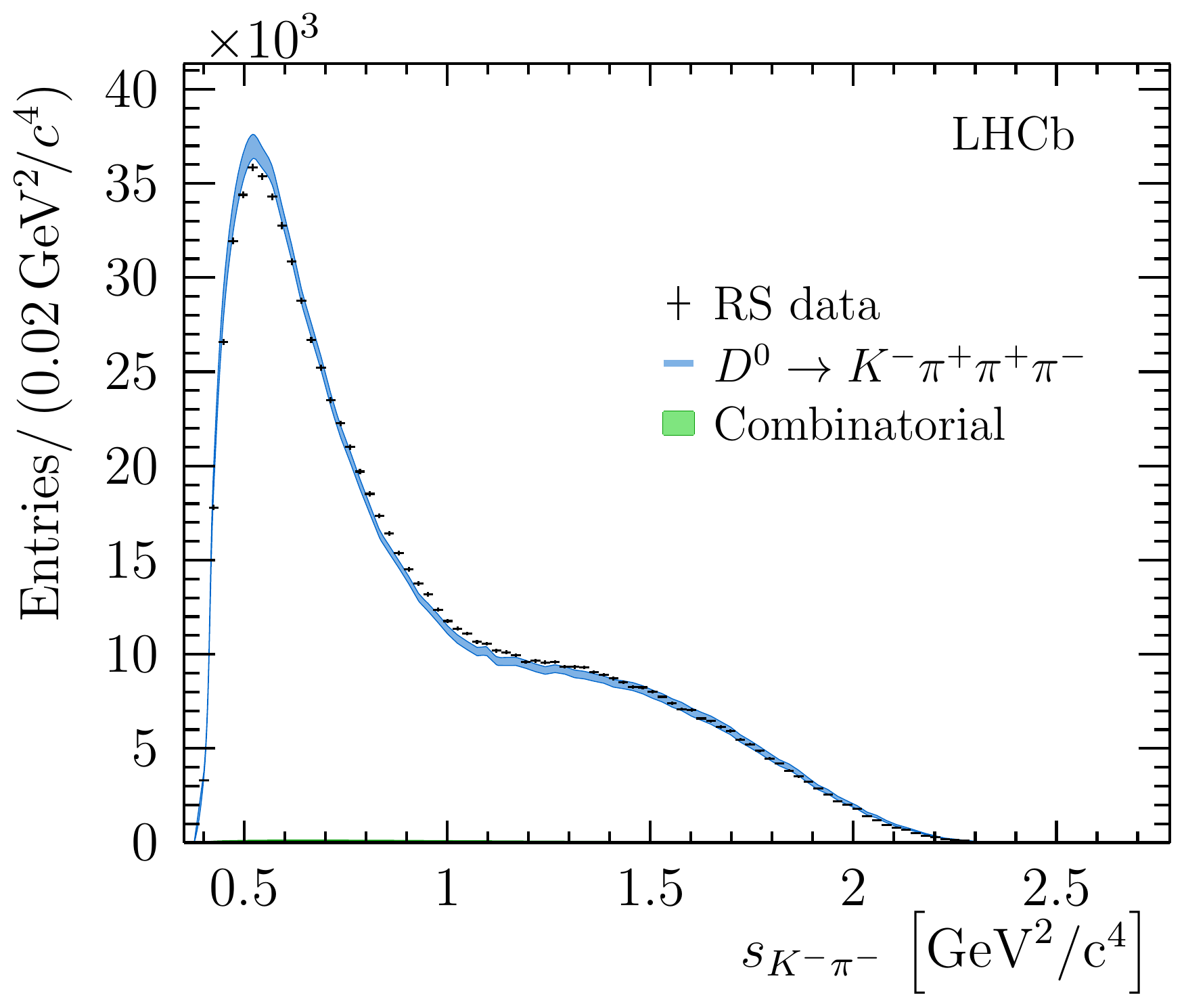}
  
  \includegraphics[width=0.48\textwidth]{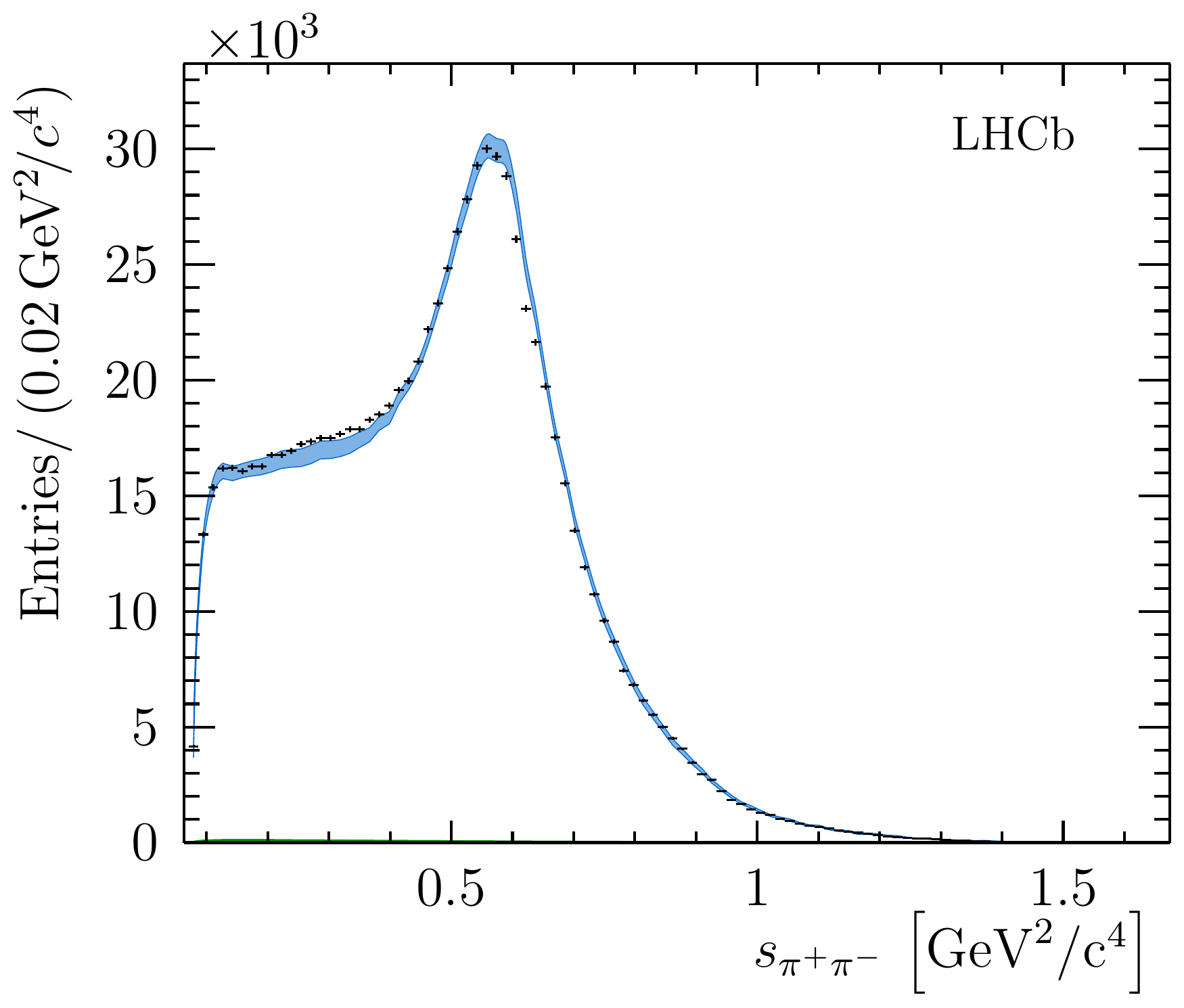} \includegraphics[width=0.48\textwidth]{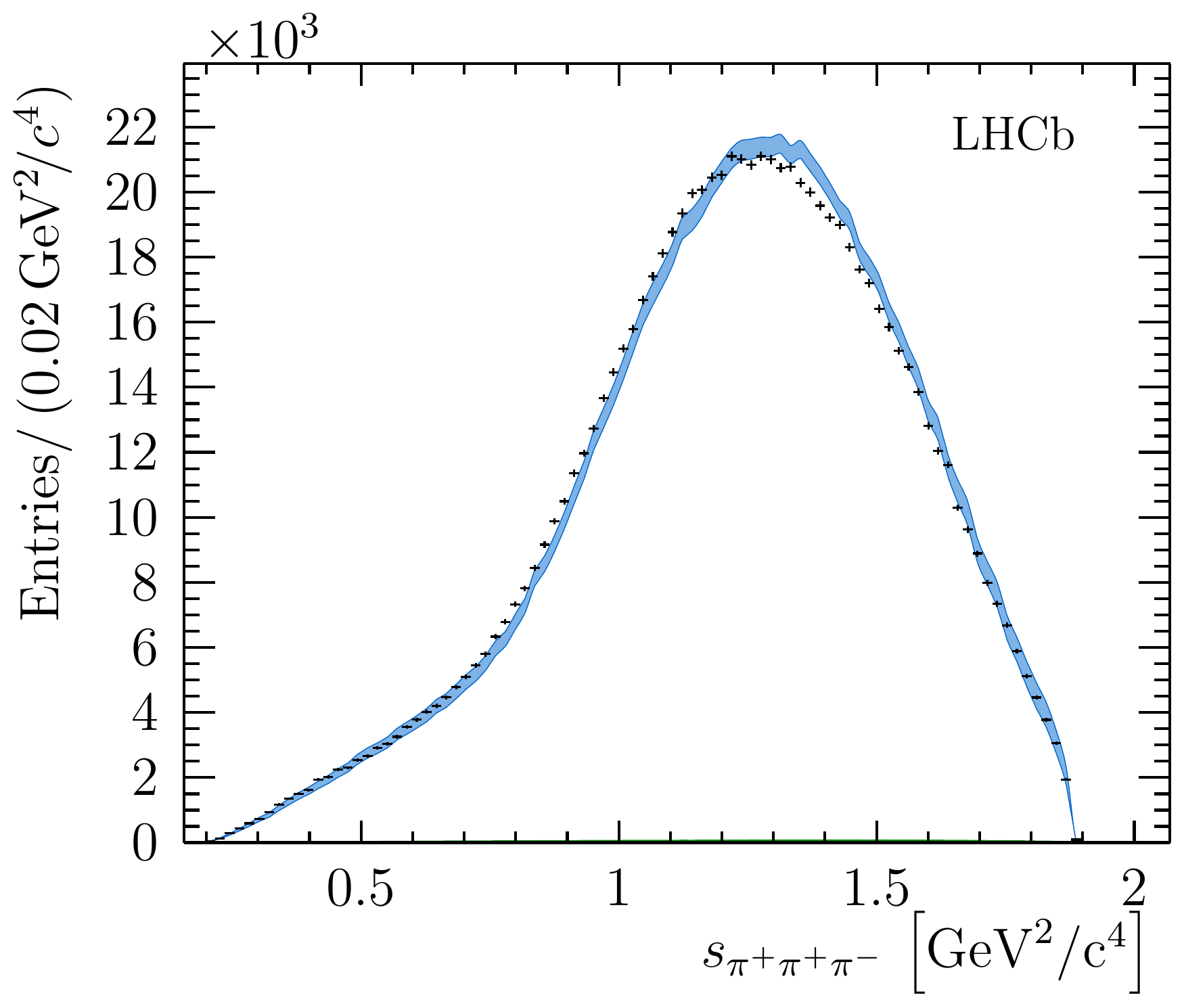}
  
  \includegraphics[width=0.48\textwidth]{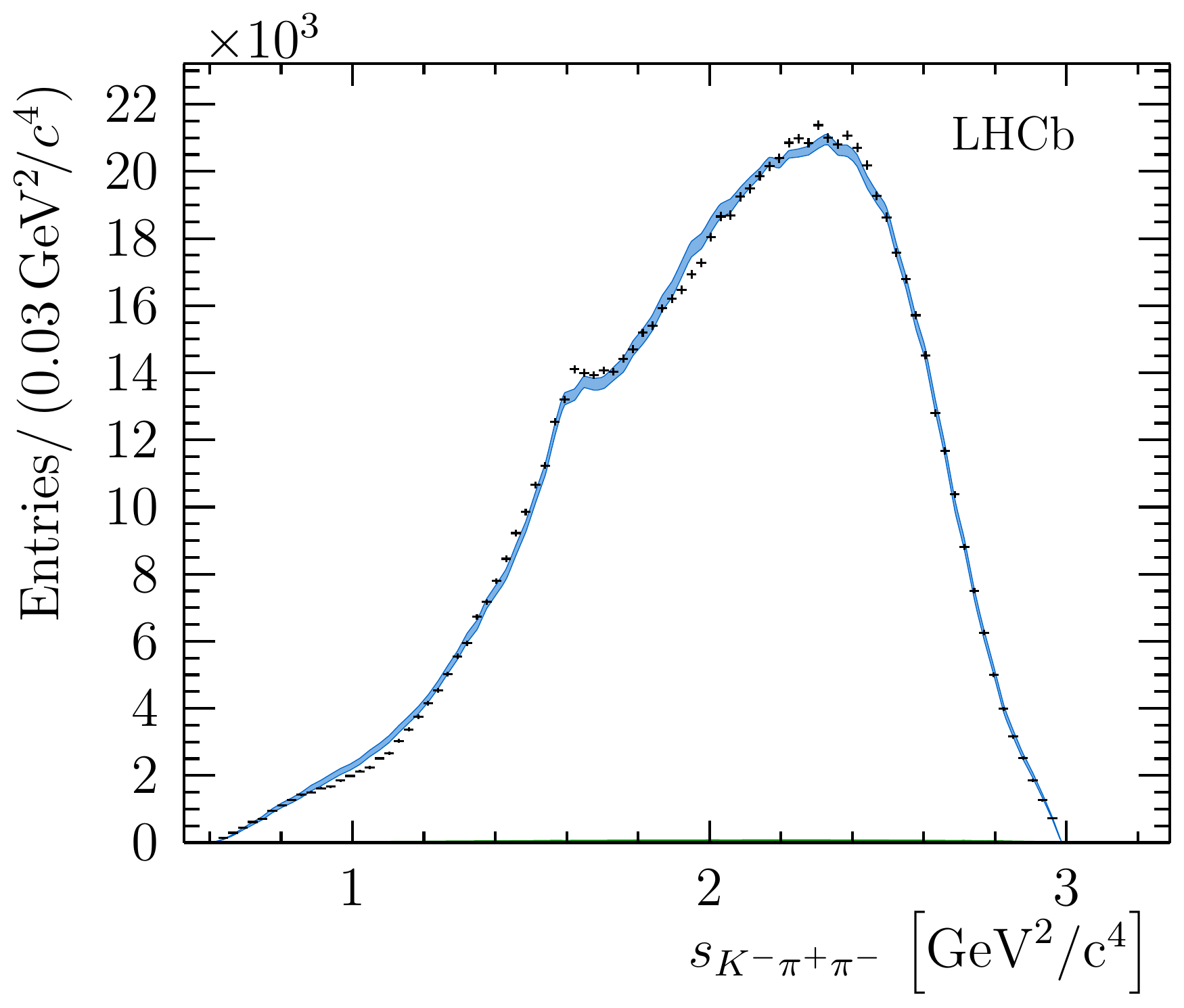} \includegraphics[width=0.48\textwidth]{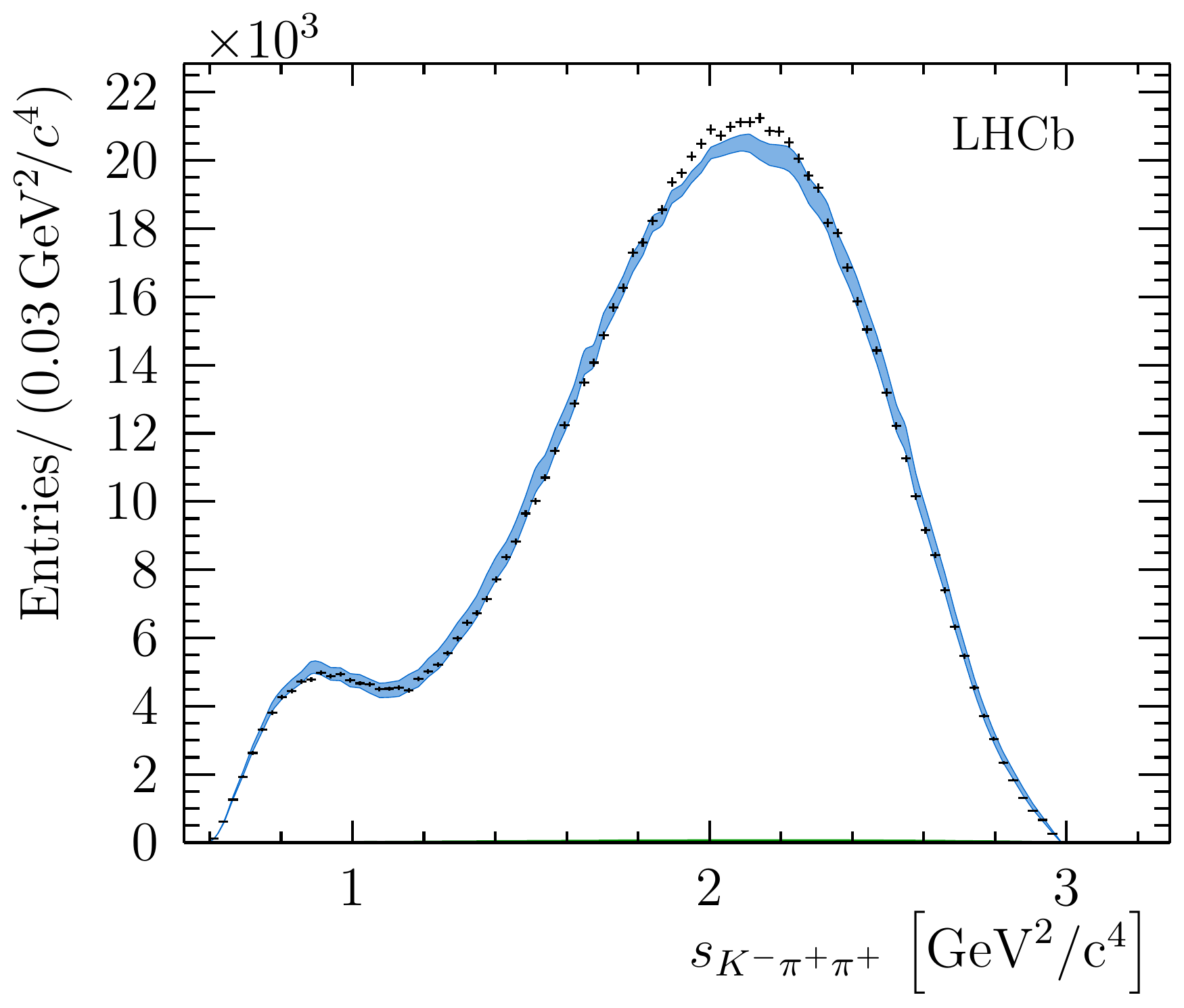}
  \caption{ Distributions for six invariant-mass observables in the RS decay \RS.
  Bands indicate the expectation from a model which excludes the decay chain $K_1(1270)^{-}\to\rho(1450)^{0}\Km$, with the width of the band indicating the total systematic uncertainty. The total background contribution, which is very low, is shown in green. 
  }
  \label{fig:RSPhase3_w1450}
\end{figure}
The results for the RS decay $\Dz\to\Km\pip\pip\pim$ are shown in this appendix for a model that excludes the amplitude $\PK_1(1270)^{-}\to\rho(1450)^{0}\Km$. 
The fit projections are shown in Fig.~\ref{fig:RSPhase3_w1450}. 
The $\chi^2$ per \dof of this fit is 1.28. 
The fit fractions and parameters are shown in Table~\ref{tb:RSparams_w1450}, and the partial fractions and parameters for the components associated with the resonances $\AONE$, $\PK_1(1270)^{-}$ and $\PK(1460)^{-}$ in Tables~\ref{tb:a1params_w1450},~\ref{tb:k1params_w1450},~\ref{tb:kparams_w1450}, respectively. 
This model would be preferred to that presented in Sect.~\ref{sec:results} if the $K_1(1270)\to\rho(1450)^{0}\Km$ decay chain is excluded by investigations of the $K_1(1270)$ resonance in other production modes.

\begin{table}
  \centering  
  \caption{ Table of fit fractions and coupling parameters and other quantities for the RS decay \RS,
  for a model excluding the decay chain $K_1(1270)^{-}\to \rho(1450)^0 \Km$. 
  Also given is the $\chi^2$ per \dof ($\nu$) for the fit. 
  The first uncertainty is statistical and the second systematic.
  Couplings are defined with respect to the coupling to the channel $\Dz\rightarrow[\Kstarb(892)^{0}\rho(770)^{0}]^{L=2}$.
  }
  \scalebox{0.9}{\begin{tabular}{l 
>{\collectcell\num}r<{\endcollectcell} @{${}\pm{}$} >{\collectcell\num}l<{\endcollectcell} @{${}\pm{}$} >{\collectcell\num}l<{\endcollectcell}
>{\collectcell\num}r<{\endcollectcell} @{${}\pm{}$} >{\collectcell\num}l<{\endcollectcell} @{${}\pm{}$} >{\collectcell\num}l<{\endcollectcell}
>{\collectcell\num}r<{\endcollectcell} @{${}\pm{}$} >{\collectcell\num}l<{\endcollectcell} @{${}\pm{}$} >{\collectcell\num}l<{\endcollectcell}
}
\toprule
 & \multicolumn{3}{c}{Fit Fraction [\%]} & \multicolumn{3}{c}{$\left|g\right|$} & \multicolumn{3}{c}{$\mathrm{arg}(g) [^o]$}\\
\midrule
$\Kstarb(892)^{0}\rho(770)^{0}$ & 7.45 & 0.09 & 0.47 & 0.200 & 0.001 & 0.014 & -26.4 & 0.4 & 1.4\\
$\left[\Kstarb(892)^{0}\rho(770)^{0}\right]^{L=1}$ & 6.03 & 0.05 & 0.25 & 0.366 & 0.002 & 0.020 & -103.1 & 0.4 & 1.7\\
$\left[\Kstarb(892)^{0}\rho(770)^{0}\right]^{L=2}$ & 8.30 & 0.09 & 0.71 & \multicolumn{3}{c}{} & \multicolumn{3}{c}{}\\
$\rho(1450)^{0}\Kstarb(892)^{0}$ & 0.11 & 0.02 & 0.06 & 0.068 & 0.006 & 0.013 & -133.9 & 5.7 & 16.2\\
$\left[\rho(1450)^{0}\Kstarb(892)^{0}\right]^{L=1}$ & 1.99 & 0.03 & 0.33 & 0.652 & 0.006 & 0.064 & 97.3 & 0.5 & 3.8\\
$\left[\rho(1450)^{0}\Kstarb(892)^{0}\right]^{L=2}$ & 0.36 & 0.03 & 0.11 & 0.581 & 0.022 & 0.088 & 8.2 & 2.2 & 16.5\\
\midrule
$\rho(770)^{0}\left[\Km\pip\right]^{L=0}$ & 1.29 & 0.04 & 0.09 & 0.318 & 0.007 & 0.012 & 69.8 & 0.9 & 6.0\\
$\quad \alpha_{3/2}$ & \multicolumn{3}{c}{} & 1.227 & 0.011 & 0.015 & -129.4 & 0.5 & 1.4\\
$\Kstarb(892)^{0}\left[\pi^{+}\pi^{-}\right]^{L=0}$ & 4.57 & 0.17 & 0.75 & \multicolumn{3}{c}{} & \multicolumn{3}{c}{}\\
$\quad f_{\pi\pi}$ & \multicolumn{3}{c}{} & 0.352 & 0.006 & 0.034 & -148.5 & 0.8 & 1.7\\
$\quad \beta_1$ & \multicolumn{3}{c}{} & 0.507 & 0.012 & 0.045 & 69.7 & 1.1 & 3.0\\
\midrule
$a_{1}(1260)^{+}K^{-}$ & 33.56 & 0.22 & 1.58 & 0.771 & 0.006 & 0.043 & -151.6 & 0.5 & 3.6\\
$K_{1}(1270)^{-}\pi^{+}$ & 4.67 & 0.05 & 0.26 & 0.260 & 0.003 & 0.007 & 90.5 & 0.9 & 1.9\\
$K_{1}(1400)^{-}\left[\Kstarb(892)^{0}\pi^{-}\right]\pi^{+}$ & 0.87 & 0.03 & 0.14 & 0.112 & 0.002 & 0.011 & -156.6 & 1.2 & 8.5\\
\midrule
$K_{2}^{*}(1430)^{-}\left[\Kstarb(892)^{0}\pi^{-}\right]\pi^{+}$ & 0.47 & 0.01 & 0.03 & 0.309 & 0.004 & 0.014 & -79.0 & 0.7 & 2.6\\
\midrule
$K(1460)^{-}\pi^{+}$ & 5.07 & 0.18 & 0.51 & 0.134 & 0.003 & 0.013 & 220.4 & 2.6 & 16.0\\
\midrule
$\left[\Km\pip\right]^{L=0}\left[\pi^{+}\pi^{-}\right]^{L=0}$ & 30.20 & 0.45 & 3.20 & \multicolumn{3}{c}{} & \multicolumn{3}{c}{}\\
$\quad \alpha_{3/2}$ & \multicolumn{3}{c}{} & 0.897 & 0.009 & 0.020 & -147.2 & 0.5 & 1.3\\
$\quad \alpha_{K\eta^\prime}$ & \multicolumn{3}{c}{} & 2.316 & 0.101 & 0.308 & -2.6 & 2.1 & 6.1\\
$\quad \beta_1$ & \multicolumn{3}{c}{} & 0.656 & 0.008 & 0.067 & 33.0 & 0.6 & 2.3\\
$\quad f_{\pi\pi}$ & \multicolumn{3}{c}{} & 0.093 & 0.001 & 0.009 & -149.6 & 0.7 & 2.7\\
\midrule
Sum of Fit Fractions & 104.94 & 0.75 & 2.72 & \multicolumn{6}{c}{}\\
$\chi^2 / \nu$ & \multicolumn{3}{c}{$41896/32702 = 1.281 $} & \multicolumn{6}{c}{}\\
\bottomrule
\end{tabular}
}
  \label{tb:RSparams_w1450}
\end{table}

\FloatBarrier 
\begin{table}
  \centering
  \caption{Table of fit fractions and coupling parameters for the component involving the \AONE meson, 
  from the fit performed on the RS decay \RS. 
  The coupling parameters are defined with respect to the $\AONE\rightarrow\rho(770)^{0}\pip$ coupling. 
  For each parameter, the first uncertainty is statistical and the second systematic. }
  \scalebox{0.9}{\begin{tabular}{l 
>{\collectcell\num}r<{\endcollectcell} @{${}\pm{}$} >{\collectcell\num}l<{\endcollectcell} @{${}\pm{}$} >{\collectcell\num}l<{\endcollectcell}
>{\collectcell\num}r<{\endcollectcell} @{${}\pm{}$} >{\collectcell\num}l<{\endcollectcell} @{${}\pm{}$} >{\collectcell\num}l<{\endcollectcell}
>{\collectcell\num}r<{\endcollectcell} @{${}\pm{}$} >{\collectcell\num}l<{\endcollectcell} @{${}\pm{}$} >{\collectcell\num}l<{\endcollectcell}
}
\toprule
  \multicolumn{10}{l}{$a_1(1260)^{+}$\quad$m_0= 1183.73 \pm 1.08 \pm 7.96 \mevcc$; $\Gamma_0=423.36\pm2.20\pm12.89\mevcc$}\\
  \rule{0pt}{3ex}    
 & \multicolumn{3}{c}{Partial Fractions [\%]} & \multicolumn{3}{c}{$\left|g\right|$} & \multicolumn{3}{c}{$\mathrm{arg}(g) [^o]$}\\
\midrule
$\quad \rho(770)^{0}\pi^{+}$ & 90.05 & 0.47 & 1.26 & \multicolumn{6}{c}{}\\
$\quad \left[\pi^{+}\pi^{-}\right]^{L=0}\pi^{+}$ & 3.08 & 0.07 & 0.21 & \multicolumn{6}{c}{}\\
$\quad \quad \beta_1$ & \multicolumn{3}{c}{} & 1.135 & 0.019 & 0.060 & -17.7 & 1.0 & 1.0\\
$\quad \quad \beta_0$ & \multicolumn{3}{c}{} & 0.312 & 0.007 & 0.016 & 157.3 & 1.4 & 2.9\\
$\quad \quad f_{\pi\pi}$ & \multicolumn{3}{c}{} & 0.159 & 0.003 & 0.011 & 176.8 & 1.0 & 2.3\\
$\quad \left[\rho(770)^{0}\pi^{+}\right]^{L=2}$ & 0.84 & 0.04 & 0.07 & 0.584 & 0.012 & 0.024 & -146.1 & 1.3 & 3.3\\
\bottomrule
\end{tabular}
}
  \label{tb:a1params_w1450}
\end{table}

\begin{table}
  \centering
  \caption{Table of fit fractions and coupling parameters for the component involving the $\KONE{1270}^{-}$ meson,
  from the fit performed on the RS decay \RS. 
  The coupling parameters are defined with respect to the $\KONE{1270}^{-}\rightarrow\rho(770)^{0}\Km$ coupling. 
  For each parameter, the first uncertainty is statistical and the second systematic. }
  \scalebox{0.9}{\begin{tabular}{l 
>{\collectcell\num}r<{\endcollectcell} @{${}\pm{}$} >{\collectcell\num}l<{\endcollectcell} @{${}\pm{}$} >{\collectcell\num}l<{\endcollectcell}
>{\collectcell\num}r<{\endcollectcell} @{${}\pm{}$} >{\collectcell\num}l<{\endcollectcell} @{${}\pm{}$} >{\collectcell\num}l<{\endcollectcell}
>{\collectcell\num}r<{\endcollectcell} @{${}\pm{}$} >{\collectcell\num}l<{\endcollectcell} @{${}\pm{}$} >{\collectcell\num}l<{\endcollectcell}
}
\toprule
  \multicolumn{10}{c}{$K_1(1270)^{-}$\quad$m_0=1285.03\pm0.47\pm1.06\mevcc$; $\Gamma_0=90.79\pm 1.12\pm2.54 \mevcc$} \\
  \rule{0pt}{3ex} 
  & \multicolumn{3}{c}{Partial Fractions [\%]} & \multicolumn{3}{c}{$\left|g\right|$} & \multicolumn{3}{c}{$\mathrm{arg}(g) [^o]$}\\
\midrule
$\quad \rho(770)^{0}K^{-}$ & 50.66 & 0.84 & 2.21 & \multicolumn{6}{c}{}\\
$\quad \Kstarb(892)^{0}\pi^{-}$ & 25.25 & 0.56 & 1.80 & 0.520 & 0.008 & 0.024 & -133.2 & 0.9 & 2.2\\
$\quad \left[K^{-}\pi^{+}\right]^{L=0}\pi^{-}$ & 5.97 & 0.29 & 0.37 & 0.390 & 0.010 & 0.018 & 95.2 & 1.4 & 3.7\\
$\quad \left[\Kstarb(892)^{0}\pi^{-}\right]^{L=2}$ & 2.73 & 0.14 & 0.24 & 0.946 & 0.028 & 0.147 & 8.8 & 1.7 & 2.4\\
$\quad \omega(782)\left[\pi^{+}\pi^{-}\right]K^{-}$ & 1.73 & 0.11 & 0.16 & 0.208 & 0.008 & 0.011 & 33.0 & 2.1 & 12.5\\
\bottomrule
\end{tabular}
}
  \label{tb:k1params_w1450}
\end{table}

\begin{table}
  \centering
  \caption{Table of fit fractions and coupling parameters for the component involving the \Kexc meson, 
  from the fit performed on the RS decay \RS.
  The coupling parameters are defined with respect to the $\Kexc\rightarrow\Kstarb(892)^{0}\pim$ coupling. 
  For each parameter, the first uncertainty is statistical and the second systematic. }
  \scalebox{0.9}{\begin{tabular}{l 
>{\collectcell\num}r<{\endcollectcell} @{${}\pm{}$} >{\collectcell\num}l<{\endcollectcell} @{${}\pm{}$} >{\collectcell\num}l<{\endcollectcell}
>{\collectcell\num}r<{\endcollectcell} @{${}\pm{}$} >{\collectcell\num}l<{\endcollectcell} @{${}\pm{}$} >{\collectcell\num}l<{\endcollectcell}
>{\collectcell\num}r<{\endcollectcell} @{${}\pm{}$} >{\collectcell\num}l<{\endcollectcell} @{${}\pm{}$} >{\collectcell\num}l<{\endcollectcell}
}
\toprule
  \multicolumn{10}{l}{$K(1460)^{-}$\quad$m_0=1571.22\pm 4.90\pm 33.76\mevcc$; $\Gamma_0=376.64\pm 7.43 \pm 25.30 \mevcc$} \\
 & \multicolumn{3}{c}{Partial Fractions [\%]} & \multicolumn{3}{c}{$\left|g\right|$} & \multicolumn{3}{c}{$\mathrm{arg}(g) [^o]$}\\
\midrule
$\quad \Kstarb(892)^{0}\pi^{-}$ & 45.73 & 1.09 & 1.95 & \multicolumn{6}{c}{}\\
$\quad \left[\pi^{+}\pi^{-}\right]^{L=0}K^{-}$ & 37.71 & 1.06 & 1.98 & \multicolumn{6}{c}{}\\
$\quad \quad f_{KK}$ & \multicolumn{3}{c}{} & 1.573 & 0.059 & 0.066 & -102.6 & 2.7 & 10.0\\
$\quad \quad \beta_1$ & \multicolumn{3}{c}{} & 0.875 & 0.032 & 0.042 & 85.3 & 2.6 & 8.3\\
$\quad \quad \beta_0$ & \multicolumn{3}{c}{} & 0.323 & 0.010 & 0.023 & 25.6 & 2.1 & 8.4\\
\bottomrule
\end{tabular}
}
  \label{tb:kparams_w1450}
\end{table}

\FloatBarrier

\end{appendices}

\addcontentsline{toc}{section}{References}
\setboolean{inbibliography}{true}
\bibliographystyle{LHCb}
\bibliography{main,LHCb-PAPER,LHCb-DP}

\ifx\mcitethebibliography\mciteundefinedmacro
\PackageError{LHCb.bst}{mciteplus.sty has not been loaded}
{This bibstyle requires the use of the mciteplus package.}\fi
\providecommand{\href}[2]{#2}
\begin{mcitethebibliography}{10}
\mciteSetBstSublistMode{n}
\mciteSetBstMaxWidthForm{subitem}{\alph{mcitesubitemcount})}
\mciteSetBstSublistLabelBeginEnd{\mcitemaxwidthsubitemform\space}
{\relax}{\relax}

\bibitem{Atwood:1996ci}
D.~Atwood, I.~Dunietz, and A.~Soni,
  \ifthenelse{\boolean{articletitles}}{\emph{{Enhanced CP violation with ${B\to
  K D^{0} (\Dzb)}$ modes and extraction of the CKM angle $\gamma$}},
  }{}\href{http://dx.doi.org/10.1103/PhysRevLett.78.3257}{Phys.\ Rev.\ Lett.\
  \textbf{78} (1997) 3257},
  \href{http://arxiv.org/abs/hep-ph/9612433}{{\normalfont\ttfamily
  arXiv:hep-ph/9612433}}\relax
\mciteBstWouldAddEndPuncttrue
\mciteSetBstMidEndSepPunct{\mcitedefaultmidpunct}
{\mcitedefaultendpunct}{\mcitedefaultseppunct}\relax
\EndOfBibitem
\bibitem{Atwood:2000ck}
D.~Atwood, I.~Dunietz, and A.~Soni,
  \ifthenelse{\boolean{articletitles}}{\emph{{Improved methods for observing CP
  violation in $\Bpm \rightarrow KD$ and measuring the CKM phase $\gamma$}},
  }{}\href{http://dx.doi.org/10.1103/PhysRevD.63.036005}{Phys.\ Rev.\
  \textbf{D63} (2001) 036005},
  \href{http://arxiv.org/abs/hep-ph/0008090}{{\normalfont\ttfamily
  arXiv:hep-ph/0008090}}\relax
\mciteBstWouldAddEndPuncttrue
\mciteSetBstMidEndSepPunct{\mcitedefaultmidpunct}
{\mcitedefaultendpunct}{\mcitedefaultseppunct}\relax
\EndOfBibitem
\bibitem{LHCb-PAPER-2016-003}
LHCb collaboration, R.~Aaij {\em et~al.},
  \ifthenelse{\boolean{articletitles}}{\emph{{Measurement of $\CP$ observables
  in $\Bpm\to\D \Kpm$ and $\Bpm\to\D\pipm$ with two- and four-body $\D$
  decays}}, }{}\href{http://dx.doi.org/10.1016/j.physletb.2016.06.022}{Phys.\
  Lett.\  \textbf{B760} (2016) 117},
  \href{http://arxiv.org/abs/1603.08993}{{\normalfont\ttfamily
  arXiv:1603.08993}}\relax
\mciteBstWouldAddEndPuncttrue
\mciteSetBstMidEndSepPunct{\mcitedefaultmidpunct}
{\mcitedefaultendpunct}{\mcitedefaultseppunct}\relax
\EndOfBibitem
\bibitem{LHCb-PAPER-2015-057}
LHCb collaboration, R.~Aaij {\em et~al.},
  \ifthenelse{\boolean{articletitles}}{\emph{{First observation of $\Dz-\Dzb$
  oscillations in $\Dz\to\Kp\pip\pim\pim$ decays and a measurement of the
  associated coherence parameters}},
  }{}\href{http://dx.doi.org/10.1103/PhysRevLett.116.241801}{Phys.\ Rev.\
  Lett.\  \textbf{116} (2016) 241801},
  \href{http://arxiv.org/abs/1602.07224}{{\normalfont\ttfamily
  arXiv:1602.07224}}\relax
\mciteBstWouldAddEndPuncttrue
\mciteSetBstMidEndSepPunct{\mcitedefaultmidpunct}
{\mcitedefaultendpunct}{\mcitedefaultseppunct}\relax
\EndOfBibitem
\bibitem{PhysRevD.89.072002}
CLEO collaboration, G.~Bonvicini {\em et~al.},
  \ifthenelse{\boolean{articletitles}}{\emph{{Updated measurements of absolute
  ${D}^{+}$ and ${D}^{0}$ hadronic branching fractions and
  ${\ensuremath{\sigma}({e}^{+}{e}^{-}\ensuremath{\rightarrow} D
  \overline{D})}$ at ${E}_{\mathrm{cm}}=3774\text{ }\text{ }\mathrm{MeV}$}},
  }{}\href{http://dx.doi.org/10.1103/PhysRevD.89.072002}{Phys.\ Rev.\
  \textbf{D89} (2014) 072002},
  \href{http://arxiv.org/abs/1312.6775}{{\normalfont\ttfamily
  arXiv:1312.6775}}\relax
\mciteBstWouldAddEndPuncttrue
\mciteSetBstMidEndSepPunct{\mcitedefaultmidpunct}
{\mcitedefaultendpunct}{\mcitedefaultseppunct}\relax
\EndOfBibitem
\bibitem{PhysRevD.45.2196}
Mark III collaboration, D.~Coffman {\em et~al.},
  \ifthenelse{\boolean{articletitles}}{\emph{Resonant substructure in
  $\overline{K}\pi\pi\pi$ decays of $\mathrm{D}$ mesons},
  }{}\href{http://dx.doi.org/10.1103/PhysRevD.45.2196}{Phys.\ Rev.\
  \textbf{D45} (1992) 2196}\relax
\mciteBstWouldAddEndPuncttrue
\mciteSetBstMidEndSepPunct{\mcitedefaultmidpunct}
{\mcitedefaultendpunct}{\mcitedefaultseppunct}\relax
\EndOfBibitem
\bibitem{Ablikim:2017eqz}
BES III collaboration, M.~Ablikim {\em et~al.},
  \ifthenelse{\boolean{articletitles}}{\emph{{Amplitude analysis of ${D^{0}\to
  K^{-} \pi^{+} \pi^{+} \pi^{-}}$}},
  }{}\href{http://dx.doi.org/10.1103/PhysRevD.95.072010}{Phys.\ Rev.\
  \textbf{D95} (2017) 072010},
  \href{http://arxiv.org/abs/1701.08591}{{\normalfont\ttfamily
  arXiv:1701.08591}}\relax
\mciteBstWouldAddEndPuncttrue
\mciteSetBstMidEndSepPunct{\mcitedefaultmidpunct}
{\mcitedefaultendpunct}{\mcitedefaultseppunct}\relax
\EndOfBibitem
\bibitem{LHCb-DP-2014-002}
LHCb collaboration, R.~Aaij {\em et~al.},
  \ifthenelse{\boolean{articletitles}}{\emph{{LHCb detector performance}},
  }{}\href{http://dx.doi.org/10.1142/S0217751X15300227}{Int.\ J.\ Mod.\ Phys.\
  \textbf{A30} (2015) 1530022},
  \href{http://arxiv.org/abs/1412.6352}{{\normalfont\ttfamily
  arXiv:1412.6352}}\relax
\mciteBstWouldAddEndPuncttrue
\mciteSetBstMidEndSepPunct{\mcitedefaultmidpunct}
{\mcitedefaultendpunct}{\mcitedefaultseppunct}\relax
\EndOfBibitem
\bibitem{LHCb-DP-2012-004}
R.~Aaij {\em et~al.}, \ifthenelse{\boolean{articletitles}}{\emph{{The \lhcb
  trigger and its performance in 2011}},
  }{}\href{http://dx.doi.org/10.1088/1748-0221/8/04/P04022}{JINST \textbf{8}
  (2013) P04022}, \href{http://arxiv.org/abs/1211.3055}{{\normalfont\ttfamily
  arXiv:1211.3055}}\relax
\mciteBstWouldAddEndPuncttrue
\mciteSetBstMidEndSepPunct{\mcitedefaultmidpunct}
{\mcitedefaultendpunct}{\mcitedefaultseppunct}\relax
\EndOfBibitem
\bibitem{BBDT}
V.~V. Gligorov and M.~Williams,
  \ifthenelse{\boolean{articletitles}}{\emph{{Efficient, reliable and fast
  high-level triggering using a bonsai boosted decision tree}},
  }{}\href{http://dx.doi.org/10.1088/1748-0221/8/02/P02013}{JINST \textbf{8}
  (2013) P02013}, \href{http://arxiv.org/abs/1210.6861}{{\normalfont\ttfamily
  arXiv:1210.6861}}\relax
\mciteBstWouldAddEndPuncttrue
\mciteSetBstMidEndSepPunct{\mcitedefaultmidpunct}
{\mcitedefaultendpunct}{\mcitedefaultseppunct}\relax
\EndOfBibitem
\bibitem{Sjostrand:2007gs}
T.~Sj\"{o}strand, S.~Mrenna, and P.~Skands,
  \ifthenelse{\boolean{articletitles}}{\emph{{A brief introduction to PYTHIA
  8.1}}, }{}\href{http://dx.doi.org/10.1016/j.cpc.2008.01.036}{Comput.\ Phys.\
  Commun.\  \textbf{178} (2008) 852},
  \href{http://arxiv.org/abs/0710.3820}{{\normalfont\ttfamily
  arXiv:0710.3820}}\relax
\mciteBstWouldAddEndPuncttrue
\mciteSetBstMidEndSepPunct{\mcitedefaultmidpunct}
{\mcitedefaultendpunct}{\mcitedefaultseppunct}\relax
\EndOfBibitem
\bibitem{LHCb-PROC-2010-056}
I.~Belyaev {\em et~al.}, \ifthenelse{\boolean{articletitles}}{\emph{{Handling
  of the generation of primary events in Gauss, the LHCb simulation
  framework}}, }{}\href{http://dx.doi.org/10.1088/1742-6596/331/3/032047}{{J.\
  Phys.\ Conf.\ Ser.\ } \textbf{331} (2011) 032047}\relax
\mciteBstWouldAddEndPuncttrue
\mciteSetBstMidEndSepPunct{\mcitedefaultmidpunct}
{\mcitedefaultendpunct}{\mcitedefaultseppunct}\relax
\EndOfBibitem
\bibitem{Lange:2001uf}
D.~J. Lange, \ifthenelse{\boolean{articletitles}}{\emph{{The EvtGen particle
  decay simulation package}},
  }{}\href{http://dx.doi.org/10.1016/S0168-9002(01)00089-4}{Nucl.\ Instrum.\
  Meth.\  \textbf{A462} (2001) 152}\relax
\mciteBstWouldAddEndPuncttrue
\mciteSetBstMidEndSepPunct{\mcitedefaultmidpunct}
{\mcitedefaultendpunct}{\mcitedefaultseppunct}\relax
\EndOfBibitem
\bibitem{Allison:2006ve}
Geant4 collaboration, J.~Allison {\em et~al.},
  \ifthenelse{\boolean{articletitles}}{\emph{{Geant4 developments and
  applications}}, }{}\href{http://dx.doi.org/10.1109/TNS.2006.869826}{IEEE
  Trans.\ Nucl.\ Sci.\  \textbf{53} (2006) 270}\relax
\mciteBstWouldAddEndPuncttrue
\mciteSetBstMidEndSepPunct{\mcitedefaultmidpunct}
{\mcitedefaultendpunct}{\mcitedefaultseppunct}\relax
\EndOfBibitem
\bibitem{Agostinelli:2002hh}
Geant4 collaboration, S.~Agostinelli {\em et~al.},
  \ifthenelse{\boolean{articletitles}}{\emph{{Geant4: A simulation toolkit}},
  }{}\href{http://dx.doi.org/10.1016/S0168-9002(03)01368-8}{Nucl.\ Instrum.\
  Meth.\  \textbf{A506} (2003) 250}\relax
\mciteBstWouldAddEndPuncttrue
\mciteSetBstMidEndSepPunct{\mcitedefaultmidpunct}
{\mcitedefaultendpunct}{\mcitedefaultseppunct}\relax
\EndOfBibitem
\bibitem{LHCb-PROC-2011-006}
M.~Clemencic {\em et~al.}, \ifthenelse{\boolean{articletitles}}{\emph{{The
  \lhcb simulation application, Gauss: Design, evolution and experience}},
  }{}\href{http://dx.doi.org/10.1088/1742-6596/331/3/032023}{{J.\ Phys.\ Conf.\
  Ser.\ } \textbf{331} (2011) 032023}\relax
\mciteBstWouldAddEndPuncttrue
\mciteSetBstMidEndSepPunct{\mcitedefaultmidpunct}
{\mcitedefaultendpunct}{\mcitedefaultseppunct}\relax
\EndOfBibitem
\bibitem{Breiman}
L.~Breiman, J.~H. Friedman, R.~A. Olshen, and C.~J. Stone, {\em Classification
  and regression trees}, Wadsworth international group, Belmont, California,
  USA, 1984\relax
\mciteBstWouldAddEndPuncttrue
\mciteSetBstMidEndSepPunct{\mcitedefaultmidpunct}
{\mcitedefaultendpunct}{\mcitedefaultseppunct}\relax
\EndOfBibitem
\bibitem{Roe}
B.~P. Roe {\em et~al.}, \ifthenelse{\boolean{articletitles}}{\emph{{Boosted
  decision trees as an alternative to artificial neural networks for particle
  identification}},
  }{}\href{http://dx.doi.org/10.1016/j.nima.2004.12.018}{Nucl.\ Instrum.\
  Meth.\  \textbf{A543} (2005) 577},
  \href{http://arxiv.org/abs/physics/0408124}{{\normalfont\ttfamily
  arXiv:physics/0408124}}\relax
\mciteBstWouldAddEndPuncttrue
\mciteSetBstMidEndSepPunct{\mcitedefaultmidpunct}
{\mcitedefaultendpunct}{\mcitedefaultseppunct}\relax
\EndOfBibitem
\bibitem{AdaBoost}
R.~E. Schapire and Y.~Freund, \ifthenelse{\boolean{articletitles}}{\emph{A
  decision-theoretic generalization of on-line learning and an application to
  boosting}, }{}\href{http://dx.doi.org/10.1006/jcss.1997.1504}{J.\ Comput.\
  Syst.\ Sci.\  \textbf{55} (1997) 119}\relax
\mciteBstWouldAddEndPuncttrue
\mciteSetBstMidEndSepPunct{\mcitedefaultmidpunct}
{\mcitedefaultendpunct}{\mcitedefaultseppunct}\relax
\EndOfBibitem
\bibitem{delAmoSanchez:2010ae}
BaBar collaboration, P.~del Amo~Sanchez {\em et~al.},
  \ifthenelse{\boolean{articletitles}}{\emph{{Study of $B \to X\gamma$ decays
  and determination of $|V_{td}/V_{ts}|$}},
  }{}\href{http://dx.doi.org/10.1103/PhysRevD.82.051101}{Phys.\ Rev.\
  \textbf{D82} (2010) 051101(R)},
  \href{http://arxiv.org/abs/1005.4087}{{\normalfont\ttfamily
  arXiv:1005.4087}}\relax
\mciteBstWouldAddEndPuncttrue
\mciteSetBstMidEndSepPunct{\mcitedefaultmidpunct}
{\mcitedefaultendpunct}{\mcitedefaultseppunct}\relax
\EndOfBibitem
\bibitem{PDG2014}
Particle Data Group, K.~A. Olive {\em et~al.},
  \ifthenelse{\boolean{articletitles}}{\emph{{\href{http://pdg.lbl.gov/}{Review
  of particle physics}}},
  }{}\href{http://dx.doi.org/10.1088/1674-1137/38/9/090001}{Chin.\ Phys.\
  \textbf{C38} (2014) 090001}\relax
\mciteBstWouldAddEndPuncttrue
\mciteSetBstMidEndSepPunct{\mcitedefaultmidpunct}
{\mcitedefaultendpunct}{\mcitedefaultseppunct}\relax
\EndOfBibitem
\bibitem{PhysRev.70.15}
E.~P. Wigner, \ifthenelse{\boolean{articletitles}}{\emph{Resonance reactions
  and anomalous scattering},
  }{}\href{http://dx.doi.org/10.1103/PhysRev.70.15}{Phys.\ Rev.\  \textbf{70}
  (1946) 15}\relax
\mciteBstWouldAddEndPuncttrue
\mciteSetBstMidEndSepPunct{\mcitedefaultmidpunct}
{\mcitedefaultendpunct}{\mcitedefaultseppunct}\relax
\EndOfBibitem
\bibitem{Chung:1995dx}
S.~U. Chung {\em et~al.}, \ifthenelse{\boolean{articletitles}}{\emph{{Partial
  wave analysis in K-matrix formalism}},
  }{}\href{http://dx.doi.org/10.1002/andp.19955070504}{Annalen Phys.\
  \textbf{4} (1995) 404}\relax
\mciteBstWouldAddEndPuncttrue
\mciteSetBstMidEndSepPunct{\mcitedefaultmidpunct}
{\mcitedefaultendpunct}{\mcitedefaultseppunct}\relax
\EndOfBibitem
\bibitem{Zou:2002ar}
B.~S. Zou and D.~V. Bugg, \ifthenelse{\boolean{articletitles}}{\emph{{Covariant
  tensor formalism for partial wave analyses of $\psi$ decay to mesons}},
  }{}\href{http://dx.doi.org/10.1140/epja/i2002-10135-4}{Eur.\ Phys.\ J.\
  \textbf{A16} (2003) 537},
  \href{http://arxiv.org/abs/hep-ph/0211457}{{\normalfont\ttfamily
  arXiv:hep-ph/0211457}}\relax
\mciteBstWouldAddEndPuncttrue
\mciteSetBstMidEndSepPunct{\mcitedefaultmidpunct}
{\mcitedefaultendpunct}{\mcitedefaultseppunct}\relax
\EndOfBibitem
\bibitem{TheoreticalNuclearPhysics}
J.~Blatt and V.~Weisskopf,
  \ifthenelse{\boolean{articletitles}}{\emph{Theoretical nuclear physics},
  }{}Springer-Verlag New York (1979)\relax
\mciteBstWouldAddEndPuncttrue
\mciteSetBstMidEndSepPunct{\mcitedefaultmidpunct}
{\mcitedefaultendpunct}{\mcitedefaultseppunct}\relax
\EndOfBibitem
\bibitem{PhysRevD.61.012002}
CLEO collaboration, D.~M. Asner {\em et~al.},
  \ifthenelse{\boolean{articletitles}}{\emph{{Hadronic structure in the decay
  ${\tau^{-} \to \nu_{\tau}\pi^-\pi^{0}\pi^{0}}$ and the sign of the tau
  neutrino helicity}},
  }{}\href{http://dx.doi.org/10.1103/PhysRevD.61.012002}{Phys.\ Rev.\
  \textbf{D61} (1999) 012002}\relax
\mciteBstWouldAddEndPuncttrue
\mciteSetBstMidEndSepPunct{\mcitedefaultmidpunct}
{\mcitedefaultendpunct}{\mcitedefaultseppunct}\relax
\EndOfBibitem
\bibitem{Aubert:2008bd}
BaBar collaboration, B.~Aubert {\em et~al.},
  \ifthenelse{\boolean{articletitles}}{\emph{{Improved measurement of the CKM
  angle $\gamma$ in $B^\mp \to D^{(*)} K^{(*)\mp}$ decays with a Dalitz plot
  analysis of $D$ decays to $K^0_{S} \pi^{+} \pi^{-}$ and $K^0_{S} K^{+}
  K^{-}$}}, }{}\href{http://dx.doi.org/10.1103/PhysRevD.78.034023}{Phys.\ Rev.\
   \textbf{D78} (2008) 034023},
  \href{http://arxiv.org/abs/0804.2089}{{\normalfont\ttfamily
  arXiv:0804.2089}}\relax
\mciteBstWouldAddEndPuncttrue
\mciteSetBstMidEndSepPunct{\mcitedefaultmidpunct}
{\mcitedefaultendpunct}{\mcitedefaultseppunct}\relax
\EndOfBibitem
\bibitem{Pennington:2007se}
FOCUS collaboration, J.~M. Link {\em et~al.},
  \ifthenelse{\boolean{articletitles}}{\emph{{Dalitz plot analysis of the
  $D^{+} \to K^{-} \pi^{+} \pi^{+}$ decay in the FOCUS experiment}},
  }{}\href{http://dx.doi.org/10.1016/j.physletb.2007.06.070}{Phys.\ Lett.\
  \textbf{B653} (2007) 1},
  \href{http://arxiv.org/abs/0705.2248}{{\normalfont\ttfamily
  arXiv:0705.2248}}\relax
\mciteBstWouldAddEndPuncttrue
\mciteSetBstMidEndSepPunct{\mcitedefaultmidpunct}
{\mcitedefaultendpunct}{\mcitedefaultseppunct}\relax
\EndOfBibitem
\bibitem{Anisovich:2002ij}
V.~V. Anisovich and A.~V. Sarantsev,
  \ifthenelse{\boolean{articletitles}}{\emph{{K-matrix analysis of the ($I
  J^{PC} = 00^{++}$)-wave in the mass region below 1900 MeV}},
  }{}\href{http://dx.doi.org/10.1140/epja/i2002-10068-x}{Eur.\ Phys.\ J.\
  \textbf{A16} (2003) 229},
  \href{http://arxiv.org/abs/hep-ph/0204328}{{\normalfont\ttfamily
  arXiv:hep-ph/0204328}}\relax
\mciteBstWouldAddEndPuncttrue
\mciteSetBstMidEndSepPunct{\mcitedefaultmidpunct}
{\mcitedefaultendpunct}{\mcitedefaultseppunct}\relax
\EndOfBibitem
\bibitem{PhysRev.88.1163}
K.~M. Watson, \ifthenelse{\boolean{articletitles}}{\emph{The effect of final
  state interactions on reaction cross sections},
  }{}\href{http://dx.doi.org/10.1103/PhysRev.88.1163}{Phys.\ Rev.\  \textbf{88}
  (1952) 1163}\relax
\mciteBstWouldAddEndPuncttrue
\mciteSetBstMidEndSepPunct{\mcitedefaultmidpunct}
{\mcitedefaultendpunct}{\mcitedefaultseppunct}\relax
\EndOfBibitem
\bibitem{PhysRevD.75.052003}
{FOCUS collaboration}, J.~M. Link {\em et~al.},
  \ifthenelse{\boolean{articletitles}}{\emph{Study of the
  ${D}^{0}\ensuremath{\rightarrow}{\ensuremath{\pi}}^{-}{\ensuremath{\pi}}^{+}{\ensuremath{\pi}}^{-}{\ensuremath{\pi}}^{+}$
  decay}, }{}\href{http://dx.doi.org/10.1103/PhysRevD.75.052003}{Phys.\ Rev.\ D
  \textbf{75} (2007) 052003}\relax
\mciteBstWouldAddEndPuncttrue
\mciteSetBstMidEndSepPunct{\mcitedefaultmidpunct}
{\mcitedefaultendpunct}{\mcitedefaultseppunct}\relax
\EndOfBibitem
\bibitem{dArgent:2017gzv}
P.~d'Argent {\em et~al.}, \ifthenelse{\boolean{articletitles}}{\emph{{Amplitude
  analyses of ${\Dzit\to\pi^{+}\pi^{-}\pi^{+}\pi^{-}}$ and ${\Dzit\to K^{+}
  K^{-} \pi^{+} \pi^{-}}$ decays}},
  }{}\href{http://dx.doi.org/10.1007/JHEP05(2017)143}{JHEP \textbf{05} (2017)
  143}, \href{http://arxiv.org/abs/1703.08505}{{\normalfont\ttfamily
  arXiv:1703.08505}}\relax
\mciteBstWouldAddEndPuncttrue
\mciteSetBstMidEndSepPunct{\mcitedefaultmidpunct}
{\mcitedefaultendpunct}{\mcitedefaultseppunct}\relax
\EndOfBibitem
\bibitem{PhysRevD.32.189}
S.~Godfrey and N.~Isgur, \ifthenelse{\boolean{articletitles}}{\emph{Mesons in a
  relativized quark model with chromodynamics},
  }{}\href{http://dx.doi.org/10.1103/PhysRevD.32.189}{Phys.\ Rev.\
  \textbf{D32} (1985) 189}\relax
\mciteBstWouldAddEndPuncttrue
\mciteSetBstMidEndSepPunct{\mcitedefaultmidpunct}
{\mcitedefaultendpunct}{\mcitedefaultseppunct}\relax
\EndOfBibitem
\bibitem{Daum:1981hb}
ACCMOR collaboration, C.~Daum {\em et~al.},
  \ifthenelse{\boolean{articletitles}}{\emph{{Diffractive production of strange
  mesons at $63\,Ge\kern -0.1em V$}},
  }{}\href{http://dx.doi.org/10.1016/0550-3213(81)90114-0}{Nucl.\ Phys.\
  \textbf{B187} (1981) 1}\relax
\mciteBstWouldAddEndPuncttrue
\mciteSetBstMidEndSepPunct{\mcitedefaultmidpunct}
{\mcitedefaultendpunct}{\mcitedefaultseppunct}\relax
\EndOfBibitem
\bibitem{Brandenburg:1976pg}
G.~W. Brandenburg {\em et~al.},
  \ifthenelse{\boolean{articletitles}}{\emph{{Evidence for a new
  strangeness-one pseudoscalar meson}},
  }{}\href{http://dx.doi.org/10.1103/PhysRevLett.36.1239}{Phys.\ Rev.\ Lett.\
  \textbf{36} (1976) 1239}\relax
\mciteBstWouldAddEndPuncttrue
\mciteSetBstMidEndSepPunct{\mcitedefaultmidpunct}
{\mcitedefaultendpunct}{\mcitedefaultseppunct}\relax
\EndOfBibitem
\bibitem{Meadows:2005ag}
E791 collaboration, B.~Meadows {\em et~al.},
  \ifthenelse{\boolean{articletitles}}{\emph{{Measurement of the $K^{-}
  \pi^{+}$ S-wave system in $D^{+} \to K^{-} \pi^{+} \pi^{+}$ decays from
  Fermilab E791}}, }{}\href{http://dx.doi.org/10.1063/1.2176563}{AIP Conf.\
  Proc.\  \textbf{814} (2006) 675},
  \href{http://arxiv.org/abs/hep-ex/0510045}{{\normalfont\ttfamily
  arXiv:hep-ex/0510045}}\relax
\mciteBstWouldAddEndPuncttrue
\mciteSetBstMidEndSepPunct{\mcitedefaultmidpunct}
{\mcitedefaultendpunct}{\mcitedefaultseppunct}\relax
\EndOfBibitem
\bibitem{Aitala:2005yh}
E791 collaboration, E.~M. Aitala {\em et~al.},
  \ifthenelse{\boolean{articletitles}}{\emph{{Model independent measurement of
  S-wave $\Km\pip$ systems using $\Dp \rightarrow \PK \pi \pi $ decays from
  Fermilab E791}}, }{}\href{http://dx.doi.org/10.1103/PhysRevD.73.032004,
  10.1103/PhysRevD.74.059901}{Phys.\ Rev.\  \textbf{D73} (2006) 032004},
  \href{http://arxiv.org/abs/hep-ex/0507099}{{\normalfont\ttfamily
  arXiv:hep-ex/0507099}}, [Erratum:
  \href{https://journals.aps.org/prd/abstract/10.1103/PhysRevD.74.059901}{Phys.
  Rev. \textbf{D74} (2006) 059901} ]\relax
\mciteBstWouldAddEndPuncttrue
\mciteSetBstMidEndSepPunct{\mcitedefaultmidpunct}
{\mcitedefaultendpunct}{\mcitedefaultseppunct}\relax
\EndOfBibitem
\bibitem{doi:10.1137/0705007}
K.~E. Atkinson, \ifthenelse{\boolean{articletitles}}{\emph{On the order of
  convergence of natural cubic spline interpolation},
  }{}\href{http://dx.doi.org/10.1137/0705007}{SIAM Journal on Numerical
  Analysis \textbf{5} (1968), no.~1 89}\relax
\mciteBstWouldAddEndPuncttrue
\mciteSetBstMidEndSepPunct{\mcitedefaultmidpunct}
{\mcitedefaultendpunct}{\mcitedefaultseppunct}\relax
\EndOfBibitem
\bibitem{Bigi:2001sg}
I.~I. Bigi, \ifthenelse{\boolean{articletitles}}{\emph{{Charm physics: Like
  Botticelli in the Sistine chapel}}, }{} in {\em {KAON2001: International
  Conference on CP Violation Pisa, Italy, June 12-17}}, 2001.
\newblock \href{http://arxiv.org/abs/hep-ph/0107102}{{\normalfont\ttfamily
  arXiv:hep-ph/0107102}}\relax
\mciteBstWouldAddEndPuncttrue
\mciteSetBstMidEndSepPunct{\mcitedefaultmidpunct}
{\mcitedefaultendpunct}{\mcitedefaultseppunct}\relax
\EndOfBibitem
\bibitem{PhysRevD.47.1252}
M.~Suzuki, \ifthenelse{\boolean{articletitles}}{\emph{Strange axial-vector
  mesons}, }{}\href{http://dx.doi.org/10.1103/PhysRevD.47.1252}{Phys.\ Rev.\
  \textbf{D47} (1993) 1252}\relax
\mciteBstWouldAddEndPuncttrue
\mciteSetBstMidEndSepPunct{\mcitedefaultmidpunct}
{\mcitedefaultendpunct}{\mcitedefaultseppunct}\relax
\EndOfBibitem
\bibitem{Atwood:2003mj}
D.~Atwood and A.~Soni, \ifthenelse{\boolean{articletitles}}{\emph{{Role of
  charm factory in extracting CKM phase information via ${B \to D K}$}},
  }{}\href{http://dx.doi.org/10.1103/PhysRevD.68.033003}{Phys.\ Rev.\
  \textbf{D68} (2003) 033003},
  \href{http://arxiv.org/abs/hep-ph/0304085}{{\normalfont\ttfamily
  arXiv:hep-ph/0304085}}\relax
\mciteBstWouldAddEndPuncttrue
\mciteSetBstMidEndSepPunct{\mcitedefaultmidpunct}
{\mcitedefaultendpunct}{\mcitedefaultseppunct}\relax
\EndOfBibitem
\bibitem{Evans:2016}
T.~Evans {\em et~al.}, \ifthenelse{\boolean{articletitles}}{\emph{{Improved
  determination of the $D \to K^-\pi^+\pi^+\pi^-$ coherence factor and
  associated hadronic parameters from a combination of $e^+e^-\to \psi(3770)\to
  c\bar{c}$ and $pp \to c \bar{c} X$ data}},
  }{}\href{http://dx.doi.org/10.1016/j.physletb.2016.04.037}{Phys.\ Lett.\
  \textbf{B757} (2016) 520},
  \href{http://arxiv.org/abs/1602.07430}{{\normalfont\ttfamily
  arXiv:1602.07430}}, [Erratum:
  \href{http://www.sciencedirect.com/science/article/pii/S0370269316306840}{Phys.
  Lett. \textbf{B765} (2017) 402}]\relax
\mciteBstWouldAddEndPuncttrue
\mciteSetBstMidEndSepPunct{\mcitedefaultmidpunct}
{\mcitedefaultendpunct}{\mcitedefaultseppunct}\relax
\EndOfBibitem
\end{mcitethebibliography}

\newpage
\newpage
\centerline{\large\bf LHCb collaboration}
\begin{flushleft}
\small
R.~Aaij$^{40}$,
B.~Adeva$^{39}$,
M.~Adinolfi$^{48}$,
Z.~Ajaltouni$^{5}$,
S.~Akar$^{59}$,
J.~Albrecht$^{10}$,
F.~Alessio$^{40}$,
M.~Alexander$^{53}$,
A.~Alfonso~Albero$^{38}$,
S.~Ali$^{43}$
% ...

{\footnotesize \it
% ...
$ ^{5}$Clermont Universit{\'e}, Universit{\'e} Blaise Pascal, CNRS/IN2P3, LPC, Clermont-Ferrand, France\\
% ...
$ ^{10}$Fakult{\"a}t Physik, Technische Universit{\"a}t Dortmund, Dortmund, Germany\\
% ...
$ ^{38}$ICCUB, Universitat de Barcelona, Barcelona, Spain\\
$ ^{39}$Instituto Galego de F{\'\i}sica de Altas Enerx{\'\i}as (IGFAE), Universidade de Santiago de Compostela, Santiago de Compostela, Spain\\
$ ^{40}$European Organization for Nuclear Research (CERN), Geneva, Switzerland\\
% ...
$ ^{43}$Nikhef National Institute for Subatomic Physics, Amsterdam, The Netherlands\\
% ...
$ ^{48}$H.H. Wills Physics Laboratory, University of Bristol, Bristol, United Kingdom\\
% ...
$ ^{53}$School of Physics and Astronomy, University of Glasgow, Glasgow, United Kingdom\\
% ...
$ ^{59}$University of Cincinnati, Cincinnati, OH, United States\\
% ...
}
\end{flushleft}

\end{document}